\documentclass[%
 aip,jcp,
 amsmath,amssymb,
 reprint,floatfix%
]{revtex4-1}
\usepackage[colorlinks=true,linkcolor=blue]{hyperref}
\usepackage{graphicx}
\usepackage{dcolumn}
\usepackage{bm}
\usepackage[version=3]{mhchem} 
\usepackage{dsfont}
\usepackage[utf8]{inputenc}
\usepackage[T1]{fontenc}
\usepackage[version=3]{mhchem}
\usepackage[caption=false]{subfig}
\usepackage{simplewick}

\begin{document}

\preprint{AIP/123-QED}

\title[effective QED]{4-component relativistic Hamiltonian with effective QED potentials for molecular calculations}

\author{Ayaki Sunaga}
\affiliation{Institute for Integrated Radiation and Nuclear Science, Kyoto University, 2, Asashiro-Nishi, Kumatori-cho, Sennan-gun, Osaka 590-0494 Japan}%

\author{Maen Salman}

\author{Trond Saue}%
\altaffiliation[Author to whom correspondence should be addressed: ]{trond.saue@irsamc.ups-tlse.fr.}
\homepage{http://dirac.ups-tlse.fr/saue}
\affiliation{Laboratoire de Chimie et Physique Quantique, UMR 5626 CNRS
--- Universit{\'e} Toulouse III-Paul Sabatier, 118 route de Narbonne,
F-31062 Toulouse, France}

\date{\today}

\begin{abstract}
  We report the implementation of effective QED potentials for all-electron 4-component relativistic molecular calculations using the DIRAC code.
  The potentials are also available for 2-component calculations, being properly picture-change transformed. The latter point is important; we demonstrate
  through atomic calculations that picture-change errors are sizable. Specificially, we have implemented the Uehling potential [E. A. Uehling, Phys. Rev. 48 , 55 (1935)]
  for vacuum polarization and two effective potentials [P. Pyykkö and L.-B. Zhao, J. Phys. B 36 , 1469 (2003); V. V. Flambaum and J. S. M. Ginges, Phys. Rev. A 72 , 052115 (2005)]
  for electron self-energy. We provide extensive theoretical background for these potentials, hopefully reaching an audience beyond QED-specialists.
  We report the following sample applications: i) we confirm the conjecture of P. Pyykk{\"o} that QED effects are observable for the \ce{AuCN} molecule by directly calculating ground-state
  rotational constants $B_0$ of the three isotopomers studied by MW spectroscopy; QED brings the corresponding substitution Au-C bond length $r_s$ from 0.23 to 0.04 pm agreement with experiment, 
  ii) spectroscopic constants of van der Waals dimers \ce{M2} (M=Hg, Rn, Cn, Og): QED induces bond length expansions on the order of
  0.15(0.30) pm for row 6(7) dimers,
  iii) we confirm that there is a significant change of valence $s$ population of \ce{Pb} in the reaction \ce{PbH4 -> PbH2 + H2}, which is thereby a good candidate for observing QED effects
  in chemical reactions, as proposed in [K. G. Dyall \textit{et al.}, Chem. Phys. Lett. 348 , 497 (2001)]. We also find that whereas in \ce{PbH4} the valence $6s_{1/2}$ population
  resides in bonding orbitals, it is mainly found in non-bonding orbitals in \ce{PbH2}. QED contributes 0.32 kcal/mol to the reaction energy, thereby reducing its magnitude by -1.27 \%.
  For corresponding hydrides of superheavy flerovium, the electronic structures are quite similar. Interestingly, the QED contribution to the reaction energy is of quite similar magnitude (0.35 kcal/mol), whereas the relative change
  is significantly smaller (-0.50 \%). This curious observation can be explained by the faster increase of negative vacuum polarization over positive electron self-energy contributions
  as a function of nuclear charge.
\end{abstract}

\maketitle

\section{INTRODUCTION}\label{sec:level1}

Relativistic quantum chemistry is the proper framework for the theoretical study of heavy elements. \cite{dyall2007introduction,reiher2014relativistic,saue2011relativistic,pyykko2012relativistic,schwerdtfeger_nucphys2015} For example, the yellow color of gold, \cite{pyykko1979relativity,pyykko1988relativistic} as well as the cell potential  of the lead-acid battery\cite{ahuja2011relativity} cannot be explained without relativistic effects. Even for light elements, the fine structure of spectra is essentially due to spin-orbit (SO) interaction (e.g. Refs. \citenum{eliav1994relativistic,dutta2012ab,shee2018equation}).

Improvements in both computational power and methodology nowadays allow highly accurate electronic structure calculations including both relativistic and electron correlation effects. A next challenge for increased accuracy is the inclusion of the effects of quantum electrodynamics (QED), which in principle means going beyond the no-pair approximation.\cite{saue2003four,liu2013going,schwerdtfeger_nucphys2015} We focus on QED effects generating the Lamb shift, roughly described as follows:
\begin{itemize}
\item vacuum polarization (VP): a charge in space is surrounded by virtual electron-positron pairs and this contributes to its observed charge
\item the electron self-energy (SE): the electron drags along its electromagnetic field and this contributes to its observed mass.  
\end{itemize}
For hydrogen the splitting between the \ce{^{2}S_{1/2}} and \ce{^{2}P_{1/2}} states is a mere 4 meV,\cite{lamb_rutherford_1947} but for \ce{U^{91+}} it has grown to a whopping 468 eV.\cite{stohlker2000} 
QED effects would possibly constitute the final correction to chemistry concerning the fundamental inter-particle interactions because the next contribution, parity non-conservation (PNC) associated with the weak force, is typically ten orders of magnitude smaller.\cite{pyykko2012physics} The magnitude of QED effects has been estimated based on the ionization potential of alkali atoms, and the \textit{rule of thumb} is that QED effects reduce  relativistic effects by about one percent.\cite{pyykko1998estimated}

Calculations within the rigorous QED framework have been reported for few-electron systems, and they are in excellent agreement with experiment. Examples are the Lamb shift of Li-like uranium,\cite{persson1993new,persson1993accurate} the hyperfine coupling constant (HFCC) of few-electron atoms,\cite{puchalski2013ground,haidar2020nonrelativistic} and the anomalous g factor,\cite{aoyama2012tenth} that provides stringent tests of the accuracy of QED.

The rigorous QED approach for few-electron systems cannot be extended to many-electron systems because of the high computational cost. A more practical, but approximate approach is the introduction of effective QED potentials (effQED).\cite{uehling1935,pyykko_zhao2003,flambaum_ginges2005,shabaev_pra2013,Malyshev_PhysRevA.106.012806} In the atomic case, some codes for the calculation with effective potentials have been reported (e.g., GRASP,\cite{dyall1989grasp} QEDMOD,\cite{shabaev2015qedmod,shabaev2018qedmod} and AMBiT\cite{kahl2019ambit}). A nice illustration is the recent work by Pa\v steka and co-workers,\cite{pasteka_prl2017} which was finally able to bring the calculated ionization potential (IP) and the electron affinity (EA) of the gold atom into meV agreement with experiment, high-order electron correlation being the missing crucial ingredient.

For the case of molecules in chemistry, pioneering works have been done by Kirk Peterson’s group. They added the following parameterized model potentials to the all-electron scalar Douglas--Kroll--Hess (DKH)  Hamiltonian:\cite{shepler2005ab,shepler2007hg+,peterson2015correlation,cox2016,feng2017correlation} i) an effective SE potential in the form of a single Gaussian function, proposed by Pyykk\"{o} and Zhao (PZ),\cite{pyykko_zhao2003} and ii) five Gaussian functions fitted by Peterson’s group \cite{shepler2005ab,shepler2007hg+,peterson2015correlation} to a parameterized expression for the Uehling VP potential\cite{uehling1935} given in Ref.~\citenum{pyykko1998estimated} and corrected in Ref.\citenum{pyykko_zhao2003}. They then found that the QED effects on the dissociation energy are about 0.6 kcal/mol and 0.4 kcal/mol in closed- and open-shell Hg systems, respectively.\cite{shepler2005ab} A bond length expansion of 0.001 {\AA}  was observed for the \ce{HgBr} molecule.\cite{shepler2007hg+} Michael Dolg and co-workers have reported pseudopotentials (PPs) fitted to include QED effects.\cite{hangele2012accurate,hangele2013relativistic,hangele2014coupled} For \ce{Cn2} a bond length expansion due to QED of about 0.003 {\AA}  was reported,\cite{hangele2014coupled} in line with the effect observed by Peterson's group.\cite{shepler2007hg+} On the other hand, in Ref.~\citenum{hangele2013accuracy}, the QED effect was found to \textit{shorten} the bond length of \ce{TsH+}, \ce{LvH} and \ce{OgH+}. The reason for this opposite trend may be that the valence orbitals have p- rather than s-orbital contributions from the heavy atom.

PPs are widely used for the inclusion of relativistic effects, and generally  give accurate results for valence properties compared with all-electron calculations.\cite{Schwerdtfeger_CPC2011,dolg_cao_cr2011} However, the PP approach cannot be applied to molecular core-properties such as NMR and M{\"o}ssbauer parameters, which bars the possibility to investigate the effect of QED in the nuclear region where such effects are generated.\cite{artemyev_2016} The effQED approach promoted by the Peterson group can in principle be applied to core-properties, but it should be noted that effQED potentials were added to approximate one-component relativistic Hamiltonians without picture-change.\cite{schwerdtfeger_snijders_1990, kello1998picture,dyall_2000} To include QED effects in a more rigorous manner, it seems more appropriate to include effective QED potentials in 4-component relativistic all-electron calculations.

In this work, we report the implementation of effective QED potentials in the DIRAC code for relativistic molecular calculations.\cite{saue2020dirac} Three potentials have been implemented: the Uehling potential \cite{uehling1935} for vacuum polarization, Pyykk\"{o} and Zhao's model SE  potential,\cite{pyykko_zhao2003} as well as the effective SE potential of Flambaum and Ginges.\cite{flambaum_ginges2005} Our implementation is based on numerical routines from the GRASP atomic code\cite{dyall1989grasp} that have been grafted onto the DFT grid of DIRAC.\cite{saue:jcc2002}

As first molecular applications of our implementation we have chosen three case studies:
\begin{itemize}
\item the \ce{AuCN} molecule for which Pekka Pyykk{\"o} has suggested QED effects on the bond length.\cite{pyykkotalk}
\item  the van der Waals dimers \ce{M2} (M = Hg, Rn, Cn, Og) for which one might suspect QED effects to be on par with interaction energies. Interestingly, van der Waals forces have been described in terms of vacuum fluctuations.\cite{genet_aflb2004,simpson_qvac}
\item  the reaction energy of Pb hydrides, \ce{PbH4 -> PbH2 + H2}, suggested by Dyall \textit{et al.} as a possible candidate for a significant QED effect in chemistry.\cite{dyall_cpl2001} In addition to the Pb system, we have also calculated the heavier analogue, Fl hydrides.
\end{itemize}

Very recently, Leonid Skripnikov reported the implementation of effective QED potentials for 4-component all-electron molecular calculations, so far with a focus on transition energies.\cite{skripnikov_jcp154.201101} The initial report has been followed by applications to \ce{Ba+}, \ce{BaF}, \ce{RaF} and E120F\cite{skripnikov_jcp155.144103} as well as the five low-lying excited states of \ce{RaF}.\cite{zaitsevskii2022jcp_RaF} The implementation is to some extent complementary to ours in that it uses the effective SE potential proposed by Shabaev and co-workers.\cite{shabaev_pra2013,shabaev2015qedmod,Malyshev_PhysRevA.106.012806} Interestingly, the implementation is based on the DIRAC code as well.

The paper is organized as follows: in Sec.~\ref{sec:theory} we review the effective QED potentials that we have implemented, and in Sec.~\ref{sec:impl} we discuss the numerical integration of these potentials. This is followed by Sec.~\ref{sec:compdet} which gives the computational details of our calculations. Our results are presented in Sec.~\ref{sec:results}, followed by conclusions in Sec.~\ref{sec:conc}. We also provide an appendix with more extensive theory and reading suggestions. SI-units are used throughout this paper.

\section{Theory}\label{sec:theory}
The starting point for our work is an electronic Hamiltonian on the generic form
\begin{equation}\label{eq:elham}
H=V_{\rm{NN}}+\sum_i H_D(\boldsymbol{x}_i)+\frac{1}{2}\sum_{i\ne j}g(\boldsymbol{x}_i,\boldsymbol{x}_j)  
\end{equation}
where $V_{\rm{NN}}$ is the classical repulsion of fixed nuclei. The one-electron part
is the Dirac Hamiltonian
\begin{equation}
H_D(\boldsymbol{x}_i)=(\beta-\mathds{1}_{4})m_ec^2-i\hbar c\boldsymbol{\alpha}\cdot\boldsymbol{\nabla}_{i}-e\varphi_N(\boldsymbol{x}_i),
\end{equation}
in the electric potential $\varphi_N$ of the fixed nuclei and shifted by $-m_ec^2$ to align energies with the non-relativistic scale. In the present work, the two-electron interaction $g$ will be the instantaneous Coulomb term supplemented with the Gaunt term.\cite{gaunt:term} Further discussion of the resulting Dirac--Coulomb--Gaunt (DCG) Hamiltonian is for instance found in Ref.~\citenum{saue2011relativistic}.

Our goal is to introduce QED effects, notably electron self-energy (SE) and vacuum polarization (VP), by extending the one-electron Hamiltonian by the corresponding effective QED potentials
\begin{equation}
\begin{aligned}H_{D} & \rightarrow H_{D}-e\varphi_{{\rm {effQED}}};\\
\varphi & _{{\rm {effQED}}}=\sum_{A}\left(\varphi_{A}^{\text{SE}}+\varphi_{A}^{\text{VP}}\right).
\end{aligned}
\end{equation}
Note that the effective QED potentials are formulated as a sum over atomic contributions due to their expected short-range nature (on the order of a reduced Compton wavelength $\lambdabar=\hbar/m_ec$).\cite{artemyev_2016}

In the following we shall present the effective QED potentials selected for our implementation
with some remarks on their construction which may provide indications on their expected performance. We shall proceed within the $\hat{{\cal S}}$-matrix (scattering matrix) formalism of QED. Since we hope to address a wider audience than QED specialists, we provide a more extensive theoretical background in Appendix \ref{section:appendix}. 

\begin{figure}
\subfloat[\label{fig:SP-bound}Single-photon exchange.]{
\includegraphics[scale=0.22,trim={1cm 1cm 1cm 1cm},clip]{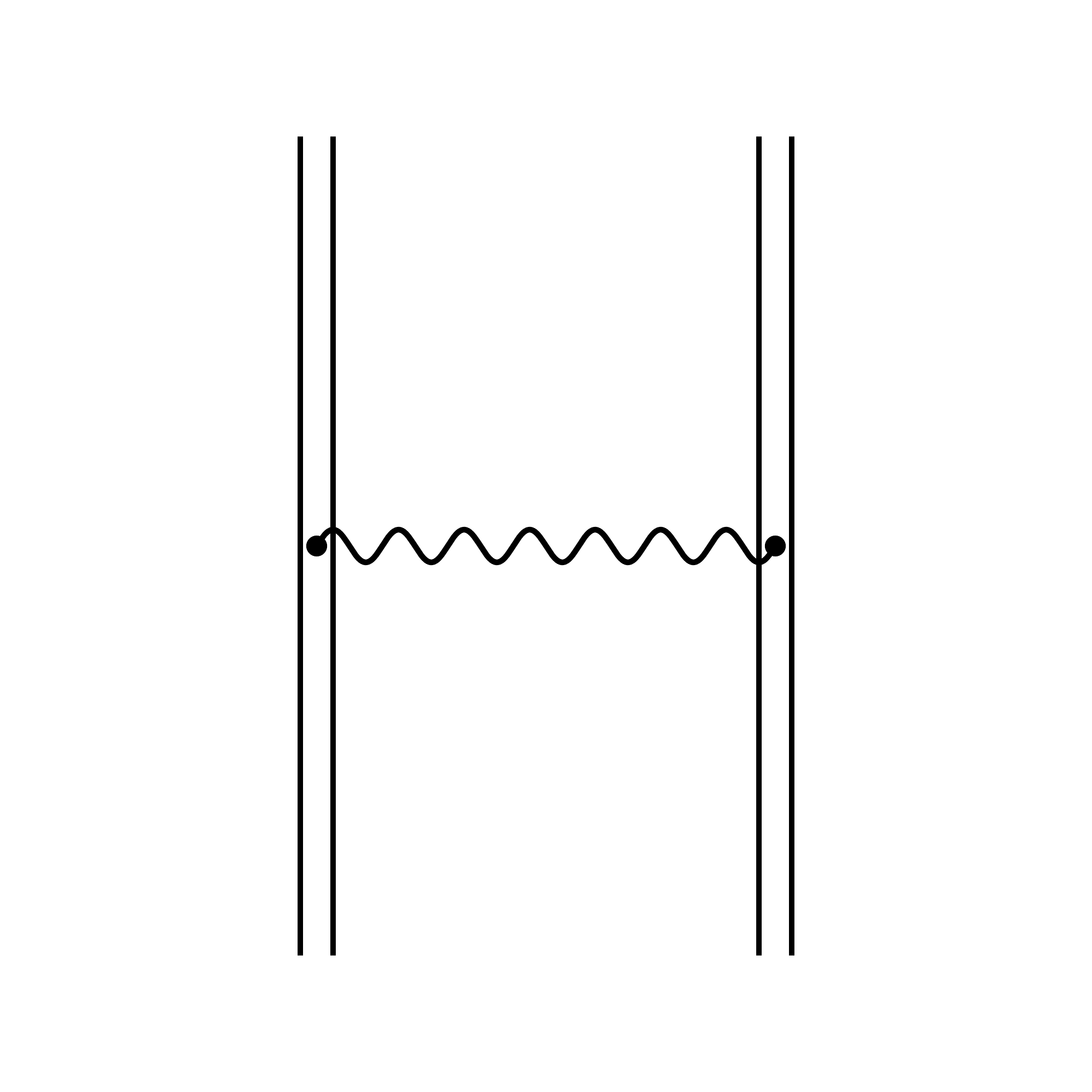}
\par}
\subfloat[\label{fig:VP-bound}Vacuum polarization.]{\includegraphics[scale=0.22,trim={1cm 1cm 1cm 1cm},clip]{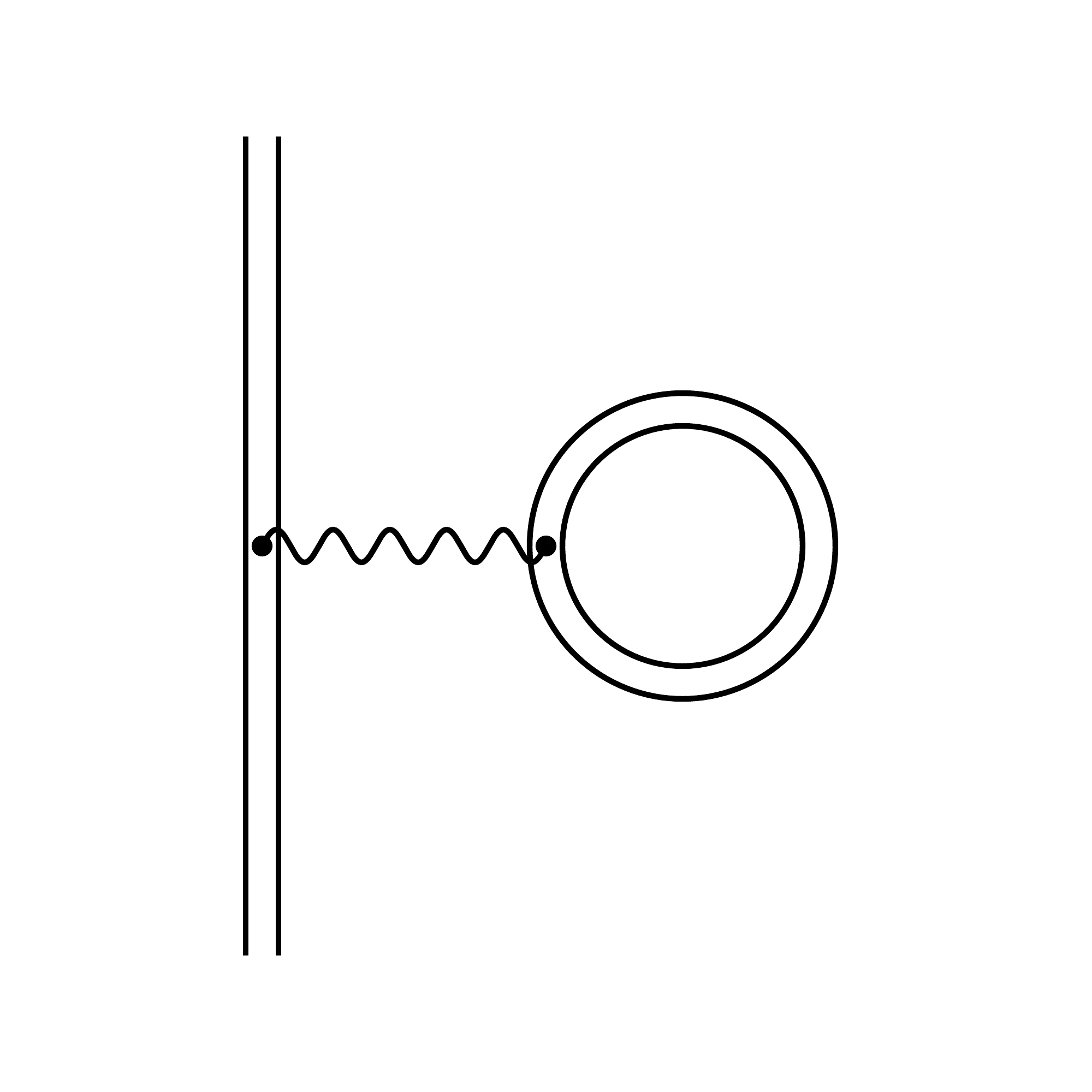}}\\
\subfloat[\label{fig:SE-bound}Self-energy.]{\includegraphics[scale=0.22,angle =90,trim={6cm 1cm 4cm 1cm},clip]{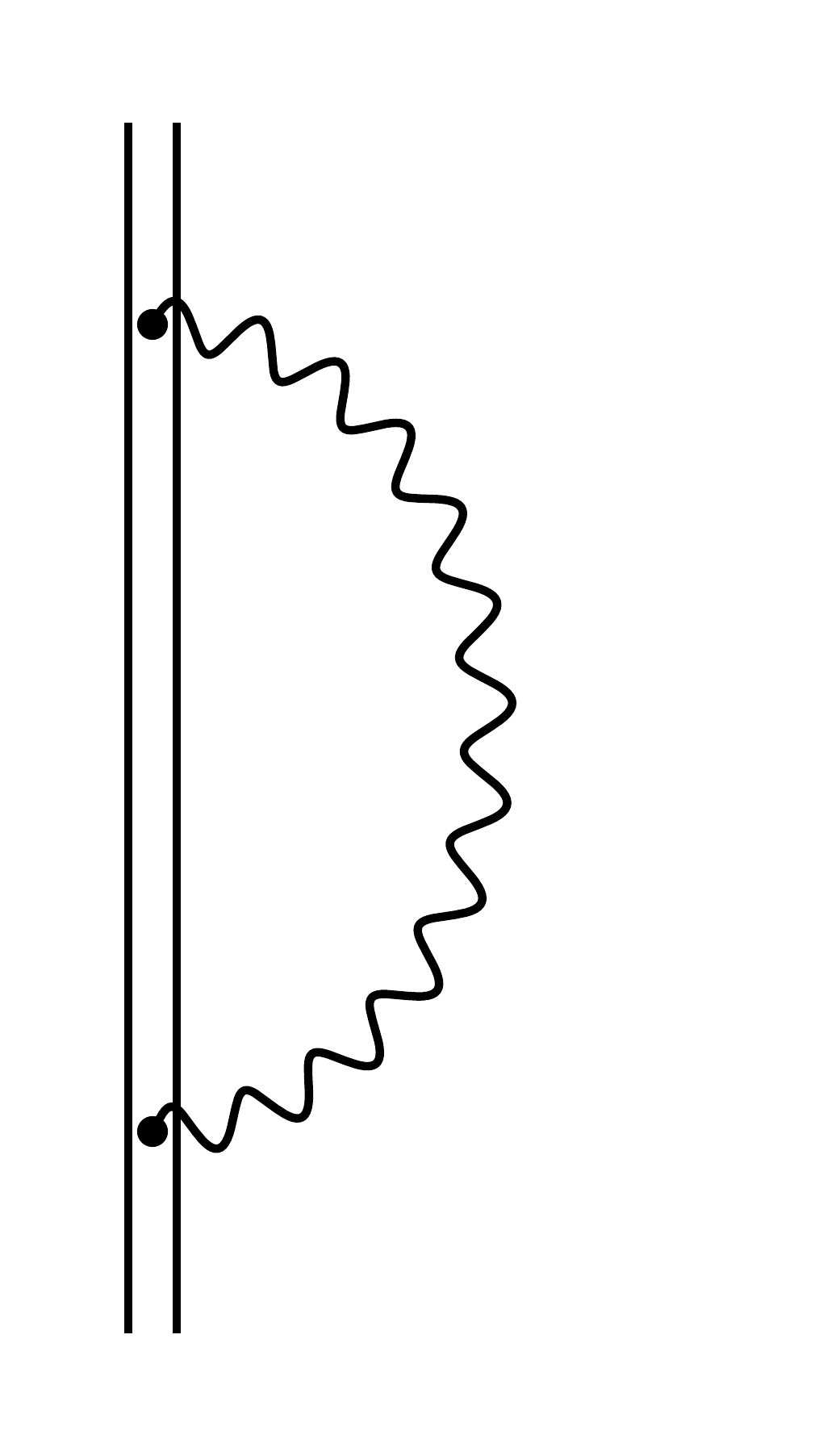}}
\par
\subfloat[\label{fig:V1-bound}Direct bubble diagram.]
{\includegraphics[scale=0.22,trim={1cm 5cm 1cm 5cm},clip]{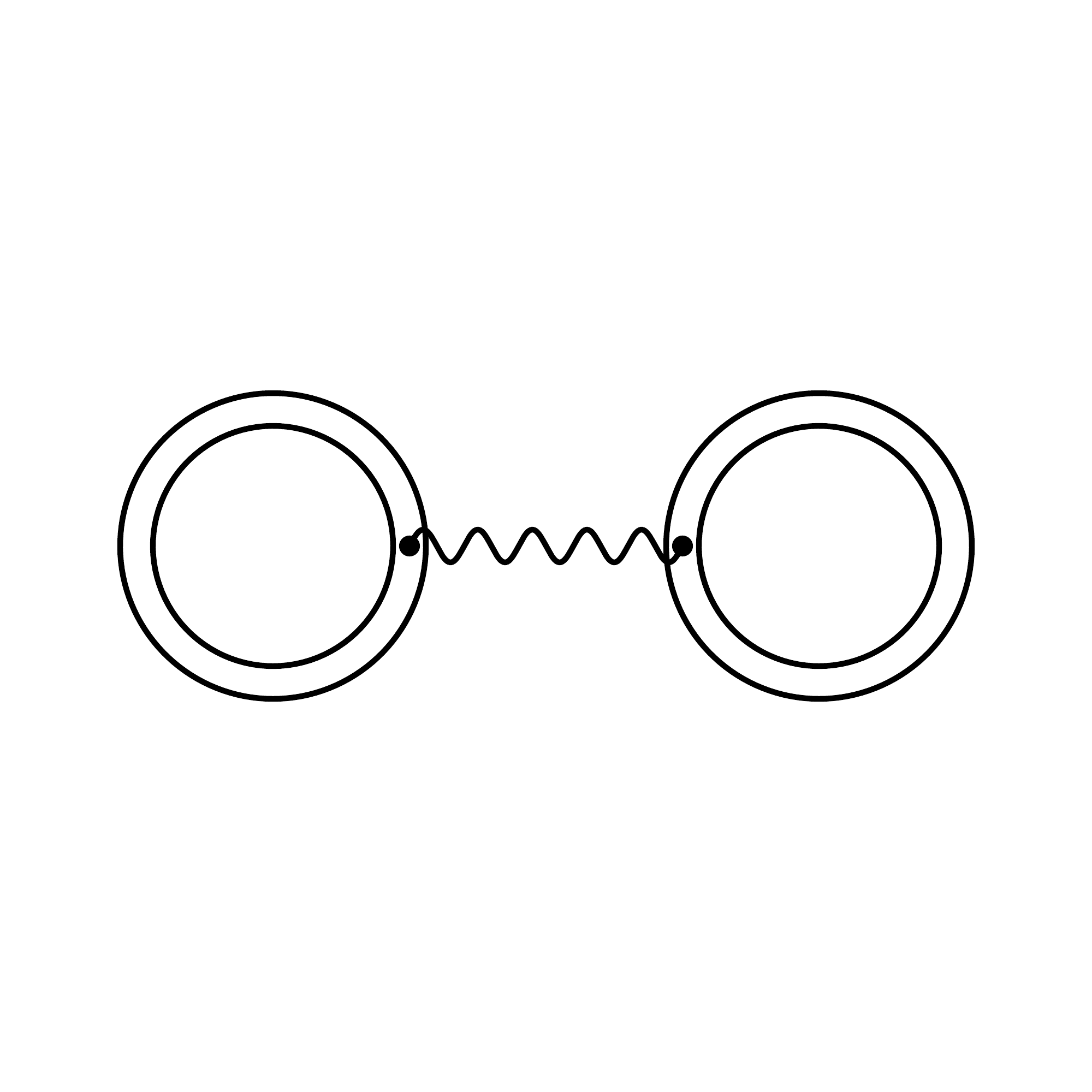}}
\subfloat[\label{fig:V2}Exchange bubble diagram.]
{\includegraphics[scale=0.22,trim={1cm 5cm 1cm 5cm},clip]{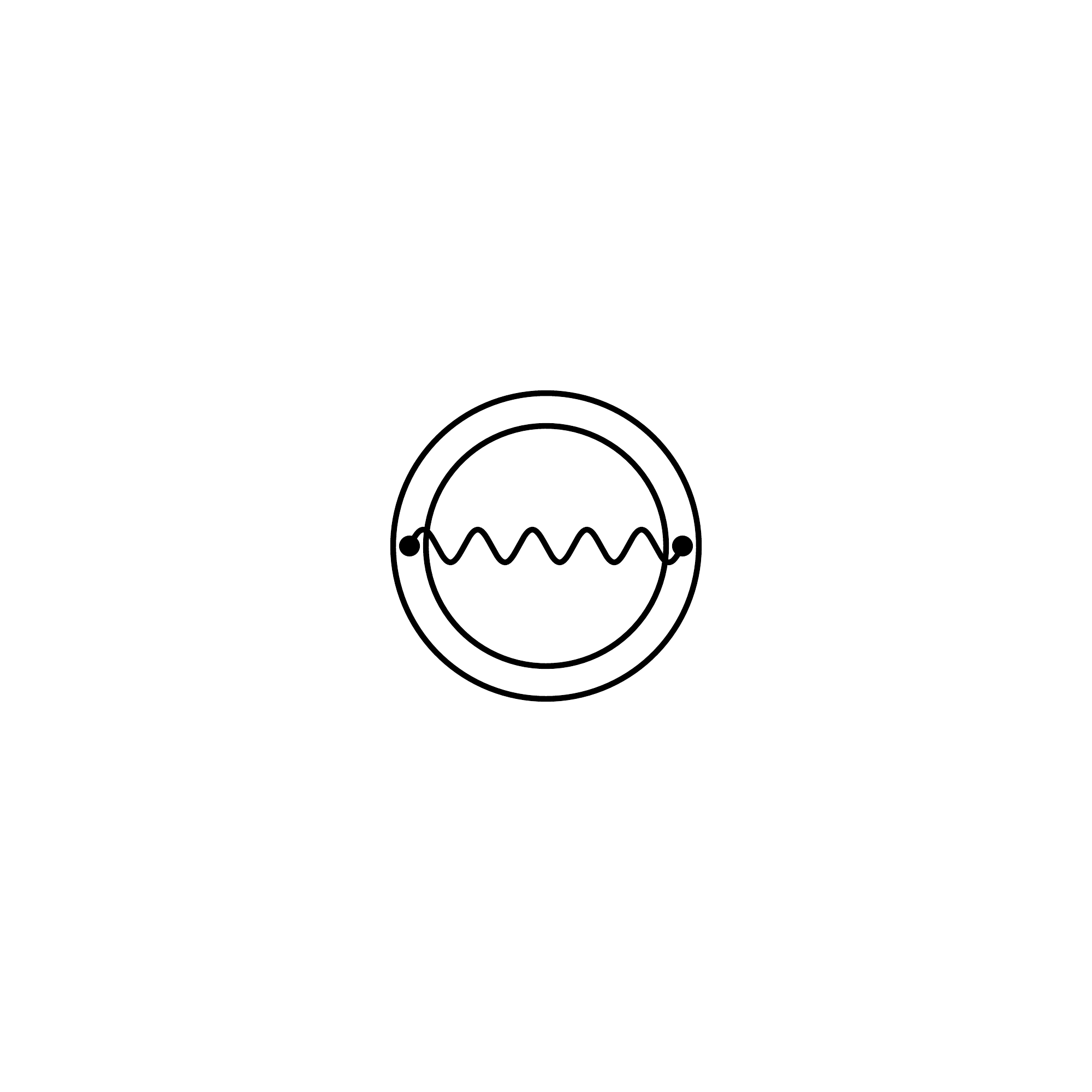}}
\caption{\label{fig:bound}The lowest-order QED corrections for a many-bound-electron
system: of order $\alpha$.}
\end{figure}

QED is the relativistic quantum field theory that describes the interaction
of electromagnetic radiation with relativistic matter (Dirac electrons). 
The interaction between electrons and photons is given by an interaction
Hamiltonian density 
\begin{equation}
\hat{{\cal H}}_{I}\left(x\right)=-ec\bar{\hat{\Psi}}\left(x\right)\gamma^{\mu}\hat{\Psi}\left(x\right)\hat{A}_{\mu}\left(x\right).\label{eqn:H_I}
\end{equation}
Here, $\hat{\Psi}\left(x\right)$ and $\bar{\hat{\Psi}}\left(x\right)$
are the quantized Dirac field operator and its corresponding adjoint, whereas $\hat{A}_{\mu}\left(x\right)$ is the quantized photon field operator. The job of these operators is to create and annihilate, at the spacetime point $x=(ct,\boldsymbol{x})$, electrons and photons, respectively. This last expression accounts (explicitly) for the coupling between electron and photon fields, and is obtained through minimal substitution of the four-gradient of the Dirac Lagrangian density, in accordance with the \textit{principle of minimal electromagnetic interaction} (term coined by Gell--Mann \cite{gell1956interpretation}). For detailed derivations and discussions, the reader may consult Schweber in Ref.~\citenum{schweber_rqft} (chapter 10), Peskin and Schroeder in Ref.~\citenum{peskin:schroeder} (chapter 4), as well as Greiner and Reinhardt in Ref.~\citenum{greiner_fieldquantization} (section 8.6).
The scattering matrix is the special case of the time-evolution operator $\hat{U}(t,t_0)$, where the initial $t_0$ and final times $t$ are at $\mp\infty$, to ensure Lorentz invariance.
Upon expansion of the $\hat{{\cal S}}$-matrix operator in the fundamental charge $e$, the $n$th-order term $\hat{{\cal S}}^{\left(n\right)}$ contains a time-ordered string of $n$ interaction Hamiltonian densities ${\cal H}_{I}$, as seen in  Eq.~\eqref{eqn:s-matrix-expansion}.
Using Wick's theorem,\cite{wick_physrev1950} a time-ordered string is converted into
a linear combination of normal-ordered ones with all possible contractions, which in turn
can be translated into the iconic Feynman diagrams.\cite{kaiser_2005_feynman} We limit attention to systems of $n$ electrons
and zero photons (photon vacuum). The latter implies that any string of normal-ordered 
photon operators $\hat{A}_{\mu}\left(x\right)$ that is not fully contracted will vanish upon taking expectation values, such that the $\hat{{\cal S}}$-matrix expansion is effectively limited
to even-ordered contributions, associated with the fine-structure constant $\alpha=e^2/4\pi\varepsilon_0\hbar c$ as expansion parameter. To lowest order in $\alpha$ appears
five Feynman diagrams, shown in Fig. \ref{fig:bound}: two of them give state-independent energy-shifts and are usually ignored within a perturbative setting, whereas the remaining three represent electron self-energy, vacuum polarization and single-photon exchange. The latter diagram describes the relativistic electron-electron interaction, mediated by photons, to lowest order and is in line with the statement of Dirac:
\begin{quote}
  Classical electrodynamics, in its accurate (restricted) relativistic form,
  teaches us that the idea of an interaction energy between particles is only an    approximation and should be replaced by the idea of each particle emitting waves, which travel outward with a finite velocity and influence the other particles in passing over them.\cite{dirac1932relativistic}
\end{quote}  
In the diagrams of Fig. \ref{fig:bound} double electron lines appear to indicate that we are working within the Bound-State QED (BSQED) framework in which the Dirac field operators are expanded in solutions of the Dirac equation in some external (contravariant) four-potential: $A^e = (\varphi^e/c,\boldsymbol{A}^e)$ (Furry picture\cite{furry1951-picture}), rather than free-particle ones. In the atomic case, this provides us with a second perturbation expansion parameter $Z\alpha$, as will be seen in the next section.

\subsection{Effective QED potentials for vacuum polarization}

The four-potential associated with the vacuum polarization effect
can be written as
\begin{equation}
\varphi_{\text{VP}}^{\mu}\left(\boldsymbol{x}_{1}\right)=\frac{e}{2\pi i}\int_{C_{F}}dz\int d^{3}x_{2}\frac{\text{Tr}\left[\gamma^{\mu}G_{A^e}\left(\boldsymbol{x}_{2},\boldsymbol{x}_{2};z\right)\gamma^{0}\right]}{4\pi\epsilon_{0}\left|\boldsymbol{x}_{1}-\boldsymbol{x}_{2}\right|},\label{eqn:VP-potential}
\end{equation}
where the complex $z$-integral is to be evaluated along the Feynman contour $C_{F}$ that goes above and below positive- and negative-energy poles, respectively, of the bound electron Green's function $G_{A^e}$. This function is related to the bound electron propagator $S_{A^e}^{F}$ by Eq.~\eqref{eqn:propagator_green}.
This VP potential leads to the following vacuum polarization energy-shift
\begin{equation}
\Delta E_{\text{VP}}^{\alpha,2}=-e\sum_i\int d^{3}x\bar{\psi}_{i}\left(\boldsymbol{x}\right)\gamma_{\mu}\psi_{i}\left(\boldsymbol{x}\right)\varphi_{\text{VP}}^{\mu}\left(\boldsymbol{x}\right).
\end{equation}

From consideration of time-reversal symmetry, one can show that in
the case of a purely scalar external potential: $A^{e}=(\varphi^{e}/c,\mathbf{0})$, vector components of the vacuum polarization four-potential vanish
\begin{equation}
\varphi_{\text{VP}}^{\mu}\left(\boldsymbol{x}\right)=0\quad\text{for}\quad\mu=1,2,3.
\end{equation}
The bound Green's function $G_{A^e}$ can be written in terms of the free Green's function
$G_{0}$ and expanded in powers of the time-independent external
potential $A^e$ (hence $Z\alpha$ in the atomic case) as shown in Eq.~\eqref{eqn:green-expansion}.
As discussed in Sec.~\ref{subsec:VP:-Vacuum-polarization}, the
first non-vanishing term of this expansion is the one that is linear
in the external potential $A^e\left(\boldsymbol{x}\right)$ (the
one-potential term). The potential of Eq.~\eqref{eqn:VP-potential} is divergent (as seen in Sec.~\ref{subsec:VP:-Vacuum-polarization}) and calls for regularization and renormalization (see Sec.~\ref{subsubsec:reg-and-ren}). After employing these techniques, one can extract the physical contribution associated with this vacuum polarization effect, and, in the point nucleus problem,  represent it by the following scalar potential\cite{uehling1935}
\begin{equation}
\varphi_{\text{Ueh.}}^{\text{point}}\left(\boldsymbol{x}\right) =\frac{Ze}{4\pi\varepsilon_{0}r_{x}}\frac{2\alpha}{3\pi}K_{1}\left(\frac{2r_{x}}{\lambdabar}\right),
\label{eqn:Uehling}
\end{equation}
that corrects the classical Coulomb potential. Here, $r_{x}\equiv\left|\boldsymbol{x}\right|$ is the radial distance and expressed in terms of the function\cite{fullerton_rinker_1976}
\begin{equation}
K_{1}\left(x\right) =\int_{1}^{\infty}d\zeta e^{-x\zeta}\left(\frac{1}{\zeta^{2}}+\frac{1}{2\zeta^{4}}\right)\sqrt{\zeta^{2}-1}.\label{eq:K1}
\end{equation}
(see also Refs.~\citenum{klarsfeld_plb1977} and \citenum{frolov2012}). This potential is named after Uehling who first calculated it in 1935 for a point charge nuclear distribution (as indicated by the superscript ``point'').
The corresponding potential for an arbitrary nuclear distribution $\rho^{\text{nuc.}}$, normalized to one, is obtained by the following convolution\cite{fullerton_rinker_1976}
\begin{equation}
\varphi^{\text{nuc.}}_{\text{Ueh.}}\left(\boldsymbol{x}\right)=\int d^{3}\boldsymbol{y}{\,}\rho^{\text{nuc.}}\left(\boldsymbol{y}\right)\varphi^{\text{point}}_{\text{Ueh.}}\left(\boldsymbol{x}-\boldsymbol{y}\right).\label{eqn:Uehling_general}
\end{equation}

In the case of a spherically symmetric nuclear charge distribution, one obtains, after angular integration\cite{fullerton_rinker_1976}
\begin{equation}
\begin{aligned}\varphi_{\text{Ueh.}}^{\text{nuc.}}\left(\boldsymbol{x}\right) & =\frac{Ze}{4\pi\varepsilon_{0}r_{x}}\lambdabar \frac{2\alpha}{3}\int_{0}^{\infty}r_{y}dr_{y}\rho^{{\rm {nuc.}}}\left(r_{y}\right)\\
\times \big[K_{0}\big( & \frac{2}{\lambdabar}|r_{x}-r_{y}|\big)-K_{0}\big(\frac{2}{\lambdabar}|r_{x}+r_{y}|\big)\bigg],
\end{aligned}\label{eq:Uehling}
\end{equation}
where appears the function
\begin{equation}
  K_{0}\left(x\right)=\int_{1}^{\infty}d\zeta e^{-x\zeta} \left(\frac{1}{\zeta^{3}}+\frac{1}{2\zeta^{5}}\right)\sqrt{\zeta^{2}-1}.\label{eq:K0}
\end{equation}
The integral functions $K_0$ and $K_1$ are related through
\begin{equation}
K_1(x)=-\frac{d}{dx}K_0(x).
\end{equation}

The Uehling potential generally represents the dominant vacuum polarization effect.\cite{persson1993accurate,beiermohrpersson1998soff}
The Feynman diagram associated with this process is presented in Fig. \ref{fig:VP-bound-1}, and
associated with the $\alpha\left(Z\alpha\right)$ perturbation order.
The higher-order vacuum polarization potentials, associated with the
Wichmann--Kroll:\cite{wichmann_kroll1956} $\alpha\left(Z\alpha\right)^{3}$ and Källén--Sabry:\cite{kallen_sab_1955} 
$\alpha^{2}\left(Z\alpha\right)$ processes, are briefly discussed at the
end of Sec.~\ref{subsec:VP:-Vacuum-polarization}. 

\subsection{Effective QED potentials for self-energy}

The energy-shift associated with the self-energy process, in which
an electron emits and absorbs a virtual photon, is given by the following
expression
\begin{equation}
\begin{aligned}\Delta E_{\text{SE}}^{\alpha,2} & =-e\sum_{i}\int d^{3}x_{1}\int d^{3}x_{2}\\
 & \times\psi_{i}^{\dagger}\left(\boldsymbol{x}_{2}\right)\varphi_{\text{SE}}\left(\boldsymbol{x}_{2},\boldsymbol{x}_{1};E_{i}\right)\psi_{i}\left(\boldsymbol{x}_{1}\right)
\end{aligned}\label{eqn:SE-deltaE}
\end{equation}

This expression probably originated from the work of Baranger \textit{et al.} \cite{baranger_physrev1953} (section II). Notice at this point that unlike the vacuum polarization effect that is represented
by a local scalar potential, the self-energy effect is represented
by a non-local matrix potential
\begin{equation}
\begin{aligned}\varphi_{\text{SE}}\left(\boldsymbol{x}_{2},\boldsymbol{x}_{1};E_{i}\right) & =-\frac{e}{2\pi i}\int_{C_{F}}dz\alpha^{\mu}G_{A^{e}}\left(\boldsymbol{x}_{2},\boldsymbol{x}_{1};z\right)\alpha_{\mu}\\
 & \times\frac{\exp\big(g\left(\boldsymbol{x}_{2},\boldsymbol{x}_{1};z-E_{i}\right)\big)}{4\pi\epsilon_{0}\left|\boldsymbol{x}_{1}-\boldsymbol{x}_{2}\right|},\\
g\left(\boldsymbol{x}_{2},\boldsymbol{x}_{1};z\right) & =+\frac{i}{\hbar}\left|\boldsymbol{x}_{1}-\boldsymbol{x}_{2}\right|\sqrt{z^{2}/c^{2}+i\epsilon}.
\end{aligned}
\label{eqn:SE-potential-1}
\end{equation}

Here, $\epsilon$ is a small positive number, and the $z$-integral is again to be evaluated along the Feynman contour $C_{F}$. This expression is obtained using the covariant Feynman gauge photon propagator. The corresponding expression obtained using Coulomb gauge photon propagator is given by Lindgren in Ref. \citenum{lindgren_rmbt2016} (section 4.6.1.2) (See also Malenfant in Ref. \citenum{se-coulomb-gauge1987malenfant}). As in the vacuum polarization case, the self-energy potential of Eq.~\eqref{eqn:SE-potential-1} is divergent (as seen in Sec.~\ref{subsec:SE:-Self-energy}), and calls for a regularization and renormalization treatment in order extract the physical (finite) correction; see Sec.~\ref{subsubsec:reg-and-ren}. 

In the next two sections, we shall assume that the non-local potential of Eq.~\eqref{eqn:SE-potential-1} can be written in terms of a local effective potential $\varphi_{\text{SE}}(\boldsymbol{x}_1)$ as
\begin{equation}
\varphi_{\text{SE}}(\boldsymbol{x}_2,\boldsymbol{x}_1;E_i) \approx \varphi_{\text{SE}}(\boldsymbol{x}_1) \delta(\boldsymbol{x}_2-\boldsymbol{x}_1),
\end{equation}
and discuss some choices of $\varphi_{\text{SE}}(\boldsymbol{x}_1)$ that are designed to reproduce some precise self-energy correction calculations, and are employed in our numerical calculations.

\subsubsection{Pyykkö and Zhao SE potential}

In Ref. \citenum{pyykko_zhao2003}, Pyykkö and Zhao (PZ) proposed a simple local
self-energy potential, of the following form
\begin{equation}
\varphi_{\text{SE}}\left(\boldsymbol{x}\right)=Be^{-\beta r_x^{2}}.
\end{equation}
The parameters $B$ and $\beta$ are quadratic nuclear charge ($Z$) dependent functions
\begin{align}
B\left(Z\right) & =-48.6116+1.53666\,Z+0.0301129\,Z^{2}\\
\beta\left(Z\right) & =-12751.3+916.038\,Z+5.7797\,Z^{2}
\end{align}
where the six decimal numbers were chosen to fit precise $29\leq Z\leq83$
atomic calculations of the renormalized self-energy correction in
all orders of $\left(Z\alpha\right)^{n\geq0}$ to the:
\begin{enumerate}
\item 2s energy-levels of the hydrogen-like systems, i.e., the renormalized
version of Eq.~\eqref{eqn:SE-deltaE}, taken from calculations
of 1) Beier \textit{et al.}\cite{beiermohrpersson1998soff}  with nuclear charges
$26\leq Z\leq110$, using a homogeneously charged sphere nuclear model,
and 2) Indelicato and Mohr\cite{indelicatomohr1998calculation} with Coulombic nuclear charges of $5\leq Z\leq90$.
\item M1 hyperfine splitting for
  lithium-like atoms taken from calculations\footnote{The $-\delta$ in Table 1 of Ref.\citenum{pyykko_zhao2003} corresponds to $100\times SE_{Li}/\Delta E^{Hfs}_{Li}$
  with values taken from Table 2 of Ref.\citenum{boucardindelicato2000}, where $SE_{Li}$ in turn comes from Ref.~\citenum{Blundell_PhysRevA.55.1857}} of Boucard and Indelicato\cite{boucardindelicato2000}
done on stable isotopes with $3\leq Z\leq92$.
\end{enumerate}

\subsubsection{Flambaum and Ginges SE potential}

The starting point for the potential proposed by Flambaum and Ginges (FG)  \cite{flambaum_ginges2005} is associated with the one-potential bound-state self-energy process, of order $\alpha(Z\alpha)$, given in Eqs.~\eqref{eqn:SE-1-potential} and \eqref{eqn:S-mat-vertex} and 
represented by Fig. \ref{fig:SE-bound-1}. However, further modeling, including
parametrization, is introduced such that the potential can account for the full self-energy process to all orders in $\left(Z\alpha\right)$ and be used in atomic calculations.

\begin{figure}
\subfloat[\label{fig:scattering-process}Classical scattering process.]{
\includegraphics[scale=0.23,trim={1cm 1cm 1cm 2.0cm},clip]{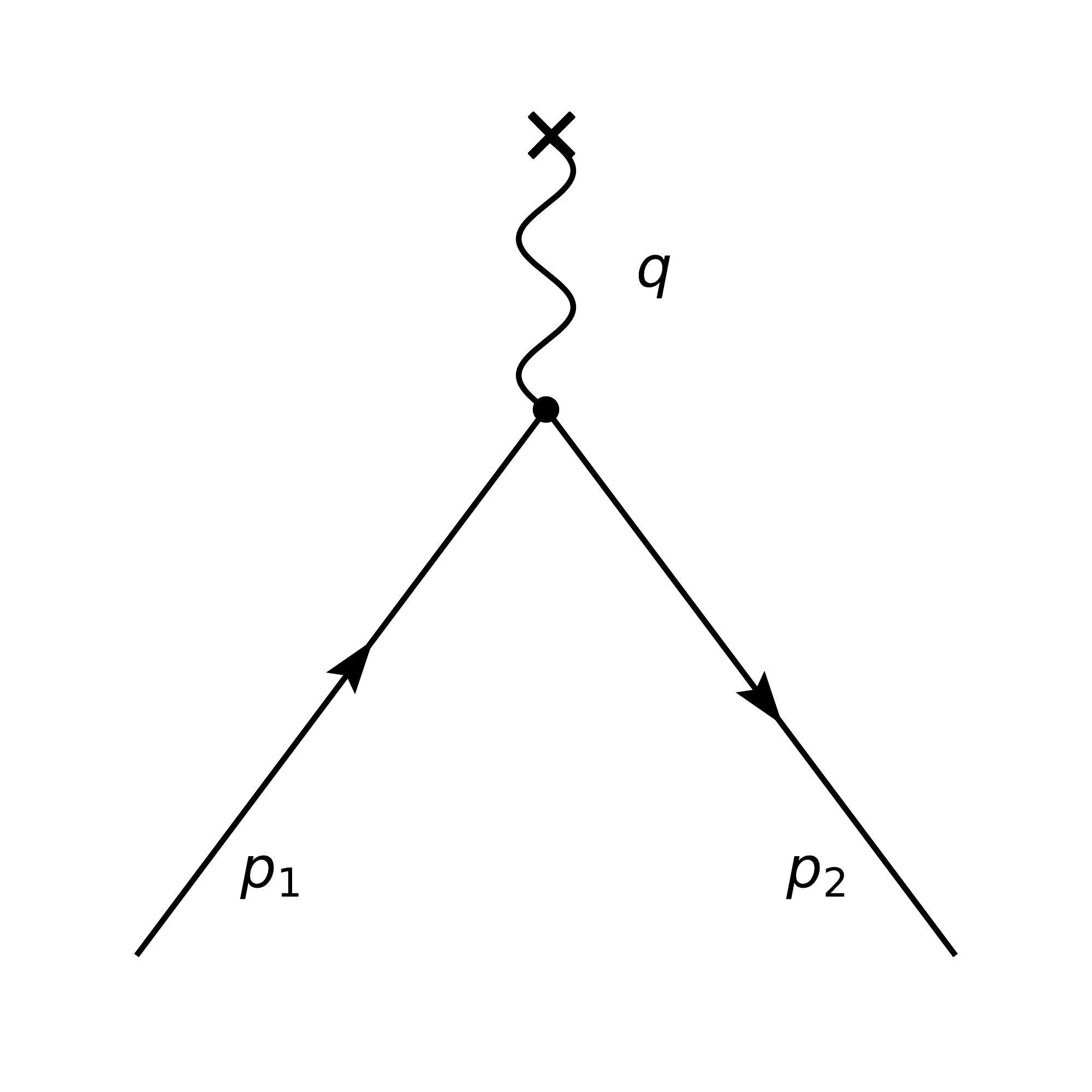}
\par}
\subfloat[\label{fig:vertex-correction}First radiative correction.]{\includegraphics[scale=0.23,trim={1cm 1cm 1cm 2cm},clip]{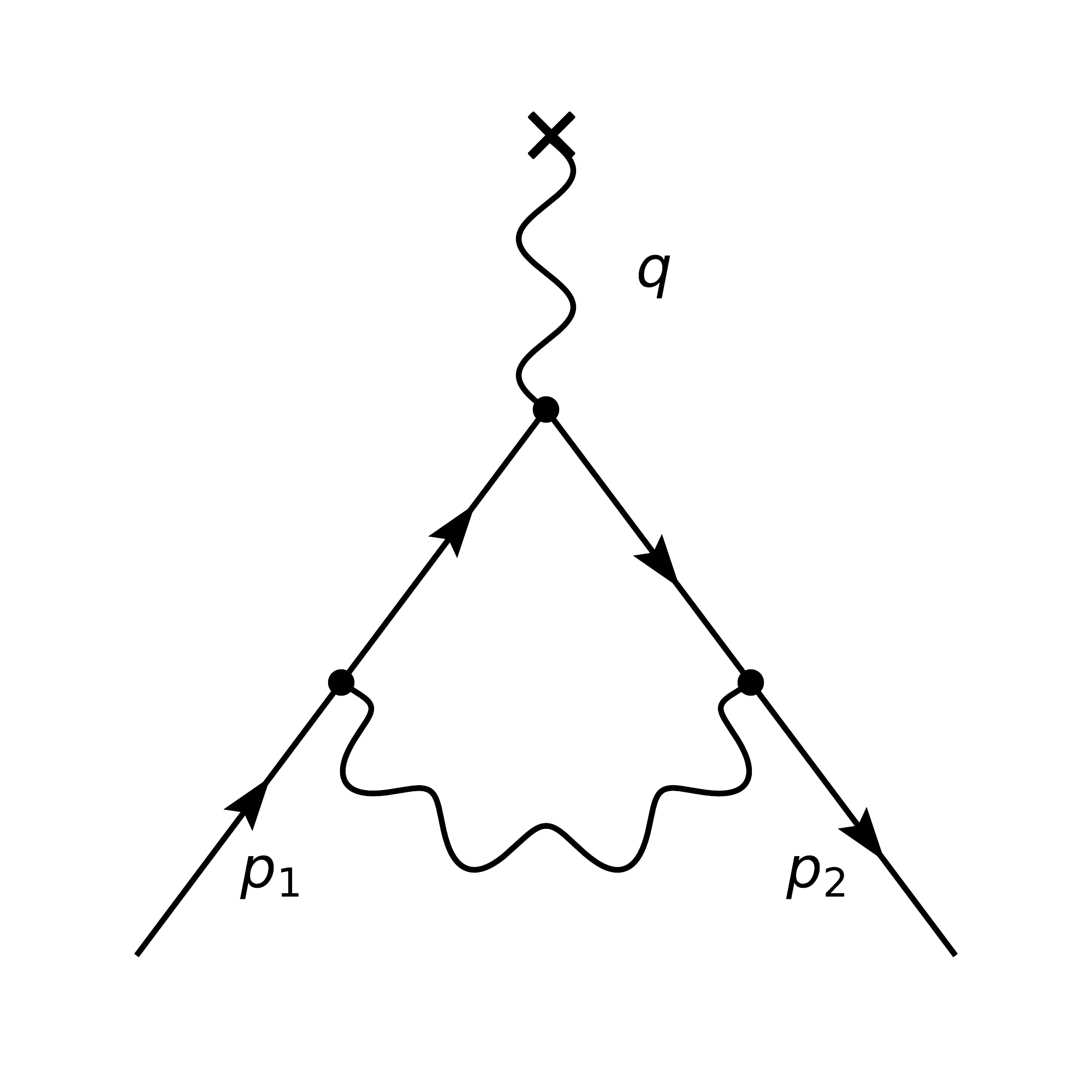}}
\caption{\label{fig:Vertex-diagrams} Momentum-space Feynman diagrams for the lowest-order scattering processes through (the  exchange of) momentum-transfer $q=p_2-p_1$.}
\end{figure}

In the evaluation of matrix elements over the operator of Eq.~\eqref{eqn:S-mat-vertex}, Flambaum and Ginges
employ free-particle solutions rather than atomic bound orbitals. 
This replacement yields the free-electron vertex-correction (VC) problem.
This terminology can be understood from consideration of the scattering of a free electron due to the interaction with a classical external potential (the vertex process).
In terms of momentum-space quantities (cf. Eq.\eqref{eq:Fourier}), including \textit{free} electron field operators, the corresponding (non-radiative) $\cal S$-matrix is given by
\begin{equation}
\begin{aligned}\hat{{\cal S}}_{\text{scattering}}^{\left(1\right)} & =-\frac{e}{i\hbar}\int\frac{d^{4}p_{2}}{\left(2\pi\hbar\right)^{4}}\int\frac{d^{4}p_{1}}{\left(2\pi\hbar\right)^{4}}\\
 & \quad\times:\bar{\hat{\Psi}}\left(p_{2}\right)\gamma^{\mu}A^e_{\mu}(p_{2}-p_{1})\hat{\Psi}\left(p_{1}\right):
\end{aligned}\label{eqn:vertex-process}
\end{equation}
(see for instance section 8.7 in Ref.~\citenum{mandl_qft2010}).
This process is represented in the left panel of Fig. \ref{fig:Vertex-diagrams}, where the wiggly line ending with a cross $\times$ describes an interaction of a free electron with the classical external potential source through the exchange of a four-momentum $q=p_2-p_1$. We note that in general a factor ($-e\gamma^\mu$) is associated with each spacetime point (vertex). The right panel of Fig. \ref{fig:Vertex-diagrams} represents one of the four lowest-order radiative
corrections to the left panel process. The corresponding $\cal S$-matrix can be combined with the one of Eq.~\eqref{eqn:vertex-process} through the substitution
\begin{equation}
\gamma^{\mu}\rightarrow\Gamma^{\mu}=\gamma^{\mu}+\Lambda^{\mu}(p_{2},p_{1}),
\end{equation}
where the vertex-correction function $\Lambda^{\mu}(p_{2},p_{1})$ is given by
Eq.\eqref{eqn:Gammap2p1}.
After a careful treatment of the divergence when $q=0$, as done in Refs.~\citenum{peskin:schroeder} (section 6.3) and \citenum{berestetskii:qed1982} (section 117), one obtains the regularized (physical) vertex-correction function  $\Lambda_{R}^{\mu}\left(p_{2},p_{1}\right)$.
Furthermore, using the fact that the vertex function is sandwiched
between free-electron (on-mass-shell) field operators, one can show
that this function can be written as
\begin{equation}
\Lambda_{R}^{\mu}\left(q\right)=\gamma^{\mu}F_{1}\left(q^{2}\right)+\frac{i}{2m_e c}\sigma^{\mu\nu}q_{\nu}F_{2}\left(q^{2}\right),\label{eqn:vertex-correction-regularized}
\end{equation}
where  $\sigma^{\mu\nu}=\frac{i}{2}\left[\gamma^{\mu},\gamma^{\nu}\right]$, and $F_{1}$ and $F_{2}$ are known as the electric and magnetic form-factors, respectively (corresponding to $f-1$ and $g$ in
Eq.~(116.6) of Ref. \citenum{berestetskii:qed1982}). The term ``form factor'' comes from diffraction physics; see for instance Ref.~\citenum{hofstadter_RevModPhys.28.214}.

Since the free-electron vertex function of Eq.~\eqref{eqn:vertex-correction-regularized}
only depends on the momentum-transfer $q=p_{2}-p_{1}$, it conveniently yields a local potential in real space. This can be clearly
seen from the following relation
\begin{equation}
\begin{aligned} & \int\frac{d^{4}p_{2}}{\left(2\pi\hbar\right)^{4}}\int\frac{d^{4}p_{1}}{\left(2\pi\hbar\right)^{4}}\bar{\hat{\Psi}}\left(p_{2}\right)\Lambda_{R}^{\mu}\left(q\right)A^e_{\mu}\left(q\right)\hat{\Psi}\left(p_{1}\right)\\
 & =\frac{1}{c}\int d^{4}x\hat{\Psi}^{\dagger}\left(x\right)\varphi_{\text{VC}}\left(x\right)\hat{\Psi}\left(x\right).
\end{aligned}
\end{equation}
When the nucleus is described as a point charge the corresponding Coulomb potential,
\begin{equation}
A^e_0(q)=\delta(q_0)\frac{2\pi\hbar^3}{c\epsilon_0}\frac{Ze}{\boldsymbol{q}^2},  
\end{equation}
generates a vertex-correction potential of the form 
\begin{equation}
  \begin{aligned}\varphi_{\text{VC}}^{\text{point}}\left(\boldsymbol{x}\right) & =\frac{\hbar^{2}}{\epsilon_{0}}\int\frac{d^{3}q}{\left(2\pi\hbar\right)^{3}}e^{+\frac{i}{\hbar}\boldsymbol{q}\cdot\boldsymbol{x}}\frac{Ze}{\boldsymbol{q}^{2}}\\ & \times\left[F_{1}\left(-\boldsymbol{q}^{2}\right)+\frac{1}{2m_ec}\boldsymbol{\gamma}\cdot\boldsymbol{q}F_{2}\left(-\boldsymbol{q}^{2}\right)\right]\\ & =\varphi_{\text{elec}}^{\text{point}}\left(\boldsymbol{x}\right)+\varphi_{\text{mag}}^{\text{point}}\left(\boldsymbol{x}\right),
\end{aligned} \label{eqn:VC-potential-real} 
\end{equation}
which splits into electric and magnetic scalar potentials. We note that due to the time-independence of the Coulomb potential, the time-like part of the 4-momentum transfer $q=p_2-p_1$ vanishes.
In terms of the variable $t=q^2=-\boldsymbol{q}^{2}$, the form factors are Hermitian analytic functions,\cite{lang_pucker_2016} that is
\begin{equation}
  F(t)=F^\ast(t^\ast).
\end{equation}
This feature, combined with these functions being radial in terms of $\boldsymbol{q}$ and the clever use of complex analysis techniques, allowed Berestetskii \textit{et al.}  to express such functions in coordinate-space using only their imaginary parts in momentum-space
\begin{equation}
F\left(\boldsymbol{x}\right)=\frac{1}{\left(2\pi\hbar\right)^{2}r_x}\int_{4m_e^{2}c^{2}}^{\infty}dt\,\Im\left[F\left(t\right)\right]\exp\left[-\frac{1}{\hbar}r_x\sqrt{t}\right].\label{eq:B114.4}
\end{equation}
(see eq.(114.4) of Ref.\citenum{berestetskii:qed1982}).
It may be noted that the lower limit of integration over $t$ is $4m_e^{2}c^{2}$, corresponding to the threshold of pair creation.\cite{eden_prsa1952}  
Expressions for  the imaginary part of the form factors can be  found in Refs.~\citenum{berestetskii:qed1982} (eq.(117.14-15)) and \citenum{barbieri1972} (eq.(2.12))
\begin{align}
\Im\left[F_{1}\left(t\right)\right] & =\frac{\alpha}{\sqrt{t\left(t-4m_e^{2}c^{2}\right)}}\bigg[2m_e^{2}c^{2}-3t/4\label{eq:F1}\\
 & \qquad+\left(t/2-m_e^{2}c^{2}\right)\log\left(\frac{t-4m_e^{2}c^{2}}{\lambda^{2}}\right)\bigg],\nonumber\\
\Im\left[F_{2}\left(t\right)\right] & =\frac{\alpha m^{2}c^{2}}{\sqrt{t\left(t-4m_e^{2}c^{2}\right)}}\label{eq:F2}.
\end{align}
Building on the work of Berestetskii \textit{et al.},\cite{berestetskii:qed1982} Flambaum and Ginges have evaluated the integral of Eq.~\eqref{eqn:VC-potential-real}, and obtained the associated real-space potentials. After the variable substitution $t=4m_e^{2}c^{2}\zeta^2$, the magnetic potential was found to be
\begin{equation}
\varphi_{\text{mag}}^{\text{point}}\left(\boldsymbol{x}\right) =\frac{\alpha\hbar}{4\pi m_{e}c}i\boldsymbol{\gamma}\cdot\boldsymbol{\nabla}_{x}\big[\frac{Ze}{4\pi\varepsilon_{0}r_{x}}
\big(K_{m}\left(\frac{2r_{x}}{\lambdabar}\right)-1\big)\big],
\label{eq:FG_mag}
\end{equation}
where we have introduced the function
\begin{equation}
  K_m(x)=\int_{1}^{\infty}d\zeta\,\frac{e^{-x\zeta}}{\zeta^{2}\sqrt{\zeta^{2}-1}},
\end{equation}
which can be recognized as the 2nd Bickley--Naylor function ${\rm Ki}_2$ (cf. Ref.~\citenum{frolov2012}). Note that the same variable $\zeta$ is employed in the Uehling potential (cf. Eqs.~\eqref{eq:K1} and \eqref{eq:K0}).
The magnetic contribution  gives the first-order correction to the magnetic moment: the
anomalous magnetic moment of the electron, first calculated by Schwinger,
see for instance Mandl in Ref.~\citenum{mandl_qft2010} (section 10.5).

On the other hand, the electric form factor yields the electric effective potential
\begin{equation}
\begin{aligned} & 
\varphi_{\text{elec}}^{\text{point}}\left(\boldsymbol{x},\lambda\right)=-\frac{\alpha}{\pi}\frac{Ze}{4\pi\varepsilon_{0}r_x}K_{e}\left(\frac{2r_{x}}{\lambdabar}\right),
\end{aligned}\label{eq:FG_HF}
\end{equation}
where we have introduced the function
\begin{align}
K_{e}(x)&=\int_{1}^{\infty}d\zeta \frac{e^{-x\zeta}}{\sqrt{\zeta^{2}-1}}\bigg\{-\frac{3}{2}+\frac{1}{\zeta^{2}}\\
&+\left(1-\frac{1}{2\zeta^{2}}\right)\left[\ln\left(\zeta^{2}-1\right)+2\ln\left(\frac{2m_e c^{2}}{\lambda}\right)\right]\bigg\}.\nonumber
\end{align}  
These self-energy effective potentials where first derived with respect to a point nucleus (Coulomb potential), and the corresponding generalized expressions for an arbitrary normalized nuclear distribution $\rho^{\text{nuc.}}$ are obtained by convolution,\cite{ginges2016atomic} as in Eq.~\eqref{eqn:Uehling_general}.

The potential of Eq.~\eqref{eq:FG_HF} is called the high-frequency term, because it contains an energy parameter $\lambda$, already present in Eq.~\eqref{eq:F1}, that prevents the obtention of low-frequency
contributions. This parameter is associated with the introduction of a small fictitious photon mass, which needs to be plugged in the photon propagator denominator in order to make the divergent (at small momenta) momentum-space integral, associated with the vertex-correction, convergent. Details concerning this problem are discussed by Greiner and Reinhardt in Ref.~\citenum{greiner_reinhardt_qed} (eq.(5.91)), Itzykson and Zuber in Ref.~\citenum{itzykson_zuber_qft_1980} (eq.(7.45)), in addition to Peskin and Schroeder in Ref.~\citenum{peskin:schroeder} (pages 195,196). We note that the remaining divergence, occurring in the limit of zero photon mass, or $\lim_{\lambda \rightarrow 0}$ is overcome by taking into account the differential cross section associated with the Bremsstrahlung effect; detailed discussions are found in Refs.~\citenum{greiner_reinhardt_qed} (pages 311-313) and \citenum{peskin:schroeder} (section 6.4). 
Flambaum and Ginges choose a somewhat different strategy, which furthermore allows them to amend the fact that the used form factors are derived for the free-electron vertex-correction of order $\left(Z\alpha\right)^{n=1}$ only and now take into
account complementary self-energy corrections $\left(Z\alpha\right)^{n}, n\neq1$ (diagrams of Figs. \ref{fig:SE-bound-0}, \ref{fig:SE-bound-2}, \ref{fig:SE-bound-3}, and higher orders).
They write the high-frequency (HF) contribution as
\begin{equation}\label{FG_HF}
\varphi_{\text{HF}}^{\text{point}}\left(\boldsymbol{x}\right)=A\left(Z,\boldsymbol{x}\right)\varphi_{\text{elec}}^{\text{point}}\left(\boldsymbol{x},\lambda\right),
\end{equation}
where $A\left(Z,\boldsymbol{x}\right)$ is a fitting function and choose a $\lambda$-value that will minimize the low-frequency contribution. They argue that $\lambda$ should be on the order of electron binding energies, that is $(Z\alpha)^2m_ec^2$. They finally define it through
  \begin{equation}
    \ln\left(\frac{2m_ec^{2}}{\lambda}\right)=2\ln\left(\frac{1}{Z\alpha}+\frac{1}{2}\right), 
  \end{equation}
though, for better performance. Flambaum and Ginges next argue that the low-frequency (LF) potential should have the range of a $1s$ orbital of hydrogen-like atoms
and therefore choose the functional form
\begin{equation}\label{eq:FG_LF}
\varphi_{\text{LF}}^{\text{point}}\left(\boldsymbol{x}\right)=-\frac{B\left(Z\right)}{e}Z^{4}\alpha^{5}m_ec^{2}e^{-Zr_x/a_{\text{0}}},
\end{equation}
where $a_{\text{0}}=\lambdabar/\alpha$ is the Bohr radius and
\begin{equation}
  B\left(Z\right)=0.074+0.35\times Z\alpha,
\end{equation}
is a second fitting function.

The fitting function of the high-frequency contribution is written as
\begin{equation}
A\left(Z,\boldsymbol{x}\right)=\Theta\left(Z,\boldsymbol{x}\right)\left(1.071-1.976 y^{2}-2.128 y^{3}+0.169 y^{4}\right),
\end{equation}
in terms of the variable $x=\left(Z-80\right)\alpha$ and a cutoff-function of the form
\begin{equation}
\Theta\left(Z,\boldsymbol{x}\right)=\frac{r_x}{r_x +0.07\left(Z\alpha\right)^{2}\lambdabar},
\end{equation}
which will dampen the contribution of $\varphi_{\text{elec}}^{\text{point}}$ at short distances where the the locality of the effective SE potential breaks down.
The coefficients of the $A$ and $B$ fitting functions above were adjusted to 
reproduce the self-energy corrections to high $s$- and $p$-states, respectively, calculated
accurately in  Refs.~\citenum{mohr_se_n_12,mohr_kim_se_n_345} for Coulombic
hydrogen-like atoms of $5\leq Z\leq110$. 
It should be added that Thierfelder and Schwerdtfeger\cite{thierfelder2010} later modified the fitting function to $A_{n}\left(Z,\boldsymbol{x}\right)$, that is, making it dependent of the principal quantum number $n$.  These potentials with $A_{n}\left(Z,\boldsymbol{x}\right)$ instead of $A\left(Z,\boldsymbol{x}\right)$ were used by Pa\v{s}teka \textit{et al.} to calculate the electron affinity and ionization
potential of gold.\cite{pasteka_prl2017} Ginges and Berengut,\cite{ginges2016atomic} on the other hand, made both fitting functions $A$ and $B$ dependent on orbital angular momentum $\ell$ and further suggest to introduce a $\kappa$-dependence as well. 
The downside of making the effective QED potentials dependent on atomic orbital quantum numbers  is that it complicates the extension of these potentials to the molecular regime.

\subsection{Atomic shift operator}\label{sec:ashift}
With the above effective QED potentials available in an atomic code (see Section \ref{sec:impl}),
we have investigated their extension to molecular calculations by adding to the
electronic Hamiltonian, Eq.~\eqref{eq:elham}, an operator on the form
\begin{equation}
\begin{aligned}V_{{\rm {ASHIFT}}} & =\sum_{i}|\psi_{i}\rangle\omega_{i}\langle\psi_{i}|,\\
\omega_{i} & =\langle\psi_{i}|-e\varphi_{{\rm {effQED}}}|\psi_{i}\rangle,
\end{aligned}\label{eqn:ashift}
\end{equation}
where $\left\{\omega_{i}\right\}$ are expectation values of the effective QED potentials taken from atomic
calculations and $\left\{\psi_{i}\right\}$ are pre-calculated atomic orbitals, in practice
limited to those that are occupied in the electronic ground state of the atoms constituting
the molecule under study, calculated in their proper basis. The import of atomic orbitals into molecular calculations
is straightforward in the case of the DIRAC code, since such functionality is already
available through projection analysis.\cite{saue:csau,dubillard_jcp2006}
There is some overlap between the spectral representation of the self-energy proposed by Dyall\cite{dyall_jcp2013} as well
as the effective SE operator proposed by Shabaev and co-workers,\cite{shabaev_pra2013} but those
approaches are based on hydrogenic orbitals.

\section{IMPLEMENTATION}\label{sec:impl}
Routines for the radiative potentials used in this work are available in the GRASP atomic code.\cite{dyall1989grasp} Routines for calculating the Uehling potential were reported
as early as 1980.\cite{mckenzie_cpc1980} McKenzie \textit{et al.} follow the approach suggested by Wayne Fullerton and Rinker.\cite{fullerton_rinker_1976} More precisely, they employ Eq.~\eqref{eq:Uehling} for the
inner grid points until a more approximate form, Eq.~(6) of Ref.~\citenum{fullerton_rinker_1976},
becomes numerically valid. The latter form is then used until the magnitude of the potential
falls below a threshold value. The effective SE potential of Flambaum and Ginges\cite{flambaum_ginges2005} was implemented more recently,\cite{thierfelder2010} as is also
the case\cite{private_thierfelder} of the effective SE potential of Pyykk{\"o} and Zhao.\cite{pyykko_zhao2003} As already mentioned, the FG potential is in principle that
associated with a point nucleus, although fitting parameters have been optimized also to
calculations with finite nuclear charge distributions. Thierfelder and Schwerdtfeger\cite{thierfelder2010} adapted these potentials to finite nuclei by replacing the
Coulomb potentials of Eqs.~\eqref{eq:FG_mag} and \eqref{eq:FG_HF} by the potentials of
finite nuclear charge distributions, and we have so far followed this approach which appears to
be a reasonable approximation, as can be inferred from Table IV of Ref.~\citenum{ginges2016atomic}.

We have adapted the GRASP effective QED potential routines to molecular calculations by using the numerical integration scheme implemented for relativistic Kohn--Sham calculations in the DIRAC molecular code.\cite{saue:jcc2002} The scheme is based on the Becke partitioning\cite{becke:partition} of the molecular volume into atomic ones for which numerical integration is carried out in spherical coordinates. Specifically, we use Lebedev angular quadrature,\cite{lebedev1999quadrature} by default setting $\ell=15$, combined with the basis-set adaptive radial grid proposed by Lindh and co-workers.\cite{lindh2001} It may be noted that the effective QED potentials presented in the previous section are all radial, with the exception of the magnetic contribution to the Flambaum--Ginges SE potential, Eq.~\eqref{eq:FG_mag}.

Due to the very local nature of the effective QED potentials\cite{artemyev_2016} one-electron integrals over a potential associated with atomic center $A$ can be well approximated by
\begin{equation}
  V^{A}_{\mu\nu}\approx \int_0^{R_A}dr_A\int_{\Omega}d\Omega_A\left[\chi_{\mu}v^{A}\chi_{\nu}\right](\mathbf{r}_A)r_A^2 ,
\end{equation}
where $\left\{\chi_{\mu}\right\}$ are Gaussian-type basis functions. 
The most delocal potential is the low-frequency contribution to the electric form factor of the Flambaum--Ginges SE potential, Eq.~\eqref{eq:FG_LF}, since it has been designed to have the range of the $1s$ orbital of a hydrogen-like atom. For low $Z$ the potential may thereby overlap significantly with neighbor centers. By default, we therefore deactivate the effective QED potentials for $Z < 19$.
We also determine the value of the upper limit of radial integration $R_A$ based on the convergence of the low-frequency term to a very conservative $10^{-50}$.

\section{COMPUTATIONAL DETAILS}\label{sec:compdet}
For all calculations we used a development version of DIRAC code;\cite{dirac22,saue2020dirac} precise version and build information is found in output files, see Ref.~\citenum{ReplicationData}. A Gaussian model\cite{visscher1997dirac} for the nuclear charge distribution was employed throughout our calculations. Unless otherwise stated, we applied the Uehling VP potential\cite{uehling1935} and the SE potential of Flambaum and Ginges,\cite{flambaum_ginges2005} added to the Dirac--Coulomb--Gaunt (DCG) Hamiltonian. For correlated calculations we employed the molecular mean-field approximation Hamiltonian (X2Cmmf) \cite{sikkema_visscher_saue_ilias_2009} based on the DCG Hamiltonian, which we denote as $^2\textrm{DCG}^M$. In this approach, the converged Fock matrix obtained with the DCG Hamiltonian, with the effective QED potentials included, is exactly transformed to two-component form, that is, without any picture-change errors.\cite{schwerdtfeger_snijders_1990, kello1998picture,dyall_2000} All basis sets were employed in uncontracted form with the small component generated by restricted kinetic balance (see Ref.~\citenum{saue2020dirac} for details). Electronic structure analysis was carried out using projection analysis\cite{dubillard_jcp2006} where Pipek--Mezey localized MOs\cite{dubillard_jcp2006,pipek_jcp1989} are expanded in intrinsic, hence polarized, atomic orbitals.\cite{knizia_jctc2013} The analysis was done at the molecular geometries optimized with respect to the employed Hamiltonian, except for DCG with effQED, where the DCG structures were employed.

For the atomic calculations reported in Table \ref{tab:atomQED} we employed Dyall v3z basis sets;\cite{Dyall4s5s6s7s,Dyall4d,dyall_2004, dyall_gomes_2010,dyall_2011,dyall2016relativistic} 
the basis set for Uue was specially optimized by Dyall for this work.\cite{private_dyall}

For van der Waals dimers, the following orbitals were correlated: $5d6s$ for Hg, $5d6s6p$ for Rn, $6d7s$ for Cn, and $6d7s7p$ for Og. We used an virtual energy cutoff of 40 $E_h$. Dyall cv3z basis sets,\cite{dyall_2004, dyall_gomes_2010, dyall_2011} designed for core-valence correlation, were employed for the Hg and Cn species, whereas Dyall acv3z basis sets,\cite{dyall_2002, dyall_2006, dyall_2012} where the Dyall cv3z basis sets have been augmented by diffuse functions, were employed for Rn and Og species. Electronic structure calculations were done at the level of coupled-cluster singles-and-doubles  with approximate triples correction (CCSD(T)) using the RELCCSD module.\cite{visscher_lee_dyall_1996} We used the counterpoise correction\cite{boys1970} to minimize  basis set superposition errors (BSSE).

For the calculations of gold cyanide, we used the CCSD(T) method for comparison with experiment. In the CCSD(T) calculation, $4f5s5p5d6s$ for Au, and all electrons of C and N were correlated, which is the same level as the previous work.\cite{hill2013jcp_aucn} Dyall ae3z and ae4z basis sets,\cite{dyall_2004, dyall_gomes_2010,dyall2016relativistic} designed for correlation of all electrons, were employed in the calculations.  We employed a virtual energy cutoff of about 50 $E_h$ and 80 $E_h$ for dyall.ae3z and dyall.ae4z, respectively, which assures that correlating \textit{h} and \textit{i} orbitals, respectively, are included. Effective QED potentials for C and N atoms were not used, as explained in Section \ref{sec:impl}. The potential energy surface (PES) was calculated in the vicinity of the equilibrium structure with a total of 49 points for each basis set, using internal coordinates $r_1$ (Au-C distance), $r_2$ (C–N distance). The bond angle was fixed at  $\theta = 180^{\circ}$. The step size for bond distances was 0.1 $a_0$. The surface fitting and determination of the equilibrium structure was carried out using the SURFIT program,\cite{surfit} with convergence $3.2 \times 10^{-10}$ or better on the gradient. In addition, to estimate the relativistic effects we employed the two-component non-relativistic (by using .NONREL keyword), 4c-scalar-relativistic,\cite{dyall1994exact,visscher_jcp2000} and the Dirac--Coulomb Hamiltonians at the density functional theory (DFT) level. In these calculations we employed the B3LYP functional \cite{stephens_jpc1994,becke1993new} and the dyall.3zp basis sets.\cite{dyall_2004, dyall_gomes_2010,dyall2016relativistic}

For the calculation of Pb and Fl hydrides, the DCG Hamiltonian with and without effective QED potentials, as well as the L\'{e}vy-Leblond (LL) \cite{levy-leblond_cmp1967,visscher_jcp2000} Hamiltonian were employed. The dyall.3zp basis sets were used for all of the elements. The B3LYP functional was used for both the projection analysis and the geometry optimization.

\section{RESULTS}\label{sec:results}
\subsection{Atomic calculations}\label{sec:atomic}
\begin{table*}
\caption{Calculated \textit{ns} orbital energies in eV of group 1 and 11 elements from AOC-HF/dyall.v3z calculations based on the NR and DC Hamiltonians. The VP (Uehling) and SE (Flambaum--Ginges) corrections have been calculated as expectation values. }\label{tab:atomQED}
\begin{ruledtabular}
\begin{tabular}{lcccccrrcc}
  & NR & DC & VP & SE & $\Delta$QED & SE/VP & SE/VP\cite{johnson_soff_atnucdattab1985} & $\Delta$QED/$\Delta$R{[}\%{]} & \tabularnewline
  \hline
Li & -5.342 & -5.343 & -1.373E-06 & 4.092E-05 & 3.955E-05 & -29.7949 & -29.7058 & -9.01 & \tabularnewline
Na & -4.955 & -4.962 & -1.536E-05 & 2.950E-04 & 2.796E-04 & -19.2057 & -18.7963 & -4.37 & \tabularnewline
K & -4.013 & -4.028 & -3.423E-05 & 5.155E-04 & 4.813E-04 & -15.0615 & -14.7030 & -3.19 & \tabularnewline
Rb & -3.752 & -3.811 & -1.309E-04 & 1.361E-03 & 1.231E-03 & -10.3981 & -10.0783 & -2.08 & \tabularnewline
Cs & -3.365 & -3.490 & -2.989E-04 & 2.304E-03 & 2.005E-03 & -7.7089 & -7.4266 & -1.61 & \tabularnewline
Fr & -1.740 & -3.611 & -1.438E-03 & 6.333E-03 & 4.895E-03 & -4.4038 & -4.3351 & -0.26 & \tabularnewline
Uue & -2.993 & -4.327 & -1.034E-02 & 2.157E-02 & 1.123E-02 & -2.0859 & -4.3351 & -0.84 & \tabularnewline
Cu & -6.480 & -6.649 & -2.355E-04 & 2.840E-03 & 2.604E-03 & -12.0606 & -11.7316 & -1.54 & \tabularnewline
Ag & -5.985 & -6.452 & -7.342E-04 & 6.448E-03 & 5.714E-03 & -8.7825 & -8.4755 & -1.22 & \tabularnewline
Au & -6.003 & -7.923 & -4.635E-03 & 2.374E-02 & 1.910E-02 & -5.1219 & -4.9912 & -1.00 & \tabularnewline
Rg & -5.441 & -11.425 & -3.251E-02 & 8.408E-02 & 5.157E-02 & -2.5863 & -2.7223 & -0.86 & \tabularnewline
\end{tabular}
\end{ruledtabular}
\end{table*}
\begin{table*}
  \caption{\label{tab:table_ene}Relativistic and QED effects on the orbital energies  $\varepsilon$($E_{\textrm{h}}$) of the Au atom at the B3LYP/dyall.3zp level. The Uehling VP potential has been combined with two different SE potentials: FG (Flambaum--Ginges) and PZ (Pyykk{\"o}--Zhao) in variational calculations. Numbers in parentheses shows the percentage-wise ratio $\Delta$QED/$\Delta$R for each combination of effective QED potentials.}
\begin{ruledtabular}
\begin{tabular}{l|rr|rr|rr}
 & NR & DCG & $\Delta$(U+FG) &  & $\Delta$(U+PZ) & \tabularnewline
\hline 
1s$_{1/2}$ & -2689.451 & -2955.841 & 6.377E00 & (-2.39) & 6.243E00 & (-2.34)\tabularnewline
2s$_{1/2}$ & -449.932 & -523.020 & 8.448E-01 & (-1.16) & 8.586E-01 & (-1.17)\tabularnewline
2p$_{1/2}$ & -432.492 & -500.523 & 5.895E-02 & (-0.09) & 5.955E-02 & (-0.09)\tabularnewline
2p$_{3/2}$ & -432.492 & -433.755 & 1.194E-01 & (-9.45) & -3.055E-02 &( 2.42)\tabularnewline
3s$_{1/2}$ & -105.753 & -123.999 & 1.862E-01 & (-1.02) & 1.924E-01 & (-1.05)\tabularnewline
3p$_{1/2}$ & -97.515 & -114.096 & 7.320E-03 & (-0.04) & 1.462E-02 & (-0.09)\tabularnewline
3p$_{3/2}$ & -97.515 & -99.311 & 2.213E-02 & (-1.23) & -8.516E-03 &( 0.47)\tabularnewline
3d$_{3/2}$ & -82.165 & -83.236 & -1.430E-02 &( 1.34) & -9.090E-03 &( 0.85)\tabularnewline
3d$_{5/2}$ & -82.165 & -80.090 & -5.817E-03 & (-0.28) & -8.587E-03 & (-0.41)\tabularnewline
4s$_{1/2}$ & -22.559 & -27.178 & 4.650E-02 & (-1.01) & 4.833E-02 & (-1.05)\tabularnewline
4p$_{1/2}$ & -18.993 & -22.979 & 1.124E-03 & (-0.03) & 3.336E-03 & (-0.08)\tabularnewline
4p$_{3/2}$ & -18.993 & -19.424 & 4.651E-03 & (-1.08) & -2.427E-03 &( 0.56)\tabularnewline
4d$_{3/2}$ & -12.440 & -12.630 & -3.439E-03 &( 1.82) & -2.307E-03 &( 1.22)\tabularnewline
4d$_{5/2}$ & -12.440 & -11.971 & -1.648E-03 & (-0.35) & -2.185E-03 & (-0.47)\tabularnewline
4f$_{5/2}$ & -3.648 & -3.228 & -2.383E-03 & (-0.57) & -1.461E-03 & (-0.35)\tabularnewline
4f$_{7/2}$ & -3.648 & -3.091 & -1.824E-03 & (-0.33) & -1.425E-03 & (-0.26)\tabularnewline
5s$_{1/2}$ & -3.253 & -4.116 & 9.003E-03 & (-1.04) & 9.430E-03 & (-1.09)\tabularnewline
5p$_{1/2}$ & -2.108 & -2.745 & -4.569E-05 &( 0.01) & 4.105E-04 & (-0.06)\tabularnewline
5p$_{3/2}$ & -2.108 & -2.139 & 5.619E-04 & (-1.82) & -5.703E-04 &( 1.85)\tabularnewline
5d$_{3/2}$ & -0.346 & -0.333 & -4.662E-04 & (-3.61) & -3.486E-04 & (-2.70)\tabularnewline
5d$_{5/2}$ & -0.346 & -0.276 & -2.881E-04 & (-0.41) & -3.206E-04 & (-0.46)\tabularnewline
6s$_{1/2}$ & -0.148 & -0.205 & 6.519E-04 & (-1.14) & 6.726E-04 & (-1.18)\tabularnewline
\end{tabular}
\end{ruledtabular}
\end{table*}
\begin{table*}
  \caption{\label{tab:table_x2c}First-order QED effects on the orbital energies  $\varepsilon$($E_{\textrm{h}}$) of the Au atom at the B3LYP/dyall.3zp level using the DCG Hamiltonian or the X2C Hamiltonian, the latter with or without picture-change (PC) transformation.  The Uehling VP potential has been combined with two different SE potentials: FG (Flambaum--Ginges) and PZ (Pyykk{\"o}--Zhao). Numbers in parentheses shows the percentage-wise ratio $\Delta$QED/$\Delta$ for each combination of effective QED potentials.}
\begin{ruledtabular}
\begin{tabular}{l|rr|rr|cc}
 & \multicolumn{2}{c|}{DCG} & \multicolumn{2}{c|}{X2C - PC} & \multicolumn{2}{c}{X2C - noPC}\tabularnewline
  & $\Delta$(U+FG) & $\Delta$(U+PZ) & $\Delta$(U+FG) & $\Delta$(U+PZ) & $\Delta$(U+FG) & $\Delta$(U+PZ)\tabularnewline
  \hline
1s$_{1/2}$ & 6.620E00 & 6.465E00 & 6.632E00 & 6.477E00 & 7.522E00 & 8.253E00\tabularnewline
2s$_{1/2}$ & 9.003E-01 & 9.022E-01 & 9.012E-01 & 9.030E-01 & 1.079E00 & 1.175E00\tabularnewline
2p$_{1/2}$ & 1.184E-01 & 1.051E-01 & 1.187E-01 & 1.047E-01 & 1.971E-01 & 3.608E-02\tabularnewline
2p$_{3/2}$ & 1.671E-01 & 4.922E-03 & 1.682E-01 & 5.005E-03 & 1.279E-01 & 6.342E-03\tabularnewline
3s$_{1/2}$ & 2.016E-01 & 2.040E-01 & 2.020E-01 & 2.043E-01 & 2.441E-01 & 2.665E-01\tabularnewline
3p$_{1/2}$ & 2.357E-02 & 2.674E-02 & 3.574E-02 & 1.357E-03 & 2.671E-02 & 1.727E-03\tabularnewline
3p$_{3/2}$ & 3.554E-02 & 1.335E-03 & 3.574E-02 & 1.357E-03 & 2.671E-02 & 1.727E-03\tabularnewline
3d$_{3/2}$ & -1.234E-03 & 1.064E-05 & -1.220E-03 & 1.066E-05 & 3.906E-03 & 3.605E-07\tabularnewline
3d$_{5/2}$ & 6.494E-03 & -1.680E-07 & 6.528E-03 & -1.675E-07 & 3.296E-03 & -1.132E-07\tabularnewline
4s$_{1/2}$ & 5.082E-02 & 5.151E-02 & 5.097E-02 & 5.166E-02 & 6.173E-02 & 6.741E-02\tabularnewline
4p$_{1/2}$ & 5.631E-03 & 6.658E-03 & 5.638E-03 & 6.637E-03 & 9.363E-03 & 2.339E-03\tabularnewline
4p$_{3/2}$ & 8.392E-03 & 3.315E-04 & 8.444E-03 & 3.370E-04 & 6.292E-03 & 4.295E-04\tabularnewline
4d$_{3/2}$ & -1.331E-04 & 2.846E-06 & -1.299E-04 & 2.851E-06 & 9.447E-04 & 9.962E-08\tabularnewline
4d$_{5/2}$ & 1.461E-03 & -4.294E-08 & 1.468E-03 & -4.277E-08 & 7.944E-04 & -2.822E-08\tabularnewline
4f$_{5/2}$ & -2.785E-04 & -1.968E-10 & -2.782E-04 & -1.974E-10 & 1.089E-05 & -1.184E-10\tabularnewline
4f$_{7/2}$ & 2.190E-04 & -9.047E-11 & 2.194E-04 & -9.111E-11 & 9.735E-06 & -9.557E-11\tabularnewline
5s$_{1/2}$ & 1.001E-02 & 1.015E-02 & 1.004E-02 & 1.018E-02 & 1.217E-02 & 1.329E-02\tabularnewline
5p$_{1/2}$ & 9.658E-04 & 1.152E-03 & 9.668E-04 & 1.149E-03 & 1.605E-03 & 4.052E-04\tabularnewline
5p$_{3/2}$ & 1.373E-03 & 5.478E-05 & 1.382E-03 & 5.572E-05 & 1.029E-03 & 7.103E-05\tabularnewline
5d$_{3/2}$ & -1.078E-05 & 2.580E-07 & -1.048E-05 & 2.582E-07 & 8.410E-05 & 9.077E-09\tabularnewline
5d$_{5/2}$ & 1.216E-04 & -3.635E-09 & 1.222E-04 & -3.618E-09 & 6.669E-05 & -2.375E-09\tabularnewline
6s$_{1/2}$ & 7.888E-04 & 8.001E-04 & 7.908E-04 & 8.020E-04 & 9.586E-04 & 1.047E-03\tabularnewline
\end{tabular}
\end{ruledtabular}
\end{table*}
In Table \ref{tab:atomQED} we show results of atomic calculations, using average-of-configuration (AOC) HF,\cite{Thyssen2004}\footnote{In these calculations the DIRAC keyword OPENFACTOR was set to one, such that orbital eigenvalues satisfies Koopmans' theorem.} that can be directly compared with Table I of Ref.~\citenum{pyykko1998estimated} and which provide estimates for the valence-level Lamb shift for group 1 and 11 metal atoms. Pyykk{\"o} \textit{et al.} focused on $ns_{1/2}$ orbital energies for estimating ionization energies, albeit, as pointed out by Thierfelder and Schwerdtfeger,\cite{thierfelder2010} for Roentgenium $(Z=111)$, the first ionization is out of the $6d_{5/2}$ orbital. 
The VP and SE contributions come with opposite sign and are dominated by the latter.\cite{johnson_soff_atnucdattab1985} However, the ratio SE/VP decreases
significantly with increasing nuclear charge and, indeed, VP is predicted to eventually overtake SE at very high nuclear charges.\cite{thierfelder2010} Pyykk{\"o} \textit{et al.} 
calculated the SE contribution as $\langle V^U\rangle\ast$(SE/VP) where (SE/VP)  is the ratio for $2s_{1/2}$ of the corresponding hydrogen-like systems, including the nuclear-size effect, tabulated for $1\le Z\le 100$ by Johnson and Soff\cite{johnson_soff_atnucdattab1985} (a more recent compilation is provided by Yerokhin and Shabaev\cite{yerokhin_jpcrd2015}). As confirmed by later calculations\cite{labzowsky_physreva.59.2707} and the numbers in Table \ref{tab:atomQED}, this is a quite reasonable approximation.

Comparing relativistic and QED effects, one sees that the latter corrects the former by about $-1\%$ for the heavier atoms. For the gold atom it is exactly so. In Table \ref{tab:table_ene} we show the effect of relativity and QED on all orbital energies of the gold atom. Two combinations of effective QED potentials have been used in variational calculations: The Uehling (U) VP potential has been combined either with the Flambaum--Ginges (FG) or Pyykk{\"o}--Zhao SE potentials. One sees that for both combinations of effective QED potentials the relativistic effects is, with very few exceptions, reduced with a few percent. For $s_{1/2}$ orbitals the difference in QED shift between the U+FG and U+PZ combinations is below 5\%; for other orbitals the difference is generally larger. We note in particular that the shifts have systematically opposite sign for $p_{3/2}$ orbitals. Not surprisingly the largest absolute shifts are observed for inner core orbitals, whereas the largest relative shift -- 0.33\% -- is seen for the $6s_{1/2}$ orbital.

In Table \ref{tab:table_x2c} we show QED shifts of orbital energies, this time obtained perturbatively as expectation values. Compared to the shifts obtained from variational inclusion of the effective QED potentials, the largest absolute deviations concern the inner core orbitals.
The smallest relative deviations are observed for $s_{1/2}$ orbitals and decreasing towards core. The largest relative deviations, on the other hand, are seen for $p$ orbitals; the very largest relative deviation concerns $5p_{1/2}$, but this can probably be attributed to noise, since the QED shift on the energy of this orbital is particularly small with both combinations of effective QED potentials.

Table \ref{tab:table_x2c} also shows perturbative QED shifts of orbital energies obtained with the X2C Hamiltonian. When the effective QED potentials have been correctly picture-changed transformed, deviations from the parent 4c calculation are below 3 \%, which clearly validates the use of these potentials in 2-component relativistic calculations. On the other hand, without picture-change, significant errors are observed; the average unsigned error for U+FG and U+PZ is 130 \% and 47 \%, respectively. This is possibly worrisome since the U+PZ combination, expressed in terms of Gaussians, have been used without picture-change in scalar DKH calculations by Peterson and co-workers.\cite{shepler2005ab,shepler2007hg+,peterson2015correlation,cox2016}


\subsection{Gold cyanide}\label{sec:AuCN}
\begin{table}
\caption{M-C bond lengths (in pm) in \ce{MCN} (M=\ce{Cu}, \ce{Ag}, \ce{Au}) from microwave (MW) spectroscopy and calculations.}\label{tab:MCN}
\begin{ruledtabular}
\begin{tabular}{lllll}
      &\ce{CuCN}   &\ce{AgCN}   &\ce{AuCN}  &\\
\hline
$r_0$ &183.231(7)  &203.324(45) &191.251(16)& MW$^a$\\
$r_s$ &183.284(4)  &203.4182(27)&191.22519(84)&MW$^a$\\
$r_e$ &182.36      &202.42      &191.05       &Calc.\cite{zaleski2008structure}\\
$r_e$ &182.65      &202.99      &190.71       &Calc.\cite{hill2013jcp_aucn}
\end{tabular}
\end{ruledtabular}
{\footnotesize $^a$\ce{CuCN}: Ref.~\citenum{grotjahn_jacs2002}.} \ce{AgCN},\ce{AuCN}: Ref.~\citenum{okabayashi2009detection}
\end{table}
In 2008 Pyykk{\"o} and co-workers reported CCSD(T)/cc-pVQZ calculations on the noble metal cyanides (\ce{MCN}, M=\ce{Cu}, \ce{Ag}, \ce{Au}).\cite{zaleski2008structure} Small-core scalar-relativistic effective core potentials (SRECP)\cite{figgen_cp2005} were used for the metal atoms and spin-orbit corrections added at the PBE-ZORA/QZ4P level. In 2013 Peterson and
co-workers reported CCSD(T)-F12/cc-pV5Z calculations on the same compounds, using the same SRECPs as the previous authors and adding a number of corrections.\cite{hill2013jcp_aucn} As seen from Table \ref{tab:MCN} the newer calculations brought the M-C bond lengths of \ce{CuCN} and \ce{AgCN} in better agreement with experiment, but increased the gap between theory and
experiment for \ce{AuCN}. This led Pyykk{\"o} to conjecture that this could be the first evidence of the effect of QED on molecular structure.\cite{pyykkotalk}
\begin{table}
  \caption{Relativistic and QED effects on the rms radius $\langle r^2\rangle^{1/2}$(pm) of the Au atom at the B3LYP/dyall.3zp level. Effective QED potentials: VP(Uehling)+SE(Flambaum--Ginges).}\label{tab:table_rms}
\begin{ruledtabular}
\begin{tabular}{ccccccc}
 & NR & DCG & DCG+QED & $\Delta$R & $\Delta$QED & $\Delta$QED/$\Delta$R(\%)\tabularnewline
 \hline 
5s$_{1/2}$ & 57.71 & 52.23 & 52.28 & -5.48 & 0.05 & -0.86\tabularnewline
5p$_{1/2}$ & 63.16 & 56.78 & 56.78 & -6.38 & 0.00 & -0.02\tabularnewline
5p$_{3/2}$ & 63.16 & 62.37 & 62.38 & -0.80 & 0.01 & -1.04\tabularnewline
5d$_{3/2}$ & 91.07 & 90.81 & 90.78 & -0.26 & -0.03 & 10.46\tabularnewline
5d$_{5/2}$ & 91.07 & 95.75 & 95.73 & 4.68 & -0.02 & -0.36\tabularnewline
6s$_{1/2}$ & 196.07 & 167.54 & 167.79 & -28.53 & 0.25 & -0.88\tabularnewline
\end{tabular}
\end{ruledtabular}
\end{table}

To possibly verify this conjecture we first carried out exploratory calculations at the B3LYP/dyall.3zp level. Table \ref{tab:table_rms} shows the effects of relativity and QED on orbital sizes of the gold atom. For the valence 6s$_{1/2}$ we observe an impressive relativistic contraction of 28.53 pm, whereas QED leads to an orbital expansion of 0.25 pm, roughly -1 \% of the relativistic effect.

We next turn to the \ce{AuCN} molecule. We first, in Table \ref{tab:AuCN_bond}, report bonding analysis in localized orbitals.\cite{dubillard_jcp2006} One finds a single $\sigma$-type Au-C bond, dominated by carbon 2s$_{1/2}$ and gold 6s$_{1/2}$, as well as a triple C-N bond. Equilibrium bond lengths $r_e$ with respect to different Hamiltonians are reported in Table \ref{tab:table_DFT1}. We see a very significant scalar relativistic bond-length contraction of 25.31 pm, on par with the 6s$_{1/2}$ orbital contraction observed in Table \ref{tab:table_rms}. When going from a spin-free (SF) Hamiltonian to the Dirac--Coulomb one, one finds a further contraction of 0.29 pm, which agrees very well with the spin-orbit correction of -0.28 pm obtained by Hill \textit{et al.} taking the same difference, albeit at the CCSD(T) level.\cite{hill2013jcp_aucn} However, this contraction is almost canceled when adding the Gaunt term, which brings spin-other-orbit interaction\cite{saue2011relativistic} and which was not considered by Hill and co-workers.\cite{hill2013jcp_aucn} At this level of theory, the total
relativistic effect on the bond length is thereby -25.38 pm. Finally, adding QED effects, we observe a bond-length extension of 0.19 pm, -0.75 \% of the relativistic effect. One may note that the QED effect is of the same order as the effect of adding the Gaunt term.\cite{thierfelder2010} In passing we note from Table \ref{tab:table_DFT1} that incorporation
of QED effects through the atomic shift operator (ASHIFT) described in Section \ref{sec:ashift} also leads to a bond extension, albeit only capturing half of the full QED effect.
\begin{table}
\caption{Gross populations obtained by projection analysis using Pipek--Mezey localized orbitals at the DCG/B3LYP/dyall.3zp level. $\langle\varepsilon\rangle$ is the expectation value in $E_h$ with respect to the Kohn--Sham matrix.}\label{tab:AuCN_bond}
\begin{ruledtabular}
\begin{tabular}{cc|cc|ccc|ccc}
 &  & \multicolumn{2}{c|}{Au} & \multicolumn{3}{c|}{C} & \multicolumn{3}{c}{N}\tabularnewline
$\omega$ & $\langle\varepsilon\rangle$ & 5d$_{5/2}$ & 6s$_{1/2}$ & 2s$_{1/2}$ & 2p$_{1/2}$ & 2p$_{3/2}$ & 2s$_{1/2}$ & 2p$_{1/2}$ & 2p$_{3/2}$\tabularnewline
\hline 
3/2 & -0.342 & -0.01 & 0.00 & 0.00 & 0.00 & 0.92 & 0.00 & 0.00 & 1.09\tabularnewline
1/2 & -0.345 & 0.00 & 0.00 & 0.00 & 0.65 & 0.28 & 0.00 & 0.64 & 0.43\tabularnewline
1/2 & -0.586 & 0.15 & \textbf{0.39} & 0.92 & 0.17 & 0.34 & 0.01 & -0.01 & -0.02\tabularnewline
1/2 & -0.781 & 0.00 & 0.01 & 0.24 & 0.12 & 0.38 & 0.15 & 0.44 & 0.65\tabularnewline
\end{tabular}
\end{ruledtabular}
\end{table}

\begin{table}
\caption{\label{tab:table_DFT1}Equilibrium bond lengths $r_\textrm{e}$ (in pm) of \ce{AuCN} calculated at the B3LYP/dyall.3zp level using various Hamiltonians. Numbers in parenthesis indicate the change with respect to the previous line, except ASHIFT, which refers to DCG. SF refers to a spin-free 4-component relativistic Hamiltonian.}
\begin{ruledtabular}
\begin{tabular}{lll}
Hamiltonian & Au-C  &C-N \tabularnewline
\hline 
NR & 218.54           & 115.71 \tabularnewline
SF & 193.23 (-25.31) & 115.54 (-0.17)\tabularnewline
DC & 192.94 (-0.29) & 115.56 (+0.02)\tabularnewline
DCG & 193.16 (+0.22) & 115.58 (+0.01)\tabularnewline
QED & 193.35 (+0.19) & 115.57 (+0.00)\tabularnewline
ASHIFT & 193.25 (+0.09)	& 115.58 (+0.00) \tabularnewline
\end{tabular}
\end{ruledtabular}
\end{table} 
\begin{table}
\caption{Final, recommended equilibrium Au-C bondlength $r_\textrm{e}$ (pm) at the $^2\textrm{DCG}^M$-CCSD(T) level for the \ce{AuCN} molecule. $\Delta$QED is the difference between the extrapolated basis set limit ae\ensuremath{\infty}z with QED and without QED.}\label{tab:table_AuCN_cc}
\begin{ruledtabular}
\begin{tabular}{l D{.}{.}{-1} D{.}{.}{-1}}
 &  \multicolumn{1}{c}{$\textrm{Au-C}$ } &  \multicolumn{1}{c}{$\textrm{C-N}$} \tabularnewline
\hline 
ae3z & 190.89 & 116.66\tabularnewline
ae4z & 190.70 & 116.28\tabularnewline
ae\ensuremath{\infty}z & 190.58 & 116.07\tabularnewline
$\Delta$T\cite{hill2013jcp_aucn} & 0.26 & -0.10\tabularnewline
$\Delta$Q\cite{hill2013jcp_aucn} & -0.09 & 0.19\tabularnewline
Final w/o QED & 190.75 & 116.16\tabularnewline
$\Delta$QED & 0.19  & 0.00 \tabularnewline
Final & 190.94 & 116.16\tabularnewline
\end{tabular}
\end{ruledtabular}
\end{table}

To obtain more accurate bond lengths, we proceeded as indicated in Table \ref{tab:table_AuCN_cc}: $^2\textrm{DCG}^M$-CCSD(T) calculations were carried out in the Dyall ae3z and ae4z basis sets and then extrapolated to the basis-set limit, \cite{halkier_cpl1998} indicated by ``ae\ensuremath{\infty}z''. We then added the triples $\Delta$T and quadruples $\Delta$Q corrections reported by Hill \textit{et al.} \cite{hill2013jcp_aucn} to obtain a Au-C bond length of 190.75 pm, very close to the value 190.71 pm reported by Peterson and co-workers. Finally, we add a QED correction of 0.19 pm, identical to what we obtained at the B3LYP/dyall.3zp level, to obtain our final value of 190.99 pm.

The devil is, however, in the details: Our Born--Oppenheimer equilibrium bond lengths $r_e$ are not directly comparable to the structural parameters extracted from the rotational spectra recorded by Okabayashi and co-workers.\cite{okabayashi2009detection} Experiment gives access to rotational constants $B_{\nu}$ for individual vibrational states. For a linear molecule like \ce{AuCN} the rotational constant, in units of frequency, is expressed as
\begin{equation}
  B=\frac{\hbar}{4\pi I_{\perp}};\quad I_{\perp}=I_{xx}=I_{yy}=\sum_Am_A z_A^2,
\end{equation}
when the molecular axis is aligned with the $z$-axis. $z_A$ is the distance of atom $A$ from the center of mass. Effective $r_0$ and substitution $r_s$ structures are both obtained by assuming identical structures for all isotopomers of the target molecule observed in experiment.\cite{gordy_mms,demaison2016equilibrium} Effective structures $r_0$ are obtained by least-square fitting of experimental ground-state inertial moments, whereas substitution structures $r_s$ are obtained from observation of how rotational constants (and center of mass) change upon single isotope substitution $A\rightarrow A^{\prime}$. For a linear molecule one has
\begin{equation}
\left|z_A\right|=\sqrt{\frac{\hbar}{4\pi\mu}\left(\frac{1}{B^{A^{\prime}}}-\frac{1}{B^{A}}\right)};\quad\frac{1}{\mu}=\frac{1}{M}+\frac{1}{\Delta m_A}.\label{eqn:substitution}
\end{equation}
where $M$ is the total mass of the parent isotopomer. In the case of \ce{AuCN} $\left|z_C\right| $ and $\left|z_N\right|$ could be estimated from corresponding single isotope substitutions. However, since gold has a single naturally occurring isotope, \ce{^{197}Au}, $\left|z_{Au}\right|$ was obtained from the definition of center of mass.\cite{private_okabayashi}

Empirically one typically finds $r_0 \ge r_s \ge r_e$, \cite{demaison2016equilibrium} which suggests that we should rather compare our recommended $r_e$=190.99 pm for Au-C with the corresponding substitution bond length $r_s$=191.22519(84) pm reported by Okabayashi and co-workers.\cite{okabayashi2009detection} However, a better comparison is provided by calculating the ground-state rotational constant $B_0$ from $B_e$. From perturbation theory, excluding Fermi resonances, the rotational constant for a given vibrational state $\nu$ of a general molecule is related to $B_\textrm{e}$ as follows:\cite{demaison2016equilibrium}
\begin{align}\label{eqn:B0exact}
B^{\xi}_{\nu} = B^{\xi}_e &-\sum_i \alpha^{\xi}_i\left(\nu_i +\frac{d_i}{2}\right) \nonumber \\
&+\frac{1}{2}\sum_{i,j}\gamma^{\xi}_{i,j}\left(\nu_i+\frac{d_i}{2}\right)\left(\nu_j+\frac{d_j}{2}\right)+\cdots.
\end{align}
Here, $\xi$ is the axis of rotation, $\alpha^{\xi}$ and $\gamma^{\xi}$ are vibration-rotation interaction constants of different orders and $d_i$ is the degeneracy of vibration mode $i$. The series generally converges rapidly, and for \ce{AuCN} a suitable expression is therefore
\begin{equation}
\label{eqn:B0alpha}
B_0 \approx B_\textrm{e} - \frac{1}{2}\left[\alpha_{100}+\alpha_{001}+2\alpha_{01^10}\right],
\end{equation}  
using the notation $\alpha_{\nu_1\nu_2\nu_3}$, where $\nu_1$ corresponds to the C-N stretch, $\nu_2$ to the doubly degenerate bending mode and $\nu_3$ to the Au-C stretch.

Hill \textit{et al.} \cite{hill2013jcp_aucn} carried out both perturbative and variational rovibrational calculations. Using their calculated potential surfaces\cite{private_peterson} we have extracted vibration-rotation interaction constants $\alpha_{\nu_1\nu_2\nu_3}$. Combined with our best equilibrium structures from Table \ref{tab:table_AuCN_cc}, we have calculated the rotational constants of the ground vibrational state of the three isotopomers of \ce{AuCN} studied by Okabayashi and co-workers.\cite{okabayashi2009detection} As can be seen from Table \ref{tab:AuCN_rot}, the inclusion of QED corrections brings about
a dramatic improvement with respect to experiment. Not surprisingly then, when we extract substitution structures by the same procedure as Okabayashi and co-workers,\cite{okabayashi2009detection} correcting for Lamb shift effects bring our calculated substitution bond lengths within 0.05 pm of the experimental ones (cf. Table \ref{tab:AuCN_subst}). We expect further refinement of the potential surface to improve agreement with experiment; we note for instance that our calculated vibration-rotation interaction constant associated with bending for the most abundant isotopomer \ce{^197Au^12C^14N} is -10.98 MHz, compared to -11.9781 MHz when extracted from experiment.\cite{hill2013jcp_aucn}


\begin{table}
\caption{Calculated and experimental rotational constants (in MHz) for \ce{AuCN}.}\label{tab:AuCN_rot}
\begin{ruledtabular}
\begin{tabular}{lccc}
 & \ce{^{197}Au^{12}C^{14}N} & \ce{^{197}Au^{13}C^{14}N} & \ce{^{197}Au^{12}C^{15}N}\tabularnewline
 \hline 
$\alpha_{1}$ & 14.55 & 13.53 & 14.06\tabularnewline
$\alpha_{2}$ & -10.98 & -10.25 & -10.59\tabularnewline
$\alpha_{3}$ & 12.17 & 11.91 & 11.30\tabularnewline
\hline 
 & \multicolumn{3}{c}{w/o QED}\tabularnewline
$B_{e}$ & 3237.5 & 3184.5 & 3086.4\tabularnewline
$B_{0}$ & 3235.1 & 3182.1 & 3084.3\tabularnewline
\hline 
 & \multicolumn{3}{c}{with QED}\tabularnewline
$B_{e}$ & 3232.8 & 3179.9 & 3082.0\tabularnewline
$B_{0}$ & 3230.4 & 3177.5 & 3079.9\tabularnewline
$B_{0}$(exp.)\cite{okabayashi2009detection} & 3230.21115(18) & 3177.20793(13) & 3079.73540(12)\tabularnewline
\end{tabular}
\end{ruledtabular}      
\end{table}

\begin{table}
\caption{Calculated and experimental substitution structure (in pm) for \ce{AuCN} }\label{tab:AuCN_subst}  
\begin{ruledtabular}
\begin{tabular}{lll}
 & $r_{s}$(Au-C) & $r_{s}$(C-N)\tabularnewline
 \hline 
w/o QED & 190.991 & 115.910\tabularnewline
with QED & 191.184 & 115.909\tabularnewline
Exp.  & 191.22519(84) & 115.86545(97)\tabularnewline
\end{tabular}
\end{ruledtabular}
\end{table}  
\subsection{van der Waals dimers}\label{sec:rare gas dimer}
As a second molecular application of our implementation we consider spectroscopic constants of dimers with van der Waals bonding (\ce{M2}, M = Hg, Rn, Cn, Og).
In Table \ref{tab:vdW dimers} we report our calculated the equilibrium bond lengths $\textit{r}_{\textrm{e}}$, harmonic frequencies $\omega_{\textrm{e}}$, anharmonic constants $\omega_{\textrm{e}}x_{\textrm{e}}$ and dissociation energies $D_{\textrm{e}}$ for these species. We see that the QED effect on bond length is on the order of 0.15 pm for row-6 dimers and approximately doubles when going to the superheavy elements; for \ce{Og2} the QED bond length extension is in line with what was reported by Hangele and Dolg using relativistic effective core potentials.\cite{hangele2014coupled} The QED effect on dissociation energies is rather small: on the order of 0.4 \% for the superheavy dimers.

\begin{table}
  \caption{Spectroscopic constants of heavy group 12 and 18 dimers obtained at the $^2\textrm{DCG}^M$-CCSD(T) level, using the U+FG combination of effective QED potentials and either dyall.cv3z (\ce{Hg2},\ce{Cn2}) or dyall.acv3z (\ce{Rn2},\ce{Og2}) basis sets. Numbers in parentheses indicate the QED effect.}\label{tab:vdW dimers}
\begin{ruledtabular}
\begin{tabular}{lrrrr}
 & $\textit{r}_{\textrm{e}}$/pm & $\omega_\textrm{e}$$/\textrm{cm}^{-1}$ & $\omega_\textrm{e}$$x_\textrm{e}/\textrm{cm}^{-1}$ & $D_{\textrm{e}}/\textrm{cm}^{-1}$ \tabularnewline
\hline 
\ce{Hg2} & 385.71 & 16.65 & 0.232 & 277.7\tabularnewline
   & (0.15) & (-0.03) & (-0.002) & (-0.02)\tabularnewline
\ce{Cn2} & 354.75 & 22.95 & 0.255 & 532.7\tabularnewline
   & (0.36) & (-0.11) & (0.001) & (-2.78)\tabularnewline
\ce{Rn2} & 463.60 & 13.79 & 0.286 & 174.9\tabularnewline
   & (0.14) & (-0.02) & (-1.E-04) & (-0.41)\tabularnewline
\ce{Og2} & 449.97 & 17.10 & 0.210 & 391.1\tabularnewline
   & (0.28) & (-0.04) & (0.001) & (-1.32)\tabularnewline
\end{tabular}
\end{ruledtabular}
\end{table}

\begin{table*}
\caption{\label{tab:Pb_Fl_opt}Optimized equilibrium structures of Pb and Fl hydrides at the B3LYP/dyall.3zp level based on the DCG Hamiltonian. $r_\textrm{e}$ and $\alpha_\textrm{e}$ refer to the X-H bond length (\AA) and H-X-H angle (degree), respectively.}
\begin{ruledtabular}
\begin{tabular}{ccccccc}
 & \multicolumn{2}{c}{\ce{PbH2}} & \multicolumn{2}{c}{\ce{FlH2}} & \ce{PbH4} & \ce{FlH4}\tabularnewline
 & $r_\textrm{e}$ & $\alpha_\textrm{e}$ & $r_\textrm{e}$ & $\alpha_\textrm{e}$ & $r_\textrm{e}$ & $r_\textrm{e}$ \tabularnewline
 \hline
NR & 1.879 & 90.83 & 2.017 & 90.85 & 1.816 & 1.959\tabularnewline
DCG & 1.845 & 91.18 & 1.920 & 93.35 & 1.756 & 1.825\tabularnewline
\end{tabular}
\end{ruledtabular}
\end{table*}
\begin{table}
\caption{Charge $Q$ and electronic configurations of Pb and Fl atoms in the title compounds obtained by projection analysis at the B3LYP/dyall.3zp level.}\label{tab:total_charge}
\begin{ruledtabular}
\begin{tabular}{lll}
     & $Q$   & atomic configuration                                                                                        \tabularnewline
\hline 
\ce{PbH2} & 0.39 & $5d_{3/2}^{4.00}$    $5d_{5/2}^{5.99}$    $6s_{1/2}^{1.86}$    $6p_{1/2}^{0.90}$  $6p_{3/2}^{0.86}$   \tabularnewline
\ce{PbH4} & 0.66 & $5d_{3/2}^{3.99}$   $ 5d_{5/2}^{5.98}$    $6s_{1/2}^{1.41}$    $6p_{1/2}^{0.85}$  $6p_{3/2}^{1.10}$  \tabularnewline
\ce{FlH2} & 0.32 & $6d_{3/2}^{3.99}$    $6d_{5/2}^{5.97}$    $7s_{1/2}^{1.91} $   $7p_{1/2}^{1.26}$  $7p_{3/2}^{0.55}$  \tabularnewline
\ce{FlH4} & 0.46 & $6d_{3/2}^{3.98}$   $6d_{5/2}^{5.94}$    $7s_{1/2}^{1.61}$    $7p_{1/2}^{1.17}$  $7p_{3/2}^{0.84}$ \tabularnewline
\end{tabular}
\end{ruledtabular}
\end{table}

\subsection{Reaction energies: Pb and Fl hydrides}\label{sec:Pb_Fl}

\begin{table*}
\caption{Gross population obtained by projection analysis of the localized bonding orbitals in the title compounds at the B3LYP level based on the DCG Hamiltonian. $<\varepsilon>$ refers to the expectation value with respect to the converged Kohn--Sham matrix (in $E_\textrm{h}$).}\label{tab:XHn_pop}
\begin{ruledtabular}
\begin{tabular}{ccccccccc|c}
 &  &  & \multicolumn{6}{c}{X} & H$_i$ \tabularnewline
 &  & $<\varepsilon>$ & $5s_{1/2}$ & $5d_{3/2}$ & $5d_{5/2}$ & $6s_{1/2}$ & $6p_{1/2}$ & $6p_{3/2}$ & $1s_{1/2}$ \tabularnewline
 \hline 
\ce{PbH4}& $\sigma_{XH_i}$ & -0.4192 && 0.00 & 0.00 & 0.35 & 0.21 & 0.27 & 1.19 \tabularnewline
\ce{PbH2} & $\sigma_{XH_i}$ & -0.3384 & 0.00 & 0.00 & 0.00 & 0.03 & 0.37 & 0.37 & 1.24 \tabularnewline
& nb  & -1.0182 & 0.13 & 0.00 & 0.00 & 1.62 & 0.14 & 0.11 & -0.03 \tabularnewline
&  & $<\varepsilon>$ & $6s_{1/2}$ & $6d_{3/2}$ & $6d_{5/2}$ & $7s_{1/2}$ & $7p_{1/2}$ & $7p_{3/2}$ & $1s_{1/2}$ \tabularnewline
\hline
\ce{FlH4} & $\sigma_{XH_i}$ & -0.4524 && 0.00 & 0.01 & 0.40 & 0.27 & 0.20 & 1.11 \tabularnewline
\ce{FlH2} & $\sigma_{XH_i}$ & -0.3469 & 0.00 & 0.00 & 0.01 & 0.03 & 0.52 & 0.25 & 1.18 \tabularnewline
& nb & -1.2235 & 0.11 & 0.04 & 0.06 & 1.57 & 0.15 & 0.04 & -0.03 \tabularnewline
\end{tabular}
\end{ruledtabular}
\end{table*}
\begin{table*}
\caption{Reaction energy of \ce{PbH4 -> PbH2 + H2} (in kcal/mol). 
$\Delta$DCG refers to the difference between DCG and NR. Other $\Delta$ refers to the difference from the DCG value.}\label{tab:reac_Pb}
\begin{ruledtabular}
\begin{tabular}{llrrr}
       &QED effect& reac. energy  & $\Delta$(kcal/mol) & $\Delta$(\%) \\ \hline
NR    &none & 16.47                   &             &       \\
DCG & none  & -8.99                   & -25.46      &       \\
         &VP      &   -9.09                   &   -0.10        &   0.41  \\
         &SE     &    -8.56                 &        0.42     &    -1.67 \\
         &VP+SE    & -8.66                   & 0.32        &   -1.27    \\
         &VP+SE(ASHIFT\footnote{Occupation of atomic fragment was $6s^26p^2$}) & -8.97                   & 0.02        & -0.07 \\ 
         &VP+SE(ASHIFT\footnote{Using the occupations of Table \ref{tab:total_charge}}) & -9.06                   & -0.08        & 0.31 \\ 
\end{tabular}
\end{ruledtabular}
\end{table*}
\begin{table*}
\caption{Reaction energy of \ce{FlH4 -> FlH2 + H2} (in kcal/mol).
$\Delta$DCG refers to the difference between DCG and NR. Other $\Delta$ refers to the difference from the DCG value.}\label{tab:reac_Fl}
\begin{ruledtabular}
\begin{tabular}{llrrr}
       &QED effect& reac. energy  & $\Delta$(kcal/mol) & $\Delta$(\%) \\ \hline
NR     &none & 9.52                    &             &       \\
DCG & none & -60.02                  & -69.54      &       \\
   &VP      &    -60.43             &        -0.41   & 0.59    \\
   &SE      &     -59.27     &       0.75 & -1.08   \\
   &VP+SE    & -59.67                  & 0.35        & -0.50 \\
   &VP+SE(ASHIFT\footnote{Occupation of the atomic fragment was $7s^27p^2$}) & -60.10                  & -0.07       & 0.10  \\ 
   &VP+SE(ASHIFT\footnote{Using the occupations of Table \ref{tab:total_charge}}) & -60.23                  & -0.21       & 0.30 \\ 
\end{tabular}
\end{ruledtabular}
\end{table*}

As a final case study, we consider the reaction energy of \ce{XH4 -> XH2 + H2} to which Dyall \textit{et al.} proposed that the Lamb shift could make a chemically significant contribution.\cite{dyall_cpl2001}  Their argument was based on the observation that QED effects are most important for $s$ orbitals, as seen in Tables \ref{tab:table_ene} and \ref{tab:table_rms}, and that this is a reaction with a significant change of the valence $s$ population of a heavy element. We have investigated this at the B3LYP/dyall.3zp level and also included the corresponding reaction involving the heavier homologue flerovium.
Optimized equilibrium structures are given in Table \ref{tab:Pb_Fl_opt}. For the
tetrahydrides we assumed $T_d$ symmetry, in line with experiment\cite{wang_jacs2003}(\ce{PbH4})
and previous calculation\cite{schwerdtfeger_jnrs2002}(\ce{FlH4}).

To monitor valence $s$ populations we carried out bonding analysis in Pipek--Mezey localized MOs.\cite{dubillard_jcp2006,pipek_jcp1989}
From Table \ref{tab:total_charge}, the change of the valence $s$ population from \ce{XH4} to \ce{XH2} is 0.45 and 0.30 for Pb and Fl systems, respectively. From Table \ref{tab:XHn_pop}, one sees that in the tetrahydrides the valence $s$ population is contained in the four $\sigma_{XH}$ bonds. In contrast, in the dihydrides the two $\sigma_{XH}$ bonds are mediated by the valence $p$ orbitals of the metals, and most of the valence $s$ population is found in a non-bonding (nb) orbital.

Turning next to Tables \ref{tab:reac_Pb} and \ref{tab:reac_Fl} we see that both reactions are endothermic at the non-relativistic level, but becomes clearly exothermic when adding relativity. For \ce{Pb} QED reduces the relativistic effect by 1.25\%. Its value is 0.32 kcal/mol, which is at the lower end of the perturbation estimate of Dyall \textit{et al.}\cite{dyall_cpl2001} For \ce{Fl} QED reduces the relativistic effect by 0.50\%. Interestingly, its value is very close to that for the \ce{Pb} reaction, despite \ce{Fl} being a much heavier atom. The reason for this unexpected result is the cancellation between the SE and VP effects. From Tables \ref{tab:reac_Pb} and \ref{tab:reac_Fl}, the ratio of VP and SE is $\sim$ 1:$-4.2$ for the Pb system, while it is $\sim$ 1:$-1.8$ for the Fl system. Discussion along these lines is also found in Refs.~\citenum{thierfelder2010,skripnikov_jcp155.144103}. Finally, we note from Tables \ref{tab:reac_Pb} and \ref{tab:reac_Fl} that the the atomic shift operator (ASHIFT), either using atomic ground state occupations or the effective atomic configuration in the molecules given in Table \ref{tab:total_charge}, is not
reliable for describing QED effects in the molecules.

%
%
\section{Conclusions}\label{sec:conc}
We have implemented effective QED potentials for relativistic molecular calculations by grafting code from the numerical atomic code GRASP onto the DFT grid of DIRAC. A general disadvantage of  numerical integration is higher computational cost than analytical evaluation, to the extent that such expressions are available, although the implementation itself is easier and considerable savings are achieved by the locality of the
effective QED potentials.

We report several applications of the new code, mostly using the the molecular mean-field approximation Hamiltonian (X2Cmmf). We demonstrate (Table \ref{tab:table_ene}) that with
proper picture-change transformation of the effective QED potentials, our 2-component
relativistic results reproduce very well 4-component reference data. On the other hand,
this transformation is mandatory since picture-change errors are sizable.

We confirm that the discrepancy between the accurate calculations of Kirk Peterson and
co-workers\cite{hill2013jcp_aucn} and experiment\cite{okabayashi2009detection} is due to
QED by directly calculating the ground-state rotational constants $B_0$ for the
isotopomers investigated in the MW experiment. We then find that QED reduces the discrepancy
of the corresponding substitution Au-C bond length $r_s$ from 0.23 to 0.04 pm  with respect to experiment.

For the rare-gas dimers \ce{Hg2} and \ce{Rn2} we find that QED increases bond lengths
by about 0.15 pm. For the superheavy homologues the bond length increase is on the order
of 0.30 pm; the effect on dissociation energies is quite small ($\sim$0.4 \%).

We have also investigated the effect of QED on the reaction energy of \ce{XH4 -> XH2 + H2}, (X=Pb,Fl). From projection analysis we do find that there is a significant change of valence
$s$ population of the metals during the reaction, in line with the proposition of
Dyall and co-workers.\cite{dyall_cpl2001} Interestingly, though, we also find that in the
tetrahydrides the valence $s$ population essentially resides in bonding orbitals, but in non-bonding ones
in the dihydrides. We find for the dissociation of lead tetrahydride that QED reduces the magnitude
of the reaction energy by 0.32 kcal/mol (-1.27 \%); for the superheavy homologue the
magnitude of the QED effect is basically the same (0.35  kcal/mol). This possibly surprising
observation is explained by the reduction of the (negative) SE/VP ratio with increasing nuclear charge.

For these metal hydrides, and also \ce{AuCN}, we have also tried a simpler approach for the incorporation of QED effects in molecular calculations in the form of an atomic shift operator,
but we find that this is not a reliable approach.

We would like to stress that our implementation of effective QED potential is general in the sense that they are available in all parts of the code. A natural continuation of our project will
therefore be to explore the impact of these potentials on molecular properties probing electron density in the vicinity of nuclei, where the QED effects are generated.
Our results so far indicate that QED effects may be more important than for the valence properties reported in the present work.
For instance, the QED effect on the parity violation energy of \ce{H2Po2} is 2.38 \%, although it depends on the choice of effective QED potentials.\cite{sunaga_mp2021}

\begin{acknowledgments}
This project has received funding from the European Research Council (ERC) under the European Union's Horizon 2020 research and innovation program (grant agreement No 101019907 HAMP-vQED), as well as from the Agence Nationale de la Recherche (ANR-17-CE29-0004-01 molQED). AS acknowledges financial support from the Japan Society for the Promotion of Science (JSPS) KAKENHI Grant No. 17J02767, 20K22553, and 21K1464, and JSPS Overseas Challenge Program for Young Researchers, Grant No. 201880193. Computing time from CALMIP (Calcul en Midi-Pyrenes), supercomputer of ACCMS (Kyoto University) and Research Institute for Information Technology, Kyushu University (General Projects) are gratefully acknowledged. We would like to thank Kirk Peterson (Washington State), Toshiaki Okabayashi (Shizuoka), Pekka Pyykk\"{o} (Helsinki), Jacinda Ginges (Brisbane) and Radovan Bast (Troms{\o}), for valuable discussions. 
\end{acknowledgments}

\section*{Author's contributions}
The overall project was conceived and supervised by TS. All programming and calculations were carried out by AS, whereas MS is the main contributor to the theory sections, notably the appendix.

\section*{Data availability}\label{sec:SIData}
The data that support the findings of this study are openly available in ZENODO at \url{https://doi.org/10.5281/zenodo.6874728}, see also Ref.~\citenum{ReplicationData}.
The DIRAC outputs contain information about the precise build of the corresponding executable and, most importantly, the git commit hash, which provides
a unique identifier of the precise version of DIRAC generating the output, hence allowing reproduction of the results. We have also included the potential surface generated by
Hill \textit{et al.}\cite{hill2013jcp_aucn} for \ce{AuCN}, kindly made available by Kirk Peterson.

\section*{Conflicts of interest}
The authors have no conflicts to disclose.

\appendix \section{Theory background} \label{section:appendix}

Since we hope to reach a wider audience than QED specialists, we provide in this Appendix a compact, yet accessible introduction (crash course) to QED that would otherwise have necessitated consulting disparate sources. 
More precisely, in this Appendix, we shall discuss the lowest-order BSQED corrections,
and show how the effective potentials associated
with these QED processes can be derived within the scattering matrix
(${\cal S}$-matrix) formalism. These effective potentials are to
be used in practical relativistic calculations in order to account
for the physics that is missing from the Dirac theory. In ``conventional''
QED one studies how the free-electron field interacts with the free
quantized electromagnetic field and/or with an potential source (the
scattering problem). On the other hand, BSQED theory studies the same problem
but with electrons that are already interacting with some time-independent
external field, i.e. their wavefunctions are solutions to the bound-state
Dirac equation instead of the free one. This is known as the Furry picture of quantum
mechanics; see for instance Refs.~\citenum{furry1951-picture} and \citenum{schweber_rqft} (section 15g).

We shall use unbold symbols for four-quantities such as spacetime points (events):  $x=\left(ct,\boldsymbol{x}\right)$, here in contravariant coordinates, where $\boldsymbol{x}$
is the spatial position vector, and the contravariant metric tensor $g^{\mu\nu}=\text{diag}\left(+1,-1,-1,-1\right)$. The gamma matrices are
defined through their anti-commutation relation
\begin{equation}
  \gamma^{\mu}\gamma^{\nu}+\gamma^{\nu}\gamma^{\mu}=2g^{\mu\nu}\mathds{1}_{4}.
\end{equation}  
In Dirac basis they are represented by $\gamma^0=\beta$ and $\boldsymbol{\gamma}=\beta\boldsymbol{\alpha}$. Following Lindgren\cite{lindgren_rmbt2016}
we shall complement the Dirac $\boldsymbol{\alpha}$ matrices with $\alpha^0=\mathds{1}_{4}$ to form a pseudo-4-vector. 
We finally note that we put hats ($\,\hat{}\,$) on quantities that contain creation/annihilation operators acting on occupation number states. Contrary to conventional QED sources, we have decided to express the formalism in full SI units.

\subsection{Electron field operator }

\label{subsec:Electron-field-operator}

The electron field operator is given by the following annihilation
expansion over all solutions of the Dirac equation
\begin{equation}
\hat{\Psi}\left(x\right)=\sum_{i}\psi_{i}\left(x\right)c_{i},\;\text{with}\;\psi_{i}\left(x\right)=\psi_{i}\left(\boldsymbol{x}\right)e^{-\frac{i}{\hbar}E_{i}t}.\label{eqn:e-field}
\end{equation}
In this expression, $c_{i}$ is the electron annihilation operator
obeying the fermionic algebra relations
\begin{equation}
\begin{aligned}\{c_{i},c_{j}^{\dagger}\} & =\delta_{ij},\,\,\, \text{and} \,\,\, \{c_{i},c_{j}\}=\{c_{i}^{\dagger},c_{j}^{\dagger}\}=0,\end{aligned}
\end{equation}
and associated with the $i$-th spatial wavefunction $\psi_{i}\left(\boldsymbol{x}\right)$
and energy-level $E_{i}$ that solve the time-independent Dirac equation in the presence of a time-independent external four-potential $A^e = (\varphi^e/c,\boldsymbol{A}^e)$
\begin{equation}
\begin{aligned}H_{D}\psi_{i}\left(\boldsymbol{x}\right) & =E_{i}\psi_{i}\left(\boldsymbol{x}\right);\qquad\text{with }\\
H_{D}&=c\boldsymbol{\alpha}\cdot\left(-i\hbar\boldsymbol{\nabla}+e\boldsymbol{A}^e\left(\boldsymbol{x}\right)\right)-e\varphi^e\left(\boldsymbol{x}\right)+\beta m_{e}c^{2}.
\end{aligned}
\label{eqn:bound-dirac}
\end{equation}
The electron vacuum state is defined to be the one that vanishes after any
annihilation:
\begin{equation}
c_{i}\big|0_{\text{e}}\big>=0,\,\,\forall i.
\end{equation}

In order to forbid the transition of positive-energy electrons to
the negative-energy continuum by the Pauli exclusion principle, and
obtain a stable atomic theory, Dirac\cite{dirac1930theory}
postulated that this continuum should be totally filled with electrons
that are not observed (Dirac sea). This means that the vacuum state
is redefined to be the state containing no positive-energy electrons
and a fully occupied negative-energy electron sea. Dirac then argued that when a negative-energy electron absorbs enough
energy ($E\geq 2m_ec^{2}$), it becomes real (observable), and leaves,
for mass- and charge-conservation reasons, a positron behind (Dirac
hole theory).\cite{dirac1931quantised} This last
reasoning allows one to define\cite{FurryOppenheimer1934}
\begin{equation}
\begin{aligned}c_{i} & =a_{i},\quad\text{for}\quad E_{i}>0,\\
c_{i} & =b_{i}^{\dagger},\quad\text{for}\quad E_{i}<0.
\end{aligned}
\label{eqn:Furry-Oppenheimer}
\end{equation}
Here, operators $a_{i}$ and $b_{i}$ are introduced to distinguish
between the particle (electron) and its hole (positron), and the second
line indicates that the annihilation of a negative-energy electron
with $\ensuremath{c_{i}}$ is equivalent to the creation of a (positive-energy)
hole (positron) with $b_{i}^{\dagger}$. The electron field operator
of Eq.~\eqref{eqn:e-field} can be written, with respect to these
definitions, as
\begin{equation}
\hat{\Psi}\left(x\right)=\sum_{E_{i}>0}\psi_{i}\left(x\right)a_{i}+\sum_{E_{i}<0}\psi_{i}\left(x\right)b_{i}^{\dagger}.\label{eqn:e-field-1}
\end{equation}
Despite its experimental success in predicting the existence of the
positron,\cite{AndersonPositiveElectron1933} 
the hole theory (its physical implications) was, shortly after its
introduction, abandoned. Many physicists including Pauli, Bohr, Weisskopf,
Heisenberg and Majorana, opposed this theory, as clearly indicated
in Refs.~\citenum{schweberQEDandTheMenWhoMadeIt1994}
(section 1.6), ~\citenum{PauliWeisskopf1998quantization}, \citenum{miller1995earlyQED} (section 4.4) and \citenum{Majorana2006}.
This opposition came mainly from the following flaws of the Dirac
hole theory: 1) the existence of a non-observable infinite negative
energy and charge and 2) for massive boson systems, whose wavefunctions
satisfy the Klein-Gordon equation, the Dirac argument would not hold,
and the existence of these bosons is not justified. Modern quantum
field theory reached the same mathematical expressions derived with
respect to Dirac's hole theory, but provided a more symmetric picture
between electrons and positrons, in which 1) one only sees electrons and positrons with positive energies, 2) the infinite negative-energy electron
sea assumption is no longer necessary, and 3) operators such as the
Hamiltonian, and charge are replaced by their normal-ordered forms.
This physical interpretation leads to the modern definition of the
vacuum state, that obeys
\begin{equation}
a_{i}\big|0_{\text{e}}\big>=b_{i}\big|0_{\text{e}}\big>=0,\,\,\forall i,
\end{equation}
and contains zero positive-energy electrons and positrons. To get a wider and more detailed vision of the historical development
of the quantum field theory, the reader may consult Weinberg in Ref.~\citenum{weinberg_1995}
(section 1.2 and chapter 5) and Ref.~\citenum{weinberg1977search}, Mehra
in Ref.~\citenum{Mehra2001} (chapter 29), Schweber in Ref.~\citenum{schweberQEDandTheMenWhoMadeIt1994},
Kragh in Ref.~\citenum{kragh1990dirac} and Weisskopf in Ref.~\citenum{Weisskopf_development}.

\subsection{Photon field operator}

\label{subsec:Photon-field-operator}

The photon field operator is written as a sum over positive and negative plane-wave Fourier modes
\begin{align}
\hat{A}_{\mu}\left(x\right) & =\hat{A}_{\mu}^{+}\left(x\right)+\hat{A}_{\mu}^{-}\left(x\right),\label{eqn:photon-field}\\
\hat{A}_{\mu}^{+}\left(x\right) & =\sum_{r=0}^{3}\sum_{\boldsymbol{k}}N_{\boldsymbol{k}} a\left(\boldsymbol{k},r\right)\varepsilon_{\mu}\left(\boldsymbol{k},r\right)e^{-ik\cdot x},\label{eqn:photon-field+}\\
\hat{A}_{\mu}^{-}\left(x\right) & =\sum_{r=0}^{3}\sum_{\boldsymbol{k}} N_{\boldsymbol{k}} a^{\dagger}\left(\boldsymbol{k},r\right)\varepsilon_{\mu}\left(\boldsymbol{k},r\right)e^{+ik\cdot x},\label{eqn:photon-field-}
\end{align}
where $N_{\boldsymbol{k}}=\sqrt{\hbar/(2\epsilon_{0}\omega_{\boldsymbol{k}}V)}$ is the normalization constant, the zeroth component four-wave vector is $k_{0}=\left|\boldsymbol{k}\right|=\omega_{\boldsymbol{k}}/c$,
$\varepsilon_{\mu}\left(\boldsymbol{k},r\right)$ are the four polarization
vectors, and $a\left(\boldsymbol{k},r\right)$ (and $a^{\dagger}\left(\boldsymbol{k},r\right)$)
is the annihilation (creation) operator that annihilates (creates) a photon with wave vector $\boldsymbol{k}$ and polarization
$r$, respectively (see Refs.~\citenum{mandl_qft2010} (eqs.(5.16a-c)) and \citenum{ohlsson_rqp2011} (section 8.4)). 
The choice of $k_{0}$ is imposed by the fact that the photon
field operator must satisfy the Maxwell equation
\begin{equation}
\square\hat{A}_{\mu}\left(x\right)=0,\quad \text{with} \quad \square = \frac{1}{c^2}\frac{\partial^2}{{\partial t}^2}-\boldsymbol{\nabla}^2,
\end{equation}
obtained after setting the Lorenz gauge condition ($\partial^\mu \hat{A}_\mu=0$). This equation leads
to the (massless) photon energy-momentum relation
\begin{equation}
k^{2}=0.
\end{equation}
The boson creation and annihilation operators do satisfy the following bosonic
commutation relations :
\begin{align}
\left[a\left(\boldsymbol{k},r\right),a^{\dagger}\left(\boldsymbol{k}^{\prime},s\right)\right] & =\delta_{rs}\zeta_{r}\delta_{\boldsymbol{k},\boldsymbol{k}^{\prime}}\\
\left[a^{\dagger}\left(\boldsymbol{k},r\right),a^{\dagger}\left(\boldsymbol{k}^{\prime},s\right)\right] & =\left[a\left(\boldsymbol{k},r\right),a\left(\boldsymbol{k}^{\prime},s\right)\right]=0.
\end{align}
Here, $\zeta_{r}$ is a function defined by the following relation
\begin{equation}
\zeta_{r}=\begin{cases}
+1 & r=0\\
-1 & r=1,2,3
\end{cases},
\end{equation}
and the polarization vectors satisfy the following completeness relation
\begin{align}
\sum_{r=0}^{3}\zeta_{r}\epsilon^{\mu}\left(\boldsymbol{k},r\right)\epsilon^{\nu}\left(\boldsymbol{k},r\right) & =-g^{\mu\nu}.
\end{align}
Finally, we note that the photon vacuum state is defined to be the
state that satisfies the following relation
\begin{equation}
\begin{aligned} & a\left(\boldsymbol{k},r\right)\left|0_{\text{p}}\right\rangle =0,\;\forall\;\boldsymbol{k},r\quad\\
 & \rightarrow\hat{A}_{\mu}^{+}\left(x\right)\left|0_{\text{p}}\right\rangle =0,\;\forall\;\mu,x.
\end{aligned}
\end{equation}
We shall now consider the interaction between the non-interacting
electron and photon fields and show how one can derive QED corrections
using perturbation theory.

\subsection{Perturbation theory}

\label{subsec:Perturbation-theory}

As in conventional perturbation theory, one wants to get the eigensolutions
of the following total Hamiltonian
\begin{equation}
\hat{H}_{S}=\hat{H}_{S}^{0}+\lambda\hat{H}_{S}^{1}.\label{eqn:Hs}
\end{equation}
The zeroth-order Hamiltonian
\begin{align}
\hat{H}_{S}^{0} & =\hat{H}_{\text{electron}}^{0}+\hat{H}_{\text{photon}}^{0},
\end{align}
represents the free electron and photon fields. 
The electronic part is given by a spatial integral over the normal-ordered Dirac Hamiltonian density
\begin{equation}
\begin{aligned}\hat{H}_{\text{\text{electron}}}^{0} & =\int d^{3}x:\hat{\Psi}^{\dagger}\left(x\right)H_{D} \left(\boldsymbol{x}\right) \hat{\Psi}\left(x\right):\\
 & =\sum_{E_{i}>0}E_{i}a_{i}^{\dagger}a_{i}-\sum_{E_{i}<0}E_{i}b_{i}^{\dagger}b_{i},
\end{aligned}
\label{eqn:H0e}
\end{equation}
where in BSQED, $H_{D}$ is the Dirac Hamiltonian in the presence of the external four-potential $A^e = (\varphi^e/c,\boldsymbol{A}^e)$, given in Eq.~\eqref{eqn:bound-dirac},
and where normal-ordering is indicated by double dots. The free photon Hamiltonian is written as an integral of
the electromagnetic Hamiltonian density
\begin{equation}
\begin{aligned} & \hat{H}_{\text{photon}}^{0}=\frac{1}{\mu_{0}}\int d^{3}x:\bigg[-(\partial^{0}\hat{A}^{\mu}(x))(\partial_{0}\hat{A}_{\mu}(x))\\
 & \qquad\qquad\qquad\qquad+\frac{1}{2}(\partial^{\nu}\hat{A}^{\mu}(x))(\partial_{\nu}\hat{A}_{\mu}(x))\bigg]:\\
 & =\sum_{\boldsymbol{k}}\sum_{r=0}^{3}\hbar\omega_{\boldsymbol{k}}\zeta_{r}a^{\dagger}\left(\boldsymbol{k},r\right)a\left(\boldsymbol{k},r\right).
\end{aligned}
\end{equation}
For further details and discussions on the photon Hamiltonian, the reader may consult  Greiner and Reinhardt in Ref.~\citenum{greiner_fieldquantization} (section 7.3) and Mandl and Shaw in Ref.~\citenum{mandl_qft2010} (chapter 5).

The perturbation Hamiltonian $\hat{H}_{S}^{1}$ complicates the problem, and prevents us from obtaining eigensolutions of the full Hamiltonian $\hat{H}_{S}$. $\lambda$ is a dimensionless
parameter that can be varied between $0$ and $1$, and which keeps track of the perturbation-order. This parameter is to be taken
to $1$ in order to account for the full perturbation by the end of
the calculation. Notice that so far our Hamiltonians have an $S$
subscript; this is made to indicate that they are in the Schr\"{o}dinger 
picture of quantum mechanics.%
{} Assuming that we know the eigensolutions of the unperturbed time-independent problem equation
\begin{align}
\begin{aligned}\hat{H}_{S}^{0}\left|\Phi_{0}^{\alpha}\right\rangle _{S} & =E_{0}^{\alpha}\left|\Phi_{0}^{\alpha}\right\rangle _{S};\\
\left|\Phi_{0}^{\alpha}\left(t\right)\right\rangle _{S} & =e^{-iE_{0}^{\alpha}t/\hbar}\left|\Phi_{0}^{\alpha}\right\rangle _{S},
\end{aligned}
\label{eq:qed_eq1}
\end{align}
where the $\alpha$ superscript labels solutions (states and associated
energy-levels), the ultimate goal is to find eigensolutions of the
perturbed problem
\begin{align}
\begin{aligned}\hat{H}_{S}\left|\Phi^{\alpha}\right\rangle _{S} & =E^{\alpha}\left|\Phi^{\alpha}\right\rangle _{S};\\
\left|\Phi^{\alpha}\left(t\right)\right\rangle _{S} & =e^{-iE^{\alpha}t/\hbar}\left|\Phi^{\alpha}\right\rangle _{S}.
\end{aligned}
\label{eq:qed_eq2}
\end{align}

Gell--Mann and Low provided a closed form of the perturbed eigensolutions $\left(E^{\alpha},\left|\Phi^{\alpha}\right\rangle _{S}\right)$ in terms of the unperturbed ones $\left(E_{0}^{\alpha},\left|\Phi_{0}^{\alpha}\right\rangle _{S}\right)$ and the time-evolution operator;\cite{gellmannandlow1951} see also Refs.~\citenum{fetter2012quantum} (pages 61-64) and  \citenum{schweber_rqft} (section 11f.). A few years later, Sucher \cite{sucher1957} provided an expression of the perturbation energy-shift that is more symmetric in time
\begin{equation}
\begin{aligned}\Delta E^{\alpha} & =E^{\alpha}-E_{0}^{\alpha}\\
 & =\lim_{\substack{\epsilon\rightarrow0\\
\lambda\rightarrow1
}
}\frac{i\epsilon\lambda}{2}\frac{\partial}{\partial\lambda}\log\left\langle \Phi_{0}^{\alpha}\left|\hat{{\cal S}}\left(\epsilon,\lambda\right)\right|\Phi_{0}^{\alpha}\right\rangle ,
\end{aligned}\label{eqn:sucher-energy}
\end{equation}
where $\epsilon$ is an energy-parameter, to be shortly discussed.
This energy-shift expression contains the ${\cal S}$-matrix operator
that is defined to be the time-evolution operator that takes the interaction
state from the very past $t=-\infty$ to the very future $t=+\infty$,
and can be written as (see Dyson in Ref.~\citenum{dyson_s-mat_1949} eq.(4))
\begin{equation}
\hat{{\cal S}}\left(\epsilon,\lambda\right)=\text{T}\bigg[\exp\big(\frac{\lambda}{i\hbar c}\int d^{4}xe^{-\frac{\epsilon}{\hbar}\left|t\right|}\hat{{\cal H}}_{I}\left(x\right)\big)\bigg].\label{eqn:S-mat-T}
\end{equation}

In this expression, $\text{T}$ stands for time-ordering, i.e., it re-orders the inside operators such that those associated with earlier times act first. In the simplest case of two operators, the time-ordering operation is defined to be
\begin{equation}
\begin{aligned}\text{T}\left[\hat{A}\left(x_{1}\right)\hat{B}\left(x_{2}\right)\right] & \equiv\Theta\left(t_{1}-t_{2}\right)\hat{A}\left(x_{1}\right)\hat{B}\left(x_{2}\right)\\
 & \pm\Theta\left(t_{2}-t_{1}\right)\hat{B}\left(x_{2}\right)\hat{A}\left(x_{1}\right),
\end{aligned}
\end{equation}
where the minus sign applies when both operators
$\hat{A}$ and $\hat{B}$ are of fermionic nature. Furthermore, the $\cal S$-matrix in Eq.~\eqref{eqn:S-mat-T} is a functional of the interaction-Hamiltonian density $\hat{{\cal H}}_{I}\left(x\right)$,
that is related to the interaction Hamiltonian $\hat{H}_{I}^{1}\left(t\right)$ by the following integral
\begin{equation}
\hat{H}_{I}^{1}\left(t\right)=\int d^{3}x\hat{{\cal H}}_{I}\left(x\right).
\end{equation} 
Recall that $\hat{H}_{I}^{1}\left(t\right)$ is the interaction-picture version of the Schr\"{o}dinger-picture interaction-Hamiltonian $\hat{H}_{S}^{1}$ of Eq.~\eqref{eqn:Hs}.

We shall note that the interaction density is multiplied by a damping factor $e^{-\frac{\epsilon}{\hbar}\left|t\right|}$, cf. Eq.~\eqref{eqn:S-mat-T}, where $\epsilon$ is a small positive quantity that has energy dimensions. This term is known as the ``adiabatic switch'' that allows the interpolation between the perturbed and unperturbed problems ($t=0,\pm\infty$), and was first introduced by Gell-Mann and Low in Ref.~\citenum{gellmannandlow1951} (Appendix \ref{section:appendix}) while extending the ${\cal S}$-matrix formalism to cover the bound-electron problem (see also Ref.~\citenum{labzowsky1993relativistic} (section 1.3)).
The scattering matrix of Eq.~\eqref{eqn:S-mat-T} may be expanded in powers of the perturbation parameter $\lambda$ as
\begin{widetext}
\begin{equation}
\begin{aligned}\hat{{\cal S}}\left(\epsilon,\lambda\right) & =\sum_{n=0}^{\infty}\hat{{\cal S}}^{\left(n\right)}\left(\epsilon,\lambda\right)\\
\hat{{\cal S}}^{\left(n\right)}\left(\epsilon,\lambda\right) & =\frac{1}{n!}\left(\frac{\lambda}{i\hbar c}\right)^{n}\int d^{4}x_{1}\ldots\int d^{4}x_{n}e^{-\frac{\epsilon}{\hbar}\left(\left|t_{1}\right|+\ldots\left|t_{n}\right|\right)}\text{T}\left[\hat{{\cal H}}_{I}\left(x_{1}\right)\ldots\hat{{\cal H}}_{I}\left(x_{n}\right)\right].
\end{aligned}
\label{eqn:s-matrix-expansion}
\end{equation}
\end{widetext}

This form of the $\hat{{\cal S}}$-matrix expansion is known as the Dyson series, and originated from the works of Dyson \cite{dyson_s-mat_1949,dyson-qed_tomonaga-schwinger-feynman1949}
and Schwinger.\cite{schwinger_qed_covariant_formulation_1948} Detailed
derivations of the time-evolution and $\hat{{\cal S}}$-matrix operators
can be found in Fetter and Walecka Ref.~\citenum{fetter2012quantum} (pages 54-58),
Mandl and Shaw Ref.~\citenum{mandl_qft2010} (section 6.2), as well as Bjorken and Drell Ref.~\citenum{bjorken1965relativistic} (section 17.2). In QED, the (perturbation) interaction-Hamiltonian density is given by
\begin{equation}
\begin{aligned}\hat{{\cal H}}_{I}\left(x\right) & =\hat{J}_{\mu}\left(x\right)\hat{A}^{\mu}\left(x\right),\\
\text{with }\hat{J}_{\mu}\left(x\right) & =-ec\bar{\hat{\Psi}}\left(x\right)\gamma_{\mu}\hat{\Psi}\left(x\right),
\end{aligned}
\label{eqn:QED-H-density}
\end{equation}
that explicitly couples the quantized electron-current field operator $\hat{J}_{\mu}$ to the photon field operator $\hat{A}^{\mu}$.
Some authors, starting with Schwinger in Ref.~\citenum{schwinger_qed_covariant_formulation_1948} (Eq.(1.14)), use the symmetrized form 
\begin{equation}
  \hat{J}_{\mu}\left(x\right) =-\frac{ec}{2}[\bar{\hat{\Psi}}_{\alpha}\left(x\right),\hat{\Psi}_{\beta}\left(x\right)][\gamma_{\mu}]_{\alpha\beta},
\end{equation}
for the electron-current field operator, but the two forms are equivalent under time-ordering (see Eq.~(29) of Ref.~\citenum{mohr1998qed}).
We recall that the electron and photon field operators are given in Eqs.~\eqref{eqn:e-field-1} and \eqref{eqn:photon-field}, respectively.
We note that the Dirac field operator with a bar on the top represents the Dirac adjoint field: $\bar{\hat{\Psi}}\left(x\right)=\hat{\Psi}^\dagger\left(x\right)\gamma^0$. At this point, the reader can see, from the last two equations, that the QED theory treats the electron-photon field (interaction) coupling perturbatively, in powers of the elementary charge $e$.  

We next consider how to expand the time-ordered product of the $\cal{\hat{S}}$-matrix, and assign each of the obtained normal-ordered terms to a specific Feynman diagram.

\subsection{Wick's theorem, field contractions and propagators}

\label{subsec:Wick's-theorem,-field}

Wick's theorem\cite{wick_physrev1950} allows writing
the time-ordered products of Eq.~\eqref{eqn:s-matrix-expansion} in
terms of normal-ordered products of all possible contractions, as given
in the following equation
\begin{align}
 & \text{T}\left[\hat{O}\left(x_{1}\right)\hat{O}\left(x_{2}\right)\hat{O}\left(x_{3}\right)\hat{O}\left(x_{4}\right)\ldots\right]\nonumber \\
= & \,:\hat{O}\left(x_{1}\right)\hat{O}\left(x_{2}\right)\hat{O}\left(x_{3}\right)\hat{O}\left(x_{4}\right)\ldots:\nonumber \\
+ & :\contraction{}{\hat{O}}{\left(x_{1}\right)}{\hat{O}}\hat{O}\left(x_{1}\right)\hat{O}\left(x_{2}\right)\hat{O}\left(x_{3}\right)\hat{O}\left(x_{4}\right)\ldots:+\ldots\nonumber \\
+ & :\contraction{}{\hat{O}}{\left(x_{1}\right)\hat{O}\left(x_{2}\right)}{\hat{O}}\contraction[1.5ex]{\hat{O}\left(x_{1}\right)}{\hat{O}}{\left(x_{2}\right)\hat{O}\left(x_{3}\right)}{\hat{O}}\hat{O}\left(x_{1}\right)\hat{O}\left(x_{2}\right)\hat{O}\left(x_{3}\right)\hat{O}\left(x_{4}\right)\ldots:+\ldots
\end{align}

Contracted operators are moved next to each other, noting that under normal-ordering (fermionic) bosonic operators can be permuted as if they (anti)commuted.
A contraction is represented by a line that links two operators and is defined to
be the vacuum expectation value of the time-ordered product
\begin{equation}
\contraction{}{\hat{O}}{\left(x_{1}\right)}{\hat{O}}\hat{O}\left(x_{1}\right)\hat{O}\left(x_{2}\right)\equiv\big<0\big|\text{T}\big[\hat{O}\left(x_{1}\right)\hat{O}\left(x_{2}\right)\big]\big|0\big>.
\end{equation}
Since our QED interaction-Hamiltonian density contains electron and photon operators, the time-ordered product in our ${\cal S}$-matrix of Eq.~\eqref{eqn:s-matrix-expansion} will be expanded with two types of contractions: electronic and photonic. The contraction of two electron field operators (of Eq.~\eqref{eqn:e-field-1}) components $\alpha$ and $\beta$ is defined with respect to the last formula by
\begin{equation} 
\begin{aligned}
\contraction{}{\hat{\Psi}}{_\alpha\left(x_{1}\right)}{\bar{\hat{\Psi}}}\hat{\Psi}_\alpha\left(x_{1}\right)\bar{\hat{\Psi}}_\beta \left(x_{2}\right) & \equiv \big<0_{\text{e}}\big|\text{T}\big[\hat{\Psi}_\alpha\left(x_{1}\right)\bar{\hat{\Psi}}_\beta \left(x_{2}\right)\big]\big|0_{\text{e}}\big> \\ & =i\hbar\left[S_{A^e}^{F}\left(x_1,x_2\right)\right]_{\alpha\beta} , \label{eqn:electron-contraction} 
\end{aligned} 
\end{equation}
where $\left[S_{A^e}^{F}\left(x,y\right)\right]_{\alpha\beta}$ is the $\alpha,\beta$ matrix component of the Feynman electron propagator, which in turn
satisfies the Dirac propagator equation
\begin{equation}
\left[\gamma^{\mu}\left(i\hbar\partial_{\mu}+eA^e_{\mu}\left(\boldsymbol{x}\right)\right)-m_e c\right]S_{A^e}^{F}\left(x,y\right)=\mathds{1}_{4}\delta\left(x-y\right);\label{eqn:electron-contraction-equation}
\end{equation}
cf. Eq.~\eqref{eqn:bound-dirac}. Furthermore, one can show that the following identities
\begin{equation}
\begin{aligned}\contraction{}{\bar{\hat{\Psi}}}{_{\beta}\left(x_{2}\right)}{\hat{\Psi}}\bar{\hat{\Psi}}_{\beta}\left(x_{2}\right)\hat{\Psi}_{\alpha}\left(x_{1}\right) & =-\contraction{}{\hat{\Psi}}{_{\alpha}\left(x_{1}\right)}{\bar{\hat{\Psi}}}\hat{\Psi}_{\alpha}\left(x_{1}\right)\bar{\hat{\Psi}}_{\beta}\left(x_{2}\right),\\
\contraction{}{\hat{\Psi}}{_{\alpha}\left(x_{1}\right)}{\hat{\Psi}}\hat{\Psi}_{\alpha}\left(x_{1}\right)\hat{\Psi}_{\beta}\left(x_{2}\right) & =\contraction{}{\bar{\hat{\Psi}}}{_{\alpha}\left(x_{1}\right)}{\bar{\hat{\Psi}}}\bar{\hat{\Psi}}_{\alpha}\left(x_{1}\right)\bar{\hat{\Psi}}_{\beta}\left(x_{2}\right)=0,
\end{aligned}
\end{equation}
hold. These relations show that the only non-zero contractions are between electron field operators and their adjoints. 
The free Feynman electron propagator $S_{0}^{F}\left(x,y\right)$, corresponding to the case $A_{\mu}\left(\boldsymbol{x}\right)=0_\mu$, can be written as
\begin{equation} \begin{aligned} & S_{0}^{F}\left(y,x\right)=\lim_{\epsilon\rightarrow0}\int\frac{d^{4}p}{\left(2\pi\hbar\right)^{4}}e^{-\frac{i}{\hbar}p\cdot\left(y-x\right)}S_{0}^{F}\left(p\right),\\ & \quad\text{with}\quad S_{0}^{F}\left(p\right)=\text{\ensuremath{\frac{\gamma^{\mu}p_{\mu}+m_e c}{p^{2}-{m_e}^{2}c^{2}+i\epsilon}}} ,\end{aligned} \label{eqn:Free-electron-propagator}
\end{equation}
where $ S_{0}^{F}\left(p\right)$ is the Fourier transformed free-electron propagator. The role of the small positive number $\epsilon$ is to shift
energy-poles (at the energy-momentum relation) with respect to the Feynman prescription.

Similarly, the contraction of two photon operators (of Eq.~\eqref{eqn:photon-field}) is defined by the
following expression
\begin{equation} 
\begin{aligned}
\contraction{}{\hat{A}}{_\mu\left(x_{1}\right)}{\hat{A}}\hat{A}_\mu\left(x_{1}\right)\hat{A}_\nu \left(x_{2}\right) &\equiv \big<0_{\text{p}}\big|\text{T}\big[\hat{A}_\mu\left(x_{1}\right)\hat{A}_\nu \left(x_{2}\right)\big]\big|0_{\text{p}}\big>  \\ & =i\hbar D_{\mu\nu}^{F}\left(x_1,x_2\right), \end{aligned} 
\label{eqn:photon-contraction} 
\end{equation}
where $D_{\mu\nu}^{F}\left(x,y\right)$ is the photon propagator in
the Feynman gauge, is given by the following expression
\begin{equation} \begin{aligned} & D_{\mu\nu}^{F}\left(x,y\right)=\lim_{\epsilon\rightarrow0}\int\frac{d^{4}p}{\left(2\pi\hbar\right)^{4}}e^{-\frac{i}{\hbar}p\cdot\left(x-y\right)}D_{\mu\nu}^{F}\left(p\right),\\ & \quad\text{with}\quad D_{\mu\nu}^{F}\left(p\right)=g_{\mu\nu}D^{F}\left(p\right)=-\frac{\hbar^{2}}{c\epsilon_{0}}\frac{g_{\mu\nu}}{p^{2}+i\epsilon}
\end{aligned} \label{eqn:photon-propagator} \end{equation}
and satisfies the Maxwell Green's-type equation
\begin{equation}
\partial_{\sigma}\partial^{\sigma}D_{\nu\theta}^{F}\left(x,y\right)=\frac{g_{\nu\theta}}{\epsilon_{0}c}\delta\left(x-y\right).\label{eqn:propagator_equation}
\end{equation}
This equation is obtained after imposing the Lorenz gauge condition,
otherwise this propagator will not be invertible; see Schwartz in Ref.~\citenum{schwartz_qft} (section 8.5).
We should finally note that the $F$ superscript on both propagators is added to indicate that these are Feynman propagators. This means that when writing the propagators as Fourier transforms, the energy-integrals are to be taken along the Feynman contour. Different choices of paths (contours) lead to different propagators (retarded and advanced), but they all satisfy the corresponding Dirac and Maxwell equations.

\subsection{Bound electron propagator expansion}
\label{subsec:Bound-electron-propagator-expans}
The bound Feynman propagator $S_{A^e}^{F}\left(x_{2},x_{1}\right)$
of Eq.~\eqref{eqn:electron-contraction} can be expanded in powers of
the external potential as (Refs.~\citenum{itzykson_zuber_qft_1980} eq.(2-119) and 
~\citenum{feynman_1949_positron} eq.(16)):
\begin{equation}
\begin{aligned} & S_{A^e}^{F}\left(x_{2},x_{1}\right)=S_{0}^{F}\left(x_{2},x_{1}\right)\\
 & -\int d^{4}x_{3}S_{0}^{F}\left(x_{2},x_{3}\right)eA^e_{\mu}\left(\boldsymbol{x}_{3}\right)\gamma^{\mu}S_{0}^{F}\left(x_{3},x_{1}\right)+\ldots \, \, ,
\end{aligned}
\label{eqn:propagator-expansion}
\end{equation}
and written in terms of the free-electron propagator $S_{0}^{F}\left(x_{2},x_{1}\right)$ of Eq.~\eqref{eqn:Free-electron-propagator}.
It is worth noting that the bound Feynman propagator can be related
to the bound Dirac Green's function $G_{A^e}$ by the relation of Ref.~\citenum{mohr1998qed} (eq.(32)):
\begin{equation}
S_{A^e}^{F}\left(x_{2},x_{1}\right)=\frac{1}{i\hbar}\frac{1}{2\pi i}\int_{C_{F}}dz\ G_{A^e}\left(\boldsymbol{x}_{2},\boldsymbol{x}_{1};z\right)\gamma^{0}e^{-\frac{i}{\hbar}z\left(t_{2}-t_{1}\right)},\label{eqn:propagator_green}
\end{equation}
This Green's function satisfies
\begin{equation}
\left[H_{D}\left(\boldsymbol{x}_{2}\right)-z\right]G_{A^e}\left(\boldsymbol{x}_{2},\boldsymbol{x}_{1};z\right)=\mathds{1}_{4}\delta\left(\boldsymbol{x}_{2}-\boldsymbol{x}_{1}\right);
\end{equation}
cf. Eqs.~\eqref{eqn:bound-dirac} and \eqref{eqn:electron-contraction-equation}.
Using Eqs.~\eqref{eqn:propagator-expansion} and \eqref{eqn:propagator_green},
and integrating over time variables, one obtains the potential expansion associated with the Green's function
\begin{equation}
\begin{aligned} & G_{A^e}\left(\boldsymbol{x}_{2},\boldsymbol{x}_{1};z\right)=G_{0}\left(\boldsymbol{x}_{2},\boldsymbol{x}_{1};z\right)\\
 & +  ec \int d^{3}x_{3}G_{0}\left(\boldsymbol{x}_{2},\boldsymbol{x}_{3};z\right)A^e_{\mu}\left(\boldsymbol{x}_{3}\right)\alpha^{\mu}G_{0}\left(\boldsymbol{x}_{3},\boldsymbol{x}_{1};z\right)+\ldots
\end{aligned}
\label{eqn:green-expansion}
\end{equation}
where the free Dirac Green's function is given by: $G_{0}=\lim_{A^e\rightarrow0}G_{A^e}$.
These two expansions are known as the potential expansion, where consecutive
terms are known as the zero- one- and many-potential terms. The main
utility of this expansion is that it allows the isolation of ultraviolet
divergent integrals encountered when evaluating loop integrals, as
done by Baranger \textit{et al.},\cite{baranger_physrev1953}
and later by many authors working within BSQED theory.
\subsection{No-photon BSQED energy-shifts}
\label{subsec:No-photon-BSQED-energy-shifts}

Using the ${\cal S}$-matrix expansion of Eq.~\eqref{eqn:s-matrix-expansion}, one can expand Sucher's
energy-shift expression of Eq.~\eqref{eqn:sucher-energy} in powers of the interaction-Hamiltonian density and write, following Mohr
in Ref.~\citenum{mohr_qed_review1989} (eqs.(18) and (31)), 
\begin{equation} \begin{aligned}\Delta E^{\alpha} & =\lim_{\substack{\epsilon\rightarrow0\\\lambda\rightarrow1}}\frac{i\epsilon\lambda}{2}\bigg[\left\langle \Phi_{0}^{\alpha}\left|\hat{{\cal S}}^{\left(1\right)}\left(\epsilon,\lambda\right)\right|\Phi_{0}^{\alpha}\right\rangle \\ & +2\left\langle \Phi_{0}^{\alpha}\left|\hat{{\cal S}}^{\left(2\right)}\left(\epsilon,\lambda\right)\right|\Phi_{0}^{\alpha}\right\rangle \\ & -\left\langle \Phi_{0}^{\alpha}\left|\hat{{\cal S}}^{\left(1\right)}\left(\epsilon,\lambda\right)\right|\Phi_{0}^{\alpha}\right\rangle ^{2}+{\cal O}\left(\lambda^{3}\right)\bigg], \end{aligned}  \label{eqn:sucher-energy2} 
\end{equation}
where $\hat{{\cal S}}^{\left(n\right)}$ is given in Eq.~\eqref{eqn:s-matrix-expansion}.
We shall now consider a system of $n$ electrons and zero photons
(photon vacuum), represented by the following electron-photon state, labeled by $\alpha$:
\begin{equation}
\left|\Phi_{0}^{\alpha}\right\rangle =\left|n_{\text{e}}^{\alpha},0_{\text{p}}^{\alpha}\right\rangle .\label{eqn:ne-0p}
\end{equation}

We remind the reader that the electron field operators entering in our expressions describe non-interacting electrons, in the presence of an external potential, as also seen in the zeroth-order electron Hamiltonian of Eq.~\eqref{eqn:H0e}. As already pointed out in Section \ref{sec:theory}, the electron-electron interaction arises from terms describing exchange of virtual photons between bound electrons.

Since the photon state is chosen to be the vacuum one, this means
that any string of photon operators that is not fully contracted,
will vanish under the photon vacuum expectation value of Eq.~\eqref{eqn:sucher-energy}. Following this
reasoning one concludes that the first non-vanishing QED correction comes from the second-order $\hat{{\cal S}}^{\left(2\right)}$-matrix
\begin{equation}
\begin{aligned}\hat{{\cal S}}^{\left(2\right)}\left(\epsilon,\lambda\right) & =-\frac{\lambda^{2}}{2\hbar^{2}c^{2}}\int d^{4}x_{1}\int d^{4}x_{2}\\
 & \times e^{-\frac{\epsilon}{\hbar}\left(\left|t_{1}\right|+\left|t_{2}\right|\right)}\text{T}\left[\hat{{\cal H}}_{I}\left(x_{1}\right)\hat{{\cal H}}_{I}\left(x_{2}\right)\right].
\end{aligned}
\label{eqn:s2}
\end{equation}

Using Wick's theorem, we expand the electron and photon time-ordered
products, and replace operator contractions by corresponding propagators,
following the contraction definitions of Eqs.~\eqref{eqn:electron-contraction} and \eqref{eqn:photon-contraction}. Furthermore, using the symmetry properties of the photon propagator of Eq.~\eqref{eqn:photon-propagator}:
\begin{align}
D_{\mu\nu}^{F}\left(x,y\right)=D_{\mu\nu}^{F}\left(y,x\right)=D_{\nu\mu}^{F}\left(x,y\right),
\end{align}
the second-order ${\cal S}$-matrix of Eq.~\eqref{eqn:s2} can be shown to reduce to the following expression
\begin{equation}
\begin{aligned}\hat{{\cal S}}^{\left(2\right)}\left(\epsilon,\lambda\right) & =-\frac{\lambda^{2}e^{2}}{2\hbar^{2}}\int d^{4}x_{1}\int d^{4}x_{2}\\
 & \times e^{-\frac{\epsilon}{\hbar}\left(\left|t_{1}\right|+\left|t_{2}\right|\right)}\hat{F}\left(x_{1},x_{2}\right),
\end{aligned}
\label{eqn:S2}
\end{equation}
where the operator $\hat{F}\left(x_{1},x_{2}\right)$ contains the following five QED corrections to the non-interacting problem:
\begin{widetext}
\begin{equation}
\begin{aligned} & \hat{F}\left(x_{1},x_{2}\right)\\
 & =i\hbar \, \, \, D_{\mu_{1}\mu_{2}}^{F}\left(x_{1},x_{2}\right):\bar{\hat{\Psi}}(x_{1})\gamma^{\mu_{1}}\hat{\Psi}(x_{1})\bar{\hat{\Psi}}(x_{2})\gamma^{\mu_{2}}\hat{\Psi}(x_{2}): & \text{SP}\\
 & +2\hbar^{2}D_{\mu_{1}\mu_{2}}^{F}\left(x_{1},x_{2}\right)\text{Tr}\big[S_{A^e}^{F}(x_{2},x_{2})\gamma^{\mu_{2}}\big]:\bar{\hat{\Psi}}(x_{1})\gamma^{\mu_{1}}\hat{\Psi}(x_{1}): & \text{VP}\\
 & -2\hbar^{2}D_{\mu_{1}\mu_{2}}^{F}\left(x_{1},x_{2}\right):\bar{\hat{\Psi}}(x_{1})\gamma^{\mu_{1}}S_{A^e}^{F}(x_{1},x_{2})\gamma^{\mu_{2}}\hat{\Psi}(x_{2}): & \text{SE}\\
 & -i\hbar^{3}\,D_{\mu_{1}\mu_{2}}^{F}\left(x_{1},x_{2}\right)\text{Tr}\big[S_{A^e}^{F}\left(x_{1},x_{1}\right)\gamma^{\mu_{1}}\big]\text{Tr}\big[S_{A^e}^{F}\left(x_{2},x_{2}\right)\gamma^{\mu_{2}}\big] & \text{D1}\\
 & +i\hbar^{3}\,D_{\mu_{1}\mu_{2}}^{F}\left(x_{1},x_{2}\right)\text{Tr}\big[S_{A^e}^{F}\left(x_{2},x_{1}\right)\gamma^{\mu_{1}}S_{A^e}^{F}\left(x_{1},x_{2}\right)\gamma^{\mu_{2}}\big] & \text{D2}
\end{aligned}
\label{eqn:T=00005BHH=00005D}
\end{equation}
\end{widetext}

Finally, using Sucher's energy expression of Eq.~\eqref{eqn:sucher-energy2}, the second-order energy-shift becomes
\begin{align}
\begin{aligned}\Delta E^{\alpha,2} & =\lim_{\substack{\epsilon\rightarrow0\\\lambda\rightarrow1}}i\epsilon\lambda\big<\Phi_{0}^{\alpha}\big|\hat{{\cal S}}^{\left(2\right)}\left(\epsilon,\lambda\right)\big|\Phi_{0}^{\alpha}\big>\\
 & =\Delta E_{\text{SP}}^{\alpha,2}+\Delta E_{\text{VP}}^{\alpha,2}+\Delta E_{\text{SE}}^{\alpha,2}\\
 & +\Delta E_{\text{D1}}^{\alpha,2}+\Delta E_{\text{D2}}^{\alpha,2}.
\end{aligned}
\label{eqn:sucherr-alpha-2}
\end{align}

Each of these terms will be discussed in the next sections, and is represented by a Feynman diagram in Fig. \ref{fig:bound}. The elements of these diagrams are the
following:
\begin{enumerate}
\item Double external lines represent bound-electrons, i.e., with wavefunctions
and energies satisfying the interacting Dirac equation of Eq.~\eqref{eqn:bound-dirac}, in the presence of a classical time-independent external potential $A^e\left(\boldsymbol{x}\right)$.
\item Double internal lines represent a virtual bound-electron propagation
between the two vertices $S_{A^e}^{F}\left(x_{2},x_{1}\right)$ and
arise from a single contraction of two electron field operators.
\item Internal wiggly-lines connecting two vertices represent propagations
of virtual-photons $D_{\mu_{2}\mu_{1}}^{F}\left(x_{2},x_{1}\right)$
and come from a single contraction of two photon field operators.
\end{enumerate}
The last two contributions from Eq.~\eqref{eqn:T=00005BHH=00005D} correspond to fully contracted products, and they are thus free of creation and annihilation operators. This means that their corresponding energy-shifts 
$\Delta E_{\text{D1}}^{\alpha,2}$ and $\Delta E_{\text{D2}}^{\alpha,2}$
are state-independent and hence do not contribute to energy-differences. 
They are therefore discarded from further consideration; see for instance Mohr in Ref.~\citenum{mohr_qed_review1989}. On the other hand, the
first three contributions correspond to partially contracted products,
associated with the follwing physical processes:
\subsubsection{SP: Single-photon exchange}
\label{subsec:SP:-Single-photon-exchange}

This process, coming from the SP term in Eq.~\eqref{eqn:T=00005BHH=00005D}
and represented in Fig. \ref{fig:SP-bound}, describes electron-electron
interaction in its lowest-order, where an electron feels the existence
of the other electron through the exchange of a single virtual-photon.
After integrating over times $t_{1}$ and $t_{2}$ in Eq.~\eqref{eqn:S2},
and taking limits in Eq.~\eqref{eqn:sucherr-alpha-2}, this correction
yields an instantaneous direct interaction-term, in addition to a
retarded exchange interaction-term, analogous to the direct and exchange
terms in the Hartree-Fock theory:
\begin{widetext}
\begin{equation}
\begin{aligned} & \Delta E_{\text{SP}}^{\alpha,2}\\
 & =\frac{e^{2}}{2} \sum_{i,j}\int d^{3}x_{1}\int d^{3}x_{2}\bar{\psi}_{i}\left(\boldsymbol{x}_{1}\right)\gamma^{\mu}\psi_{i}\left(\boldsymbol{x}_{1}\right)\frac{1}{4\pi\epsilon_{0}\left|\boldsymbol{x}_{1}-\boldsymbol{x}_{2}\right|}\bar{\psi}_{j}\left(\boldsymbol{x}_{2}\right)\gamma_{\mu}\psi_{j}\left(\boldsymbol{x}_{2}\right) &  & \text{Direct}\\
 & -\frac{e^{2}}{2} \sum_{i,j}\int d^{3}x_{1}\int d^{3}x_{2}\bar{\psi}_{i}\left(\boldsymbol{x}_{1}\right)\gamma^{\mu}\psi_{j}\left(\boldsymbol{x}_{1}\right)\frac{e^{+\frac{i}{c\hbar}\left|E_{i}-E_{j}\right|\left|\boldsymbol{x}_{1}-\boldsymbol{x}_{2}\right|}}{4\pi\epsilon_{0}\left|\boldsymbol{x}_{1}-\boldsymbol{x}_{2}\right|}\bar{\psi}_{j}\left(\boldsymbol{x}_{2}\right)\gamma_{\mu}\psi_{i}\left(\boldsymbol{x}_{2}\right) &  & \text{Exchange}
\end{aligned}
\label{eqn:DeltaE_SP}
\end{equation}
\end{widetext}
as noted by Mohr in Ref.~\citenum{mohr_qed_highz_few_e_1985} (section IV). Notice that for $\mu=0$ , and $\mu=1,2,3$ these integrals account for the Coulomb and Gaunt interaction, respectively. On the other hand, if we used the Coulomb gauge photon propagator instead of the Feynman one, we would get the retarded Breit interaction, as noted by Lindgren in Ref.~\citenum{lindgren1992} (page 262).
\subsubsection{VP: Vacuum polarization}

\label{subsec:VP:-Vacuum-polarization}

This process, presented in Fig. \ref{fig:VP-bound}, accounts
for the instantaneous interaction of a bound-electron with the electron-positron pair cloud, polarized by the presence of a classical potential source. After plugging the VP term of Eq.~\eqref{eqn:T=00005BHH=00005D} in the second-order scattering matrix expression, one can use Sucher's
formula of Eq.~\eqref{eqn:sucher-energy2} to write the energy-shift associated with the vacuum polarization process as Ref.~\citenum{schweber_rqft} (chapter 15 eq.(205)):
\begin{equation}
\Delta E_{\text{VP}}^{\alpha}=-e\sum_{i}\int d^{3}x_{1}\bar{\psi}_{i}\left(\boldsymbol{x}_{1}\right)\gamma_{\mu}\psi_{i}\left(\boldsymbol{x}_{1}\right)\varphi_{\text{VP}}^{\mu}\left(\boldsymbol{x}_{1}\right).
\label{eqn:VP-energy} \end{equation}

We note that the vacuum polarization effect is local, i.e., it can be written as an expectation value of a local vacuum polarization four-potential
\begin{equation}
\varphi_{\text{VP}}^{\mu}\left(\boldsymbol{x}_{1}\right)=ie\hbar\int d^{3}x_{2}\frac{\text{Tr}\left[\gamma^{\mu}S_{A^e}^{F}\left(x_{2},x_{2}\right)\right]}{4\pi\epsilon_{0}\left|\boldsymbol{x}_{1}-\boldsymbol{x}_{2}\right|}.
\end{equation}

\begin{figure}
\begin{centering}
\subfloat[\label{fig:VP-bound-0}$\alpha\left(Z\alpha\right)^{0}$.]{\begin{centering}
\includegraphics[scale=0.2]{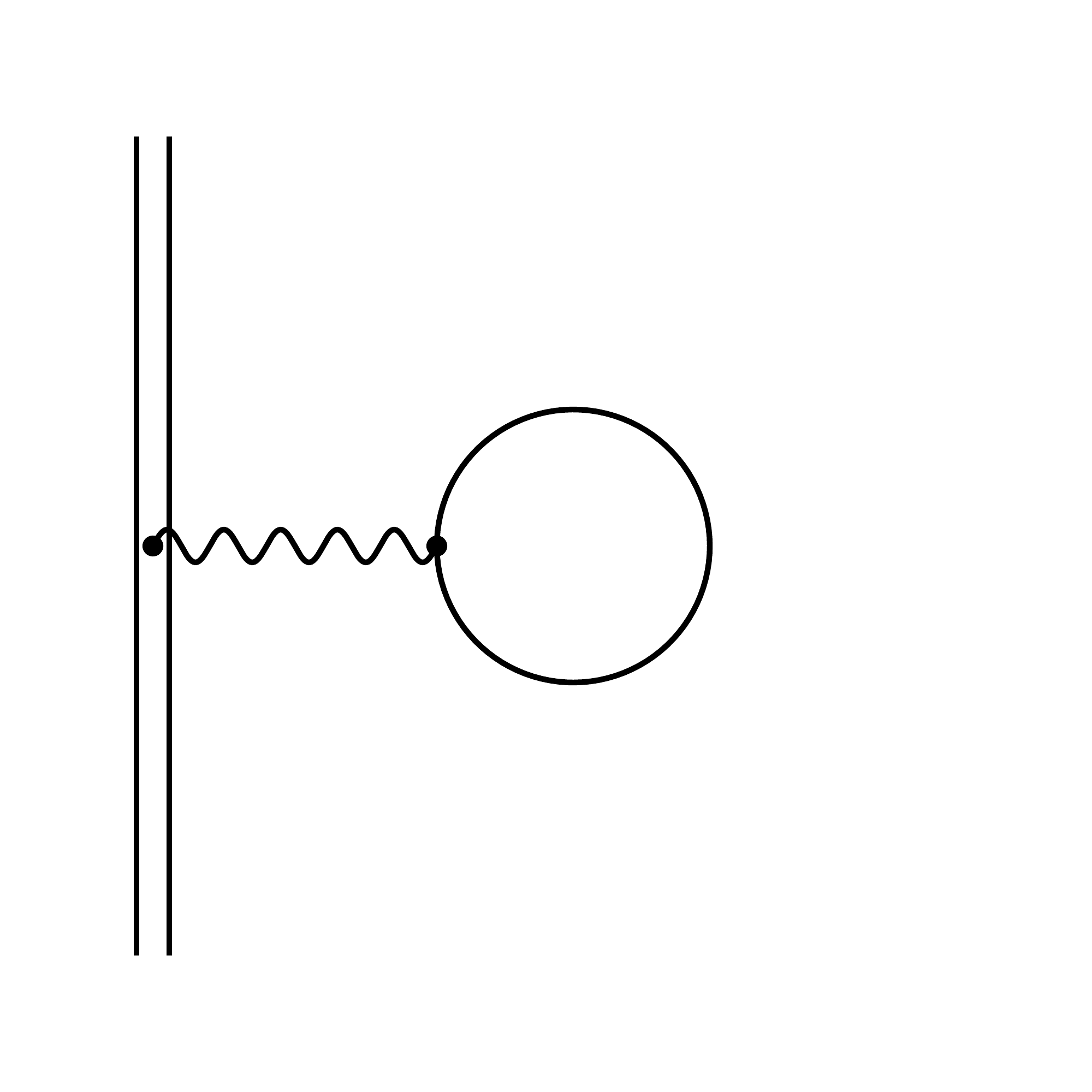}
\par\end{centering}
}\subfloat[\label{fig:VP-bound-1}$\alpha\left(Z\alpha\right)^{1}$.]{\begin{centering}
\includegraphics[scale=0.2]{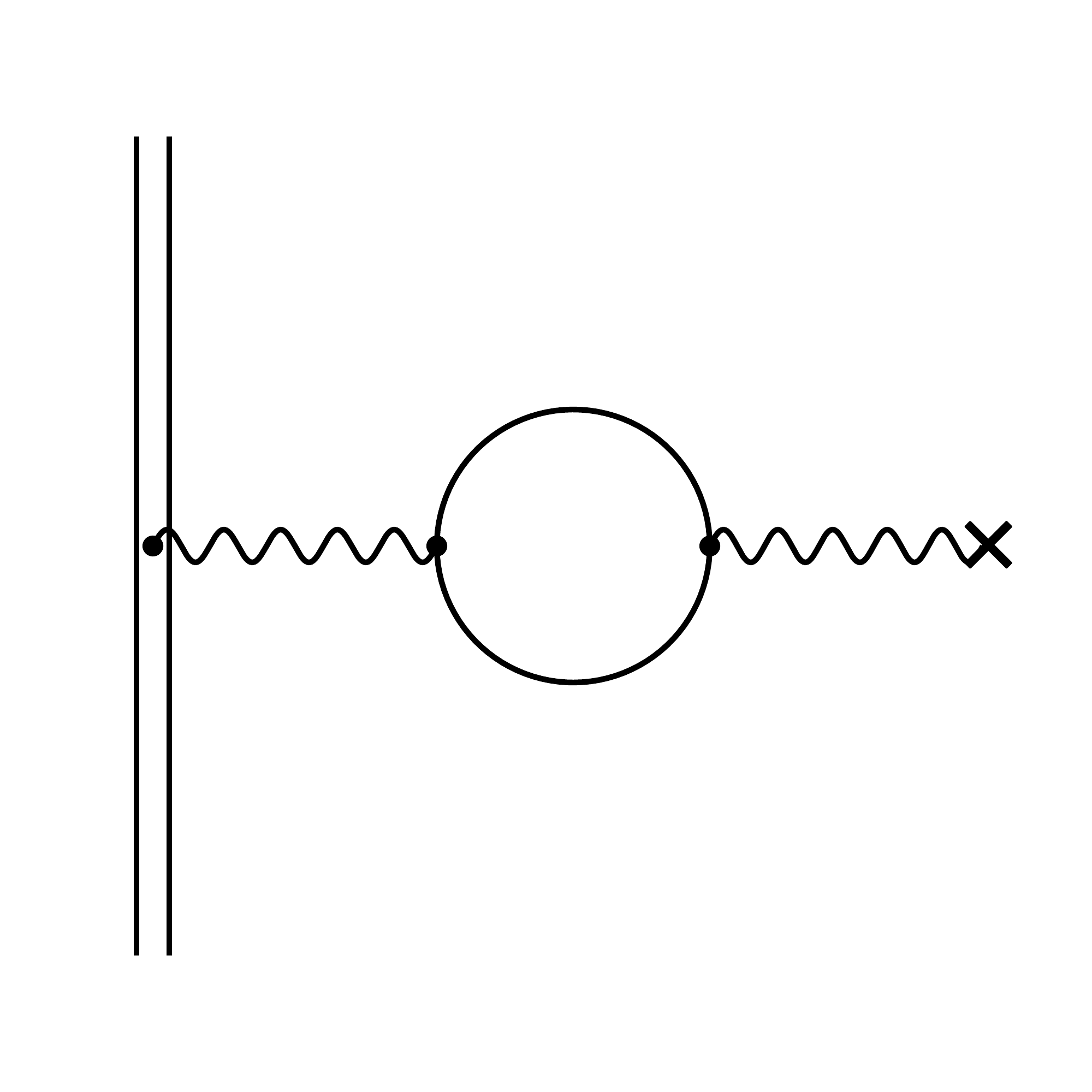}
\par\end{centering}}
\par\end{centering}
\centering{}\subfloat[\label{fig:VP-bound-2}$\alpha\left(Z\alpha\right)^{2}$.]{\begin{centering}
\includegraphics[scale=0.2]{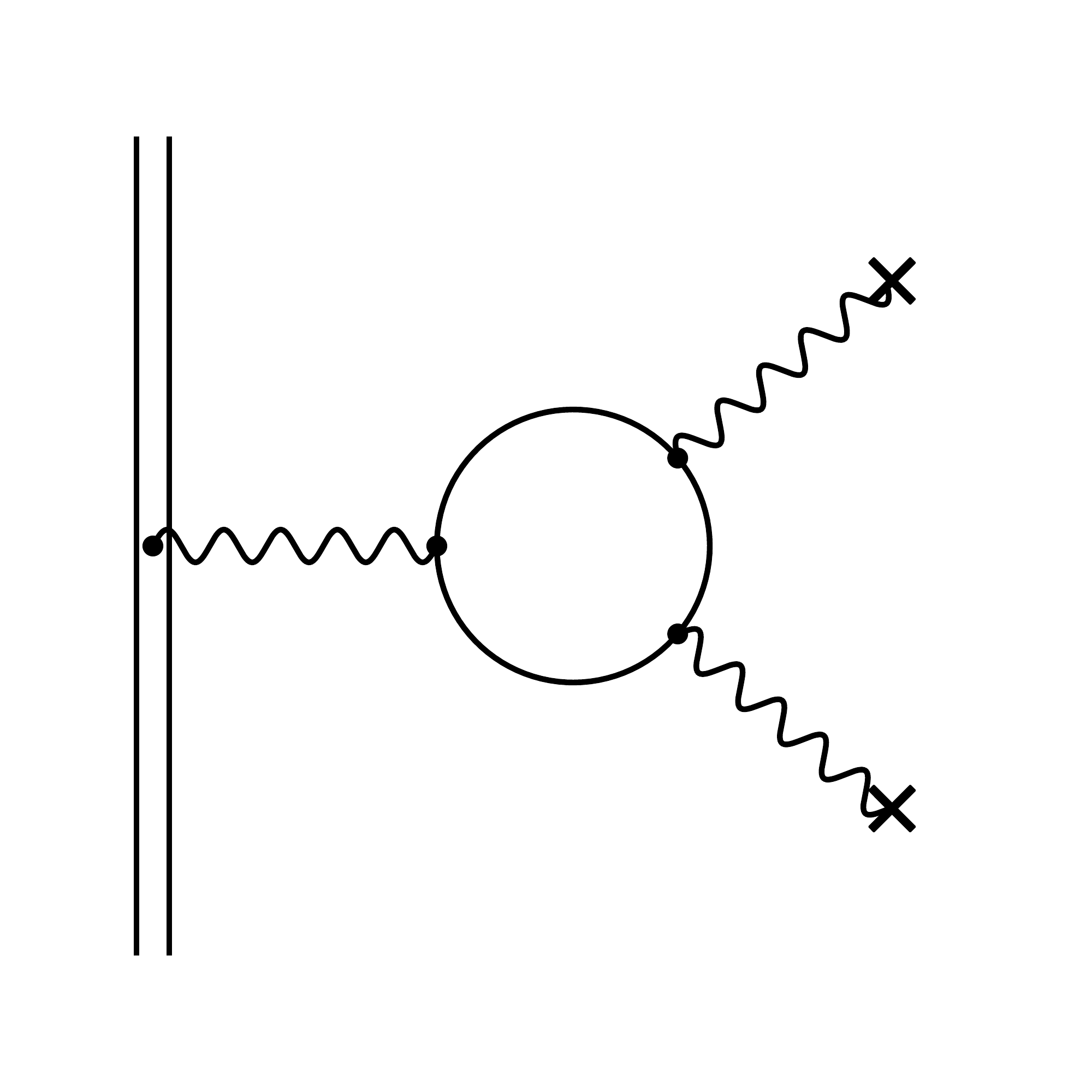}
\par\end{centering}
}\subfloat[\label{fig:VP-bound-3}$\alpha\left(Z\alpha\right)^{3}$.]{\centering{}
\includegraphics[scale=0.2]{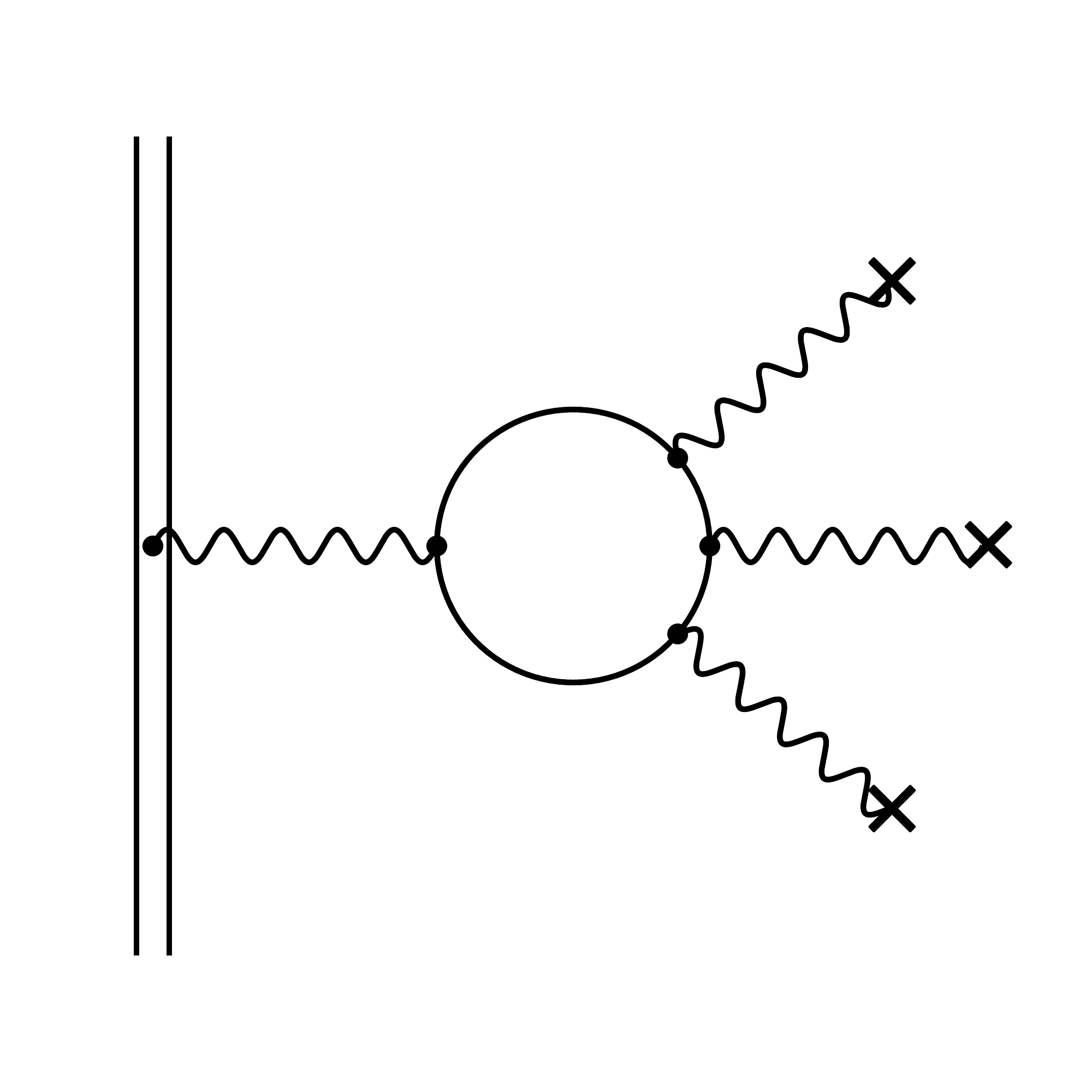}}\caption{\label{fig:VP-bound-expansion}First four bound-state vacuum polarization processes, obtained after expanding the bound propagator in powers of the external potential. A wiggly line ending with a cross $\times$ indicates an interaction with the external field.}
\end{figure}

The energy expression of Eq.~\eqref{eqn:VP-energy} (as well as the last potential) is divergent due to the fact that
\begin{equation}
\lim_{x_{1}\rightarrow x_{2}}S_{A^e}^{F}\left(x_{2},x_{1}\right)=\infty,
\end{equation}
as mentioned in Ref.~\citenum{indelicato1992coordinate}. The isolation of the divergent terms in this expression can be done by expanding the
propagator inside the trace using Eq.~\eqref{eqn:propagator-expansion},
and write the energy as
\begin{equation}
\Delta E_{\text{VP}}^{\alpha,2}=\Delta E_{\text{VP},0}^{\alpha,2}+\Delta E_{\text{VP},1}^{\alpha,2}+\Delta E_{\text{VP},2}^{\alpha,2}+\ldots
\end{equation}
where $\Delta E_{\text{VP},i}^{\alpha,2}$ represents the term that
corresponds to an $i$ number of interactions with the external potential
$\left(Z\alpha\right)^{i}$. The first four terms are presented in
Figs. \ref{fig:VP-bound-0} to \ref{fig:VP-bound-3}. Notice that
the double-line loop is replaced by a single-line one. This is made
to indicate that the propagators between these vertices are the free ones
$S_{0}^{F}$, instead of the bound-ones $S_{A^e}^{F}$. Using Furry's
theorem,\cite{furry1937theorem} that is based on a charge conjugation
symmetry argument, one can show that any diagram containing a free-electron loop with an odd number of vertices does not contribute.
This means that the above energy expression reduces to
\begin{equation}
\Delta E_{\text{VP}}^{\alpha,2}=\Delta E_{\text{VP},1}^{\alpha,2}+\Delta E_{\text{VP},3}^{\alpha,2}+\ldots
\end{equation}

A naive estimation of the degree of divergence of a QED integral can be done by calculating the superficial degree of divergence $S$ that simply counts overall momentum powers of the integral in question (in momentum-space):
\begin{equation}
S\equiv4-N_{e}-2N_{p},
\end{equation}
where 4 are the spacetime dimensions and $N_{e}$ and $N_{p}$ are the number of electron and photon propagators, respectively, in the loop in question; see for instance Refs.~\citenum{peskin:schroeder} (section 10.1) and \citenum{itzykson_zuber_qft_1980} (sections 7-1-4 and 8-1-3). The integral is said to be superficially divergent if $S\ge0$. The possible cases are:
\begin{equation}
\begin{aligned}S\leq0 &  &  & \text{convergence}\\
S=0 &  &  & \text{logarithmic divergence}\\
S=1 &  &  & \text{linear divergence}\\
S=2 &  &  & \text{quadratic divergence}
\end{aligned}
\end{equation}

This naive estimation usually overestimates the effective (divergence),
that we shall call $E$, and this can be seen after further analysis of
the integral in question. As a consequence, some superficially divergent
integral can be effectively less divergent, or hopefully convergent.
In the next Table \ref{tab:S-and-E-VP}, we list the superficial and
effective divergences of the first vacuum polarization terms. The
reader should notice that with higher-order terms, more propagators
are included in the momentum-space integral, meaning that more denominator
powers are added, and as a consequence, the integral becomes less
divergent.

\begin{table}
\begin{centering}
\begin{tabular}{|c|c|c|}
\hline 
Terms & $S$ & $E$\tabularnewline
\hline 
\hline 
$\alpha\left(Z\alpha\right)^{1}$ & $2$ & $0$\tabularnewline
\hline 
$\alpha\left(Z\alpha\right)^{3}$ & $0$ & $<0$\tabularnewline
\hline 
$\alpha\left(Z\alpha\right)^{5}$ & $<0$ & $<0$\tabularnewline
\hline 
$\vdots$ & $\vdots$ & $\vdots$\tabularnewline
\hline 
\end{tabular}\caption{\label{tab:S-and-E-VP}Superficial and effective degrees of divergence
for the bound-state vacuum polarization contributions.}
\par\end{centering}
\end{table}

We shall now focus on the first non-vanishing vacuum polarization
contribution $\Delta E_{\text{VP},1}^{\alpha}$. As seen in Table
\ref{tab:S-and-E-VP}, this term is of superficial quadratic divergence,
but it is, effectively, only logarithmic. This can be shown using the Ward identity,
as mentioned by Peskin and Schroeder in Ref.~\citenum{peskin:schroeder} (section 7.5).
After the employment of regularization, followed by renormalization (discussed in Sec.~\ref{subsubsec:reg-and-ren}), one may  extract the physical contribution 
out of the divergent $\Delta E_{\text{VP},1}^{\alpha,2}$.
In the case where the Hamiltonian is invariant under
time-reversal symmetry, i.e., if the external vector potential $\boldsymbol{A}^e\left(\boldsymbol{x}\right)$
vanishes, cf. Greiner in Ref.~\citenum{greiner2000relativistic} (eqs.(12.52-53)), then
only the time-component potential $\varphi_{\text{VP},1}^{0}$ survives,
and one obtains the Uehling potential,\cite{uehling1935} given in Eq.\eqref{eqn:Uehling}. 

For detailed discussions and derivations of the one-potential bound-state vacuum polarization correction, the reader may consult the calculation of Greiner and Reinhardt in Ref.~\citenum{greiner_reinhardt_qed} (section 5.2) where the authors used Pauli--Villars regularization, in addition to Peskin and Schroeder in Ref.~\citenum{peskin:schroeder} (section 7.5) who used dimensional-regularization to treat the occurring divergences; see also Mandl and Shaw in Ref.~\citenum{mandl_qft2010} (section 10.4), in addition to Schwartz in Ref.~\citenum{schwartz_qft} (section 16.2.2). Contrary to the conventional momentum-space approach to evaluate QED corrections, Indelicato and Mohr in Ref.~\citenum{indelicato1992coordinate} (section B) considered the vacuum polarization problem in coordinate space, and derived the physical Uehling contribution using coordinate-space Pauli--Villars regularization.

The second non-vanishing vacuum polarization effect, associated with
$\Delta E_{\text{VP},3}^{\alpha,2}$ and presented in Fig. \ref{fig:VP-bound-3},
is known as the Wichmann--Kroll effect.\cite{wichmann_kroll1956} 
As seen in Table \ref{tab:S-and-E-VP} and noted by Gyulassy,\cite{gyulassy_phd_1974}
this contribution is free of divergences. Wichmann and Kroll calculated
the effective potential associated with the $\Delta E_{\text{VP},3}^{\alpha,2}$
correction in Laplace space. On the other hand, in Ref.~\citenum{blomqvist1972} (section 4), Blomqvist has evaluated the inverse Laplace-transform, and
obtained the real-space potential expression for a point nuclear charge
distribution. The last reference presents a relatively complex analytical
expression for this $\alpha\left(Z\alpha\right)^{3}$ potential, and
this motivated Fainshtein \textit{et al.} \cite{fainshtein_1991}
to provide an approximation that facilitates the numerical computation, 
yet conserving precision.

We finally note that in the fourth-order BSQED correction, one finds
the Källén--Sabry potentials \cite{kallen_sab_1955} of order $\alpha^{2}\left(Z\alpha\right)$
that can be obtained by expanding the bound propagators of Ref.~\citenum{mohr1998qed} (fig. 25 diagrams b,c VPVP).
In order to make this momentum-space potential usable in practical
calculations, in Ref.~\citenum{blomqvist1972} (section 3) Blomqvist derived
its real-space version for a point nucleus distribution, whereas Wayne Fullerton and Rinker generalized this result to account for an extended nuclear charge distribution; see Ref.~\citenum{fullerton_rinker_1976} (eq.(9)). We finally note that the latter authors provided a good approximation of the corresponding potential, in order to render the numerical evaluation more practical.

\subsubsection{SE: Self-energy}

\label{subsec:SE:-Self-energy}
\begin{figure}
\begin{centering}
\subfloat[\label{fig:SE-bound-0}$\alpha\left(Z\alpha\right)^{0}$.]{\begin{centering}
\includegraphics[scale=0.2]{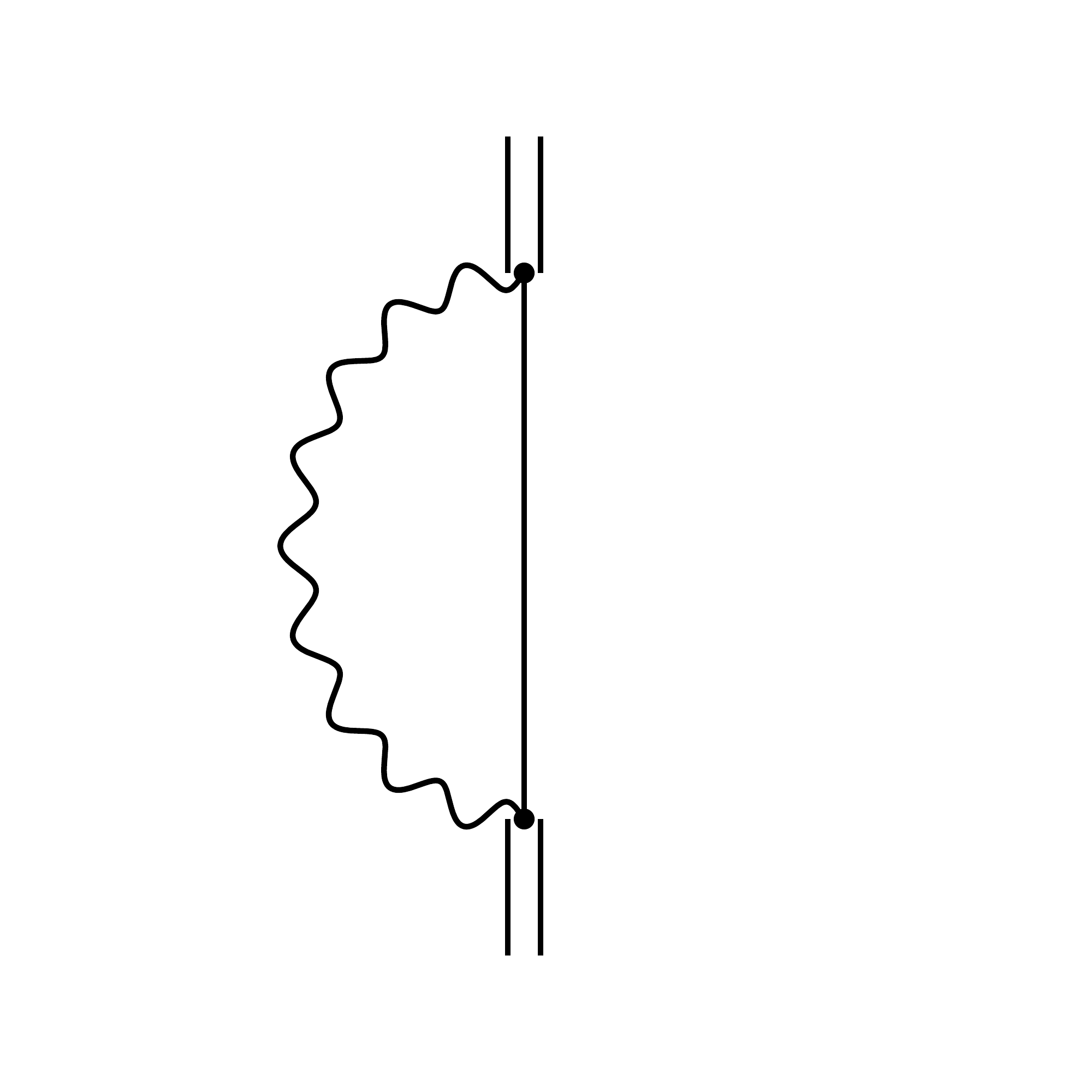}
\par\end{centering}
}\subfloat[\label{fig:SE-bound-1}$\alpha\left(Z\alpha\right)^{1}$.]{\centering{}
\includegraphics[scale=0.2]{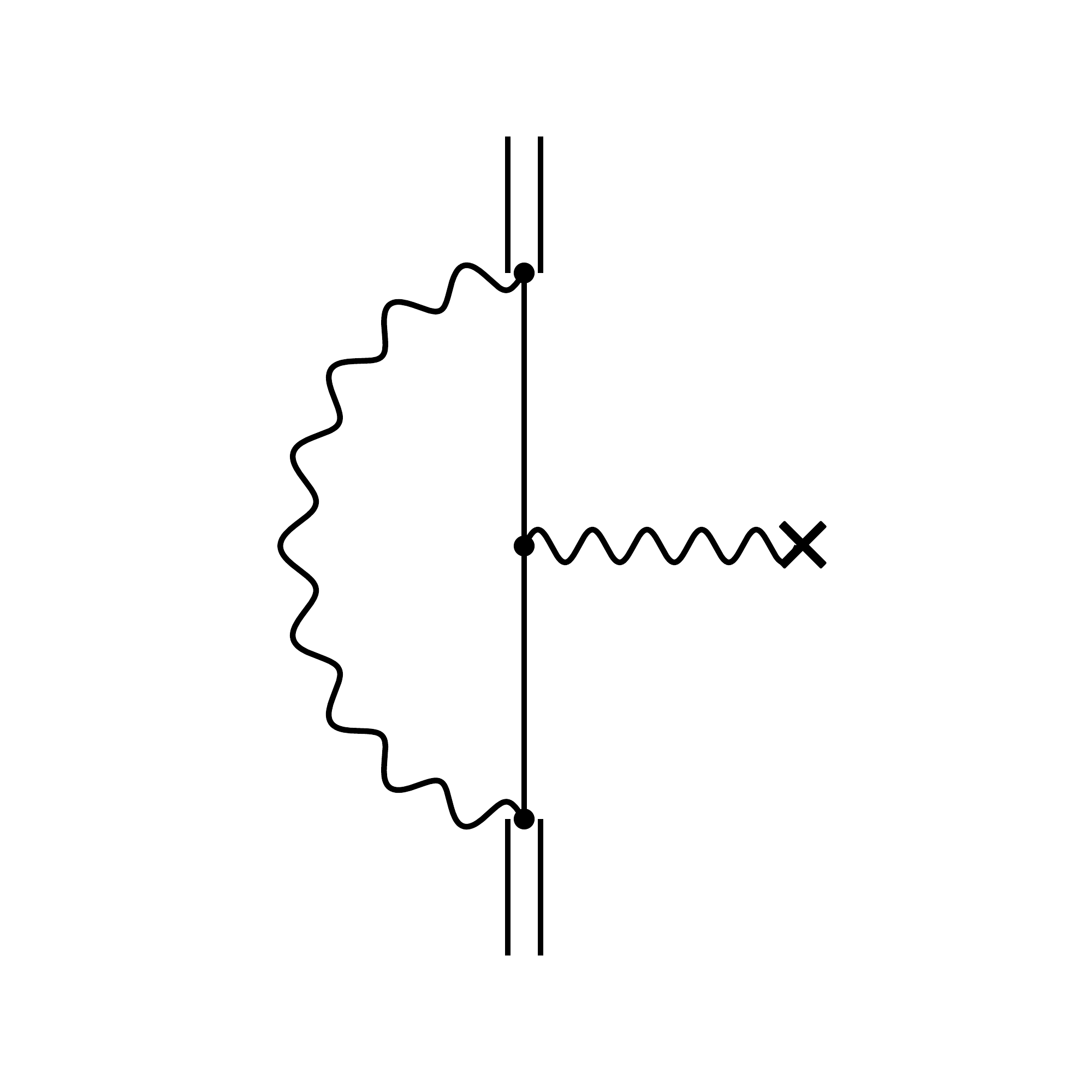}}
\par\end{centering}
\centering{}\subfloat[\label{fig:SE-bound-2}$\alpha\left(Z\alpha\right)^{2}$.]{\centering{}\includegraphics[scale=0.2]{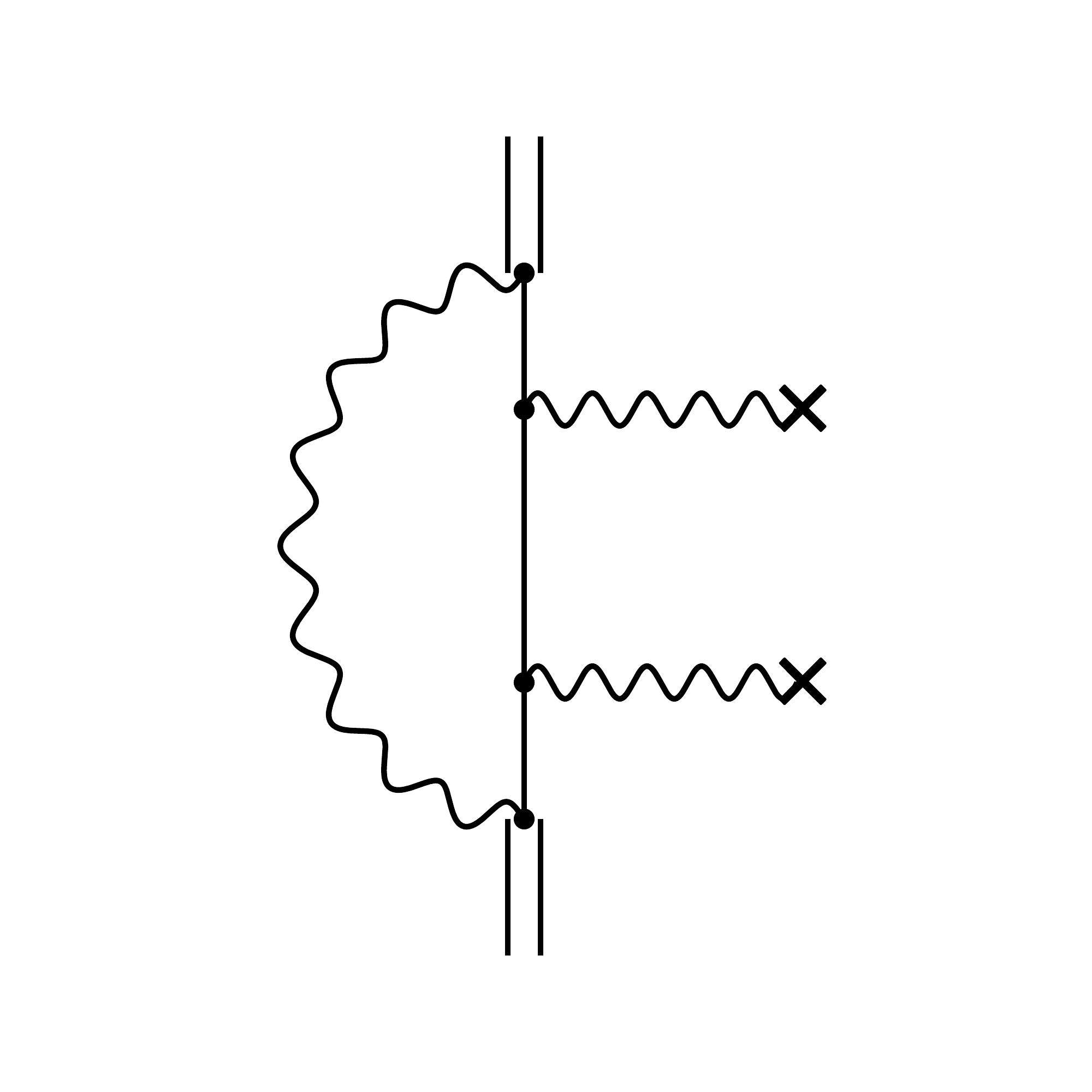}}\subfloat[\label{fig:SE-bound-3}$\alpha\left(Z\alpha\right)^{3}$.]{\centering{}\includegraphics[scale=0.2]{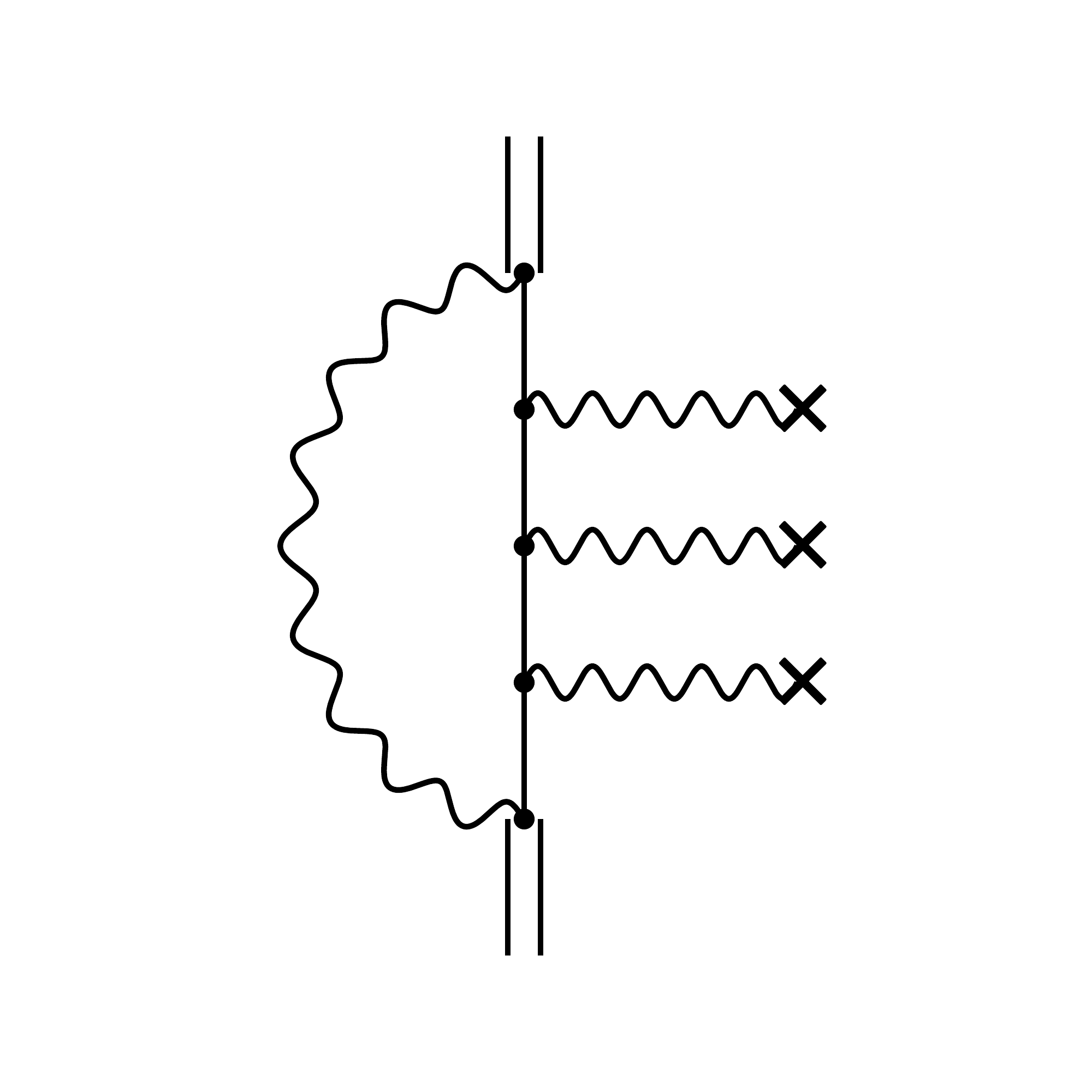}}\caption{\label{fig:SE-bound-expansion}First four bound-state self-energy processes, obtained after expanding the bound propagator in powers of the external potential.}
\end{figure}
The self-energy process, presented in Fig. \ref{fig:SE-bound},
is the dominant radiative QED correction in electronic atoms, as  seen in the work of Johnson and Soff of Ref.~\citenum{johnson_soff_atnucdattab1985} (fig.2). This process describes the interaction of the bound-electron with
itself, by emitting and absorbing a virtual-photon. The first calculation
for this correction was made in 1947 by Hans Bethe
in a purely non-relativistic framework,\cite{bethe-paper_1947} where he used a renormalization 
technique (by subtracting the free self-energy) to render the integral
less divergent, and introduced a reasonable virtual-photon energy cutoff at $E=m_ec^{2}$. This simple calculation gave hope in digging for the physical Lamb shift in the frustrating non-physical divergences in the QED theory. Using Sucher's energy formula of Eq.~\eqref{eqn:sucher-energy2}, the SE term of Eq.~\eqref{eqn:T=00005BHH=00005D} leads to the following energy-shift
\begin{equation}
\begin{aligned}\Delta E_{\text{SE}}^{\alpha,2} & =-e\sum_{i}\int d^{3}x_{1}\int d^{3}x_{2}\\
 & \times\psi_{i}^{\dagger}\left(\boldsymbol{x}_{2}\right)\varphi_{\text{SE}}\left(\boldsymbol{x}_{2},\boldsymbol{x}_{1};E_{i}\right)\psi_{i}\left(\boldsymbol{x}_{1}\right)
\end{aligned}\label{eqn:SE-energy-shift}
\end{equation}

where the self-energy potential is given by
\begin{equation}
\begin{aligned} & \varphi_{\text{SE}}\left(\boldsymbol{x}_{2},\boldsymbol{x}_{1};E_i \right)=-\frac{e}{2\pi i}\int_{C_{F}}dz\alpha^{\mu}G_{A^e}\left(\boldsymbol{x}_{2},\boldsymbol{x}_{1};z\right)\alpha_{\mu}\\
 & \times \frac{\exp\big(+\frac{i}{\hbar}\left|\boldsymbol{x}_{1}-\boldsymbol{x}_{2}\right|\sqrt{\left(z-E_{i}\right)^{2}/c^{2}+i\epsilon}\big)}{4\pi\epsilon_{0}\left|\boldsymbol{x}_{1}-\boldsymbol{x}_{2}\right|}.\label{eqn:SE-potential}
\end{aligned}
\end{equation}

Similar forms of this equation are provided by Schweber in Ref.~\citenum{schweber_rqft} (eq.(205)) and Mohr in Ref.~\citenum{mohr1974selfenergy} (eq.(2.6)). Notice at this point that unlike the vacuum polarization case, the
self-energy is a non-local effect, as seen from Eq.~\eqref{eqn:SE-energy-shift}, and this is the reason behind the complexity of its analytical and numerical evaluation. As in the vacuum polarization case, the self-energy potential is a divergent quantity and needs to be regularized. In order to isolate divergent terms, one can use the Green's function (propagator) expansion of Eq.~\eqref{eqn:green-expansion} and write the total energy-shift
as
\begin{equation}
\Delta E_{\text{SE}}^{\alpha,2}=\Delta E_{\text{SE},0}^{\alpha,2}+\Delta E_{\text{SE},1}^{\alpha,2}+\Delta E_{\text{SE},2}^{\alpha,2}+\ldots,\label{eqn:se-pot-expansion}
\end{equation}
where $\Delta E_{\text{SE},i}^{\alpha,2}$ represents the process
in which the internal electron interacts $i$ times with the external
potential, it is thus associated with the $\alpha\left(Z\alpha\right)^{i}$
order. The first four terms of the last expansion are represented
in Figs. \ref{fig:SE-bound-0} to \ref{fig:SE-bound-3}. The zero- and one-potential terms: $\Delta E_{\text{SE},0}^{\alpha,2}$ and $\Delta E_{\text{SE},1}^{\alpha,2}$
are known as the (free) self-energy and the vertex-correction processes.
These two contributions are logarithmically divergent (in momentum-space),
while all higher-order ones are convergent, as presented in Table 
\ref{tab:S-and-E-SE}. A coordinate-space
treatment of these quantities has been provided by Indelicato and Mohr. \cite{indelicato1992coordinate,indelicatomohr1998calculation,mohr1974selfenergy}
Using the bound propagator expansion of Eq.~\eqref{eqn:propagator-expansion},
one obtains the scattering matrices associated with these two processes:
\begin{table}
\centering{}%
\begin{tabular}{|c|c|c|}
\hline 
Terms & $S$ & $E$\tabularnewline
\hline 
\hline 
$\alpha\left(Z\alpha\right)^{0}$ & $1$ & $0$\tabularnewline
\hline 
$\alpha\left(Z\alpha\right)^{1}$ & $0$ & $0$\tabularnewline
\hline 
$\alpha\left(Z\alpha\right)^{2}$ & $<0$ & $<0$\tabularnewline
\hline 
$\vdots$ & $\vdots$ & $\vdots$\tabularnewline
\hline 
\end{tabular}\caption{\label{tab:S-and-E-SE}Superficial and effective degrees of divergence
for the bound-state self-energy contributions.}
\end{table}
\begin{widetext}
\begin{align}
\hat{{\cal S}}_{\text{SE},0}^{\left(2\right)}\left(\epsilon,\lambda\right) & =\lambda^{2}e^{2}\int d^{4}x_{1}\int d^{4}x_{2}e^{-\frac{\epsilon}{\hbar}\left(\left|t_{1}\right|+\left|t_{2}\right|\right)}D_{\mu_{1}\mu_{2}}^{F}(x_{1},x_{2}):\bar{\hat{\Psi}}(x_{1})\gamma^{\mu_{2}}S_{0}^{F}(x_{1},x_{2})\gamma^{\mu_{1}}\hat{\Psi}(x_{2}): \label{eqn:SE-0-potential}\\
\hat{{\cal S}}_{\text{SE},1}^{\left(2\right)}\left(\epsilon,\lambda\right) & =-\lambda^{2}e^{2}\int d^{4}x_{1}\int d^{4}x_{2}e^{-\frac{\epsilon}{\hbar}\left(\left|t_{1}\right|+\left|t_{2}\right|\right)}D_{\mu_{1}\mu_{2}}^{F}(x_{1},x_{2})\nonumber \\
 & \quad\quad\times:\bar{\hat{\Psi}}(x_{1})\gamma^{\mu_{1}}\int d^{4}x_{3}S_{0}^{F}(x_{1},x_{3})e A^e_{\mu}(\boldsymbol{x}_{3})\gamma^{\mu}S_{0}^{F}(x_{3},x_{2})\gamma^{\mu_{2}}\hat{\Psi}(x_{2}): \label{eqn:SE-1-potential} \end{align}
\end{widetext}
The next step is to transform these two real-space integral $\cal{S}$-matrices into Fourier-space ones. We first use the electron and photon propagators of Eqs.~\eqref{eqn:Free-electron-propagator} and \eqref{eqn:photon-propagator} and write electron field operators and the external (classical) potential in their Fourier-integral forms
\begin{equation}
\begin{aligned}
\hat{\Psi}(x) & =\int\frac{d^{4}p}{\left(2\pi\hbar\right)^{4}}e^{-\frac{i}{\hbar}p\cdot x}\hat{\Psi}(p),\\
A^e(x) & =\int\frac{d^{4}p}{\left(2\pi\hbar\right)^{4}}e^{-\frac{i}{\hbar}p\cdot x}A^e(p).
\end{aligned}\label{eq:Fourier}
\end{equation}
We finally note that variable dependence indicates in which space the corresponding physical quantity is: We use $x$ variables for spacetime points (coordinate-space), and $p$ and $q$ variables for four-momentum points (in momentum-space).
The first $\cal{S}$-matrix $\hat{{\cal S}}_{\text{SE},0}^{\left(2\right)}\left(\epsilon,\lambda\right)$ (the zero-potential bound-state self-energy) becomes
\begin{align}
\hat{{\cal S}}_{\text{SE},0}^{\left(2\right)}\left(0,1\right) & =\int\frac{d^{4}q}{\left(2\pi\hbar\right)^{4}}:\bar{\hat{\Psi}}(q)\gamma^{\mu}\Sigma(q)\gamma_{\mu}\hat{\Psi}(q):\\
\Sigma\left(q\right) & =e^{2}\int\frac{d^{4}p}{\left(2\pi\hbar\right)^{4}}S_{0}^{F}(q-p)D^{F}(p).\label{eqn:Sigmap}
\end{align}

Here $\Sigma\left(q\right)$ is the so-called self-energy matrix
function (see, for instance, Mandl and Shaw in Ref.~\citenum{mandl_qft2010} (eq.(9.20)). Notice that in the limit of large momentum $p$, the integrand behaves as $\propto \frac{1}{\gamma^\mu (q-p)_\mu} \frac{1}{p^3}$, which indicates a superficial linear divergence. However, with further investigation, one can show that this divergence is reduced by one degree, as noted by Schweber in Ref.~\citenum{schweber_rqft} (section 15a), and presented in Table \ref{tab:S-and-E-SE}. Following the same steps, one can show that the second scattering matrix, associated with the one-potential bound-state self-energy process $\hat{{\cal S}}_{\text{SE},0}^{\left(2\right)}\left(0,1\right) $ 
can be written as

\begin{widetext}
\begin{align}
\hat{{\cal S}}_{\text{SE},1}^{\left(2\right)}\left(0,1\right) & =-\frac{e}{i\hbar}\int\frac{d^{4}p_{2}}{\left(2\pi\hbar\right)^{4}}\int\frac{d^{4}p_{1}}{\left(2\pi\hbar\right)^{4}}A^e_{\mu}(p_{2}-p_{1}):\bar{\hat{\Psi}}(p_{2})\Lambda^{\mu}(p_{2},p_{1})\hat{\Psi}(p_{1}):\label{eqn:S-mat-vertex}\\
\Lambda^{\mu}(p_{2},p_{1}) & =i\hbar e^{2}\gamma^{\nu}\int\frac{d^{4}q}{\left(2\pi\hbar\right)^{4}}D^{F}(q)S_{0}^{F}(p_{2}-q)\gamma^{\mu}S_{0}^{F}(p_{1}-q)\gamma_{\nu},\label{eqn:Gammap2p1}
\end{align}
\end{widetext}
where in the last equation, $\Lambda^{\mu}\left(p_{2},p_{1}\right)$
is the so-called vertex-correction function; see, for instance, Mandl and Shaw\cite{mandl_qft2010} (eq.(9.48)). A detailed study of these momentum-space expressions and associated energy-shifts in the bound-electron problem was first considered by Snyderman in Ref.~\citenum{snyderman1991electron} (section 4)
(see also Yerokhin and Shabaev\cite{yerokhin1999shabaev}).

\subsubsection{Regularization and renormalization}
\label{subsubsec:reg-and-ren}

When computing integrals associated with QED corrections, one finds (as already seen) that some of these integrals are divergent. How can one extract the meaningful finite (physical) from the meaningless infinite? This is done through regularization and renormalization.

Regularization is a technique for rendering a divergent integral convergent, albeit still dependent on the regularization-parameter. The main regularization techniques are: sharp momentum-cutoff, Pauli--Villars,\cite{paulivillars} dimensional-regularization \cite{thooft_npb1972} and analytic continuation \cite{bollini1964} regularization. The reader may also consult Zeidler  in Ref.~\citenum{zeidler2} (chapter 2) for a general conceptual formulations of regularization schemes. The sharp momentum-cutoff regularization consists of cutting off momentum contributions higher than a some $p_{\text{max}}=\Lambda\gg m_e c$. Unfortunately, this intuitive regularization breaks Lorentz- and gauge-invariance and the solution is to use the other regularization schemes. The Pauli--Villars regularization consists of modifying the photon and electron propagators by introducing new propagators, associated with auxiliary masses (entering in propagators), for the self-energy and vacuum polarization processes. Finally, one can use  dimensional regularization, which is based on the fact that logarithmically divergent integrals (as for the divergences associated basic QED processes) are convergent if one modifies the spacetime dimensions through $d=4\rightarrow d=4-\epsilon$ where $\epsilon$ is a small positive number. In all cases, after regularization, the divergent integrals are parameterized by the regularization parameters and are still divergent in the limit $\Lambda \rightarrow \infty $ or $\epsilon \rightarrow 0$, for instance.
This is where  renormalization comes into play.

Renormalization is a mathematical technique that consists of redefining the electron mass and charge (in addition to fields) such that the divergences, that come from including the QED corrections, are eliminated: absorbed by the bare physical quantities. It is needed at this point to note that the experimentally observed mass and charge, $m_\text{exp}$ and $e_\text{exp}$ are results of experiments that already include QED corrections. On the other hand, one can imagine a world in which the QED interaction is switched off; in this world, one would measure what is known as bare mass and charge: $m_0$ and $e_0$. This distinction clearly shows that the electron mass and charge that we start with (before switching QED on: before taking it into consideration) should be the bare ones, instead of the measured ones. This awareness played a crucial role in formulating the renormalization theory. Since we do not have access to bare quantities, and since infinity is not natural (not measurable), the renormalization theory says that we are allowed to redefine our physical constants such that the bare ones absorb the emerging divergences and lead to overall values of the physical constants that correspond to the experimentally observed ones. Detailed discussions on renormalization in the quantum field theory  are provided by Collins in Ref.~\citenum{collins_1984}, Greiner and Reinhardt in Ref.~\citenum{greiner_reinhardt_qed} (chapter 5), Peskin and Schroeder in Ref.~\citenum{peskin:schroeder} (chapter 7), Itzykson and Zuber in Ref.~\citenum{itzykson_zuber_qft_1980} (section 7.1), and Huang in  Ref.~\citenum{huang2010QuantumFieldTheory} (chapter 13) and \citenum{huang2013Renormalization}.

\bibliography{article}

\end{document}